\documentclass[rmp,aps,preprint,nofootinbib,endfloats]{revtex4}

\usepackage{graphics}
\usepackage{epsfig}

\def\ä{\"{a}}
\def\ü{\"{u}}
\def\ö{\"{o}}
\def\Ä{\"{A}}
\def\Ü{\"{U}}
\def\Ö{\"{O}}

\begin{document}

\title{Advances in atomic force microscopy}

\author{Franz J. Giessibl}
\email{Franz.Giessibl@physik.uni-augsburg.de}
\affiliation{Experimentalphysik VI, Electronic Correlations
and Magnetism, Institute of Physics, Augsburg University,
86135 Augsburg, Germany}

\begin{abstract}
This article reviews the progress of atomic force microscopy (AFM)
in ultra-high vacuum, starting with its invention and covering
most of the recent developments. Today, dynamic force microscopy
allows to image surfaces of conductors \emph{and} insulators in
vacuum with atomic resolution. The mostly used technique for
atomic resolution AFM in vacuum is frequency modulation AFM
(FM-AFM). This technique, as well as other dynamic AFM methods,
are explained in detail in this article. In the last few years
many groups have expanded the empirical knowledge and deepened the
theoretical understanding of FM-AFM. Consequently, the spatial
resolution and ease of use have been increased dramatically.
Vacuum AFM opens up new classes of experiments, ranging from
imaging of insulators with true atomic resolution to the
measurement of forces between individual atoms.
\end{abstract}

\date{September 23 2002, revised version February 17 2003,
accepted for publication in Reviews of Modern Physics}
\maketitle
\tableofcontents

\section{INTRODUCTION}
\label{sec:intro}

Imaging individual atoms has been elusive until the introduction of
the Scanning Tunneling Microscope (STM) in 1981 by
\citet*{Binnig:1982}. This humble instrument has provided a
breakthrough in our possibilities to investigate matter on the atomic
scale: for the first time, the individual surface atoms of flat
samples could be made visible in real space. Within one year of its
invention, the STM has helped to solve one of the most intriguing
problems in surface science: the structure of the Si(111)-($7\times7$)
surface. The adatom layer of Si(111)-($7\times7$) was imaged with an
STM by \citet{Binnig:1983}. This image, combined with X-ray- and
electron-scattering data has helped \citet*{Takayanagi:1985} to
develop the Dimer-Adatom-Stacking fault (DAS)-model for
Si(111)-($7\times7$). G. Binnig and H. Rohrer, the inventors of the
STM were rewarded with the Nobel Prize in Physics in 1986. The
historic events about the initial steps and the rapid success of the
STM including the resolution of the silicon $7\times7$ reconstruction
have been presented in the Nobel Prize lecture of \citet{Binnig:1987}.
The spectacular spatial resolution of the STM along with its
intriguing simplicity has launched a sprawling research community with
a significant impact on surface science
\citep{Mody:2002}. A large number of metals and semiconductors have
been investigated on the atomic scale and marvelous images of the
world of atoms have been created within the first few years after the
inception of the STM. Today, the STM is an invaluable asset in the
surface scientist's toolbox.

\begin{figure}[h]
  \centering
  \includegraphics[width=8cm,clip=true]{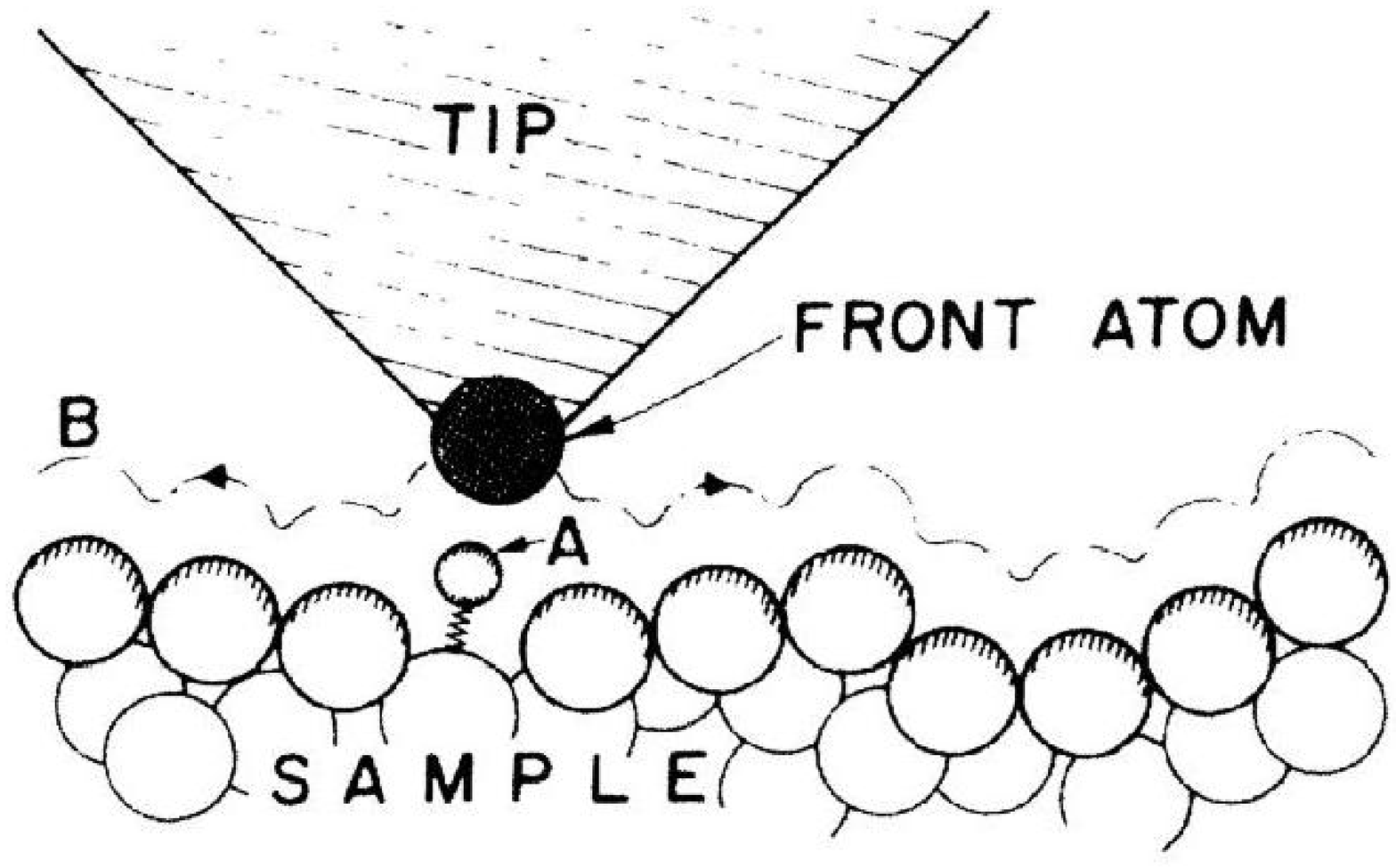}
  \caption{STM or AFM tip close to a sample (Fig. 1a of \citet{Binnig:1986b}).
  }\label{BinnigAFMtip}
\end{figure}

Despite the phenomenal success of the STM, it has a serious
limitation. The STM requires electrical conduction of the sample
material, because the STM uses the tunneling current which flows
between a biased tip close to a sample. However, early STM experiments
have shown that whenever the tip-sample distance is small enough that
a current can flow, significant forces will act collaterally to the
tunneling current. Soon it was speculated, that these forces could be
put to good use in the atomic force microscope (AFM). The force
microscope was invented by \citet{Binnig:1986a} and shortly after its
invention, \citet*{Binnig:1986b} introduced a working prototype while
Binnig and Gerber spent a sabbatical at Stanford and the IBM Research
Laboratory in Almaden, California. \citet{Binnig:1986b} were aware
that even in STM operation, significant forces between single atoms
are acting, and were confident that the AFM could ultimately achieve
true atomic resolution, see Figure \ref{BinnigAFMtip}, adapted from
\citet{Binnig:1986b}.
The STM can only image electrically conductive samples which
limits its application to imaging metals and semiconductors. But
even conductors -- except for a few special materials, like highly
oriented pyrolytic graphite (HOPG) -- cannot be studied in ambient
conditions by STM but have to be investigated in an ultra-high
vacuum (UHV). In ambient conditions, the surface layer of solids
constantly changes by adsorption and desorption of atoms and
molecules. UHV is required for clean and well defined surfaces.
Because electrical conductivity of the sample is not required in
AFM, the AFM can image virtually any flat solid surface without
the need for surface preparation. Consequently, thousands of AFMs
are in use in universities, public and industrial research
laboratories all over the world. The most of these instruments are
operated in ambient conditions.

For studying surfaces on the atomic level, an ultra-high vacuum
environment is required, where it is more difficult to operate an AFM.
In addition to the experimental challenges of STM, the AFM faces four
more substantial experimental complications which are summarized in
section \ref{section_AFM_challenges}. While \citet*{Binnig:1986b} have
anticipated true atomic resolution capability of the AFM from the
beginning, it has taken five years before atomic resolution on inert
surfaces could be demonstrated \citep{Giessibl:1991b,Giessibl:1992b,Ohnesorge:1993},
see \ref{early_AFM}. Resolving reactive surfaces by AFM with atomic resolution
took almost a decade from the invention of the AFM. The Si(111)-($7\times7$)
surface, a touchstone of the AFMs feasibility as a tool for surface science, was resolved with atomic
resolution by dynamic AFM \citep{Giessibl:1995}. The new AFM mode has proven
to work as a standard method, and in 1997 Seizo Morita from Osaka
University in Japan initiated an international workshop about \lq
non-contact AFM\rq{}. A year later, the \lq\lq First International
Workshop on Non-contact Atomic Force Microscopy (NC-AFM)\rq\rq{} was
held in Osaka, Japan with about 80 attendants. This meeting was
followed in 1999 by the Pontresina (Switzerland) meeting with roughly
120 participants and the \lq\lq Third International
Conference on Non-contact Atomic Force Microscopy (NC-AFM)\rq\rq{}
in Hamburg, Germany in 2000 with more than 200 participants.
The fourth meeting took place in
September 2001 in Kyoto, Japan, and the 2002 conference met at
McGill University in Montreal, Canada. The next meeting is scheduled
for Ireland in Summer 2003. The proceedings for these workshops and
conferences
\citep{Morita:1999,Bennewitz:2000,Schwarz:2001,Tsukada:2002,Hoffmann:2003}
and a recent review by \citet*{Garcia:2002} are a rich source of
information for AFM and its role in surface science. Also, a
multi-author book about NCAFM has recently become available
\citep{Morita:2002}. The introduction of this book
\cite{Morita:2002bk1} covers interesting aspects of the history of the
AFM. This review can only cover a part of the field, and the author
must apologize to the colleagues whose work he was not able to treat
in the depth it deserved in this review. However, many of these
publications are listed in the bibliography and references therein.

\section{ATOMIC FORCE MICROSCOPY (AFM) PRINCIPLE}

\subsection{Relation to scanning tunneling microscopy (STM)}

The AFM is closely related to the STM, and it shares its key
components, except for the probe tip. The principle of STM is
explained very well in many excellent books and review articles,
e.g.
\citet{Binnig:1985,Binnig:1987,Binnig:1999,Chen:1993,Guentherodt:1991,Stroscio:1993,Wiesendanger:1994}
and \citet{Wiesendanger:1998}. Nevertheless, the key principle of
STM is described here because the the additional challenges faced
by AFM become apparent clearly in a direct comparison.
\begin{figure}[h]
  \centering
  \includegraphics[width=8cm,clip=true]{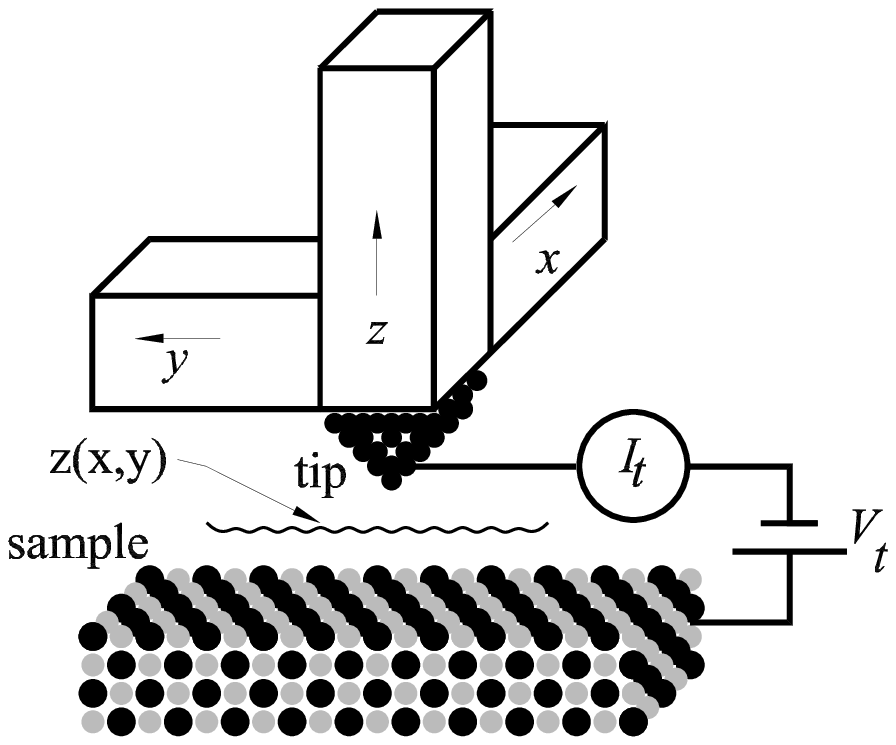}
  \caption{A scanning tunneling microscope (schematic).}\label{STM}
\end{figure}
Figure \ref{STM} shows the general setup of a scanning tunneling microscope (STM): a
sharp tip is mounted on a scanning device (\lq\lq xyz scanner\rq\rq)
which allows 3-dimensional positioning
in $x,y$ and $z$ with subatomic precision. The tunneling
tip is typically a wire that has been sharpened by chemical
etching or mechanical grinding. W, Pt-Ir or pure Ir
are often choosen as a tip material. A bias voltage $V_{t}$ is applied
to the sample and when the distance between tip and sample is in the range
of several \AA ngstr\o ms, a tunneling current $I_{t}$ flows between the tip and
sample. This current is used as the feedback signal in a $z-$feedback loop.

In the \lq\lq topographic mode\rq\rq, images are
created by scanning the tip in the $xy$-plane and recording the
$z$-position required to keep $I_t$ constant. In the
\lq\lq constant height mode\rq\rq, the probe is scanned rapidly such that
the feedback cannot follow the atomic corrugations. The atoms are then
apparent as modulations of $I_t$ which are recorded as a function of $x$ and $y$. The scanning is usually
performed in a raster fashion with a fast scanning direction (sawtooth or
sinusoidal signal) and a slow scanning direction (sawtooth signal). A
computer controls the scanning of the surface in the $xy$ plane while recording
the $z$-position of the tip (topographic mode) or $I_t$
(constant height mode). Thus, a three dimensional image $z(x,y,I_t\approx const.)$
or $I_t(x,y,z\approx const.)$ is created.

In the AFM, the tunneling tip is replaced by a force-sensing cantilever.
The tunneling tip can also be replaced by an optical near-field probe, a
microthermometer etc., giving rise to a
whole family of scanning probe microscopes, see \citet{Wickramasinghe:1989}.

\subsubsection{Tunneling current in STM}

In an STM, a sharp tip is brought close to an electrically conductive
surface that is biased at a voltage $V_{t}$. When the separation is
small enough, a
current $I_{t}$ flows between them. The typical distance between tip and
sample under these conditions is a few atomic diameters, and the transport
of electrons occurs by tunneling.
\begin{figure}[h]
  \centering
  \includegraphics[width=8cm,clip=true]{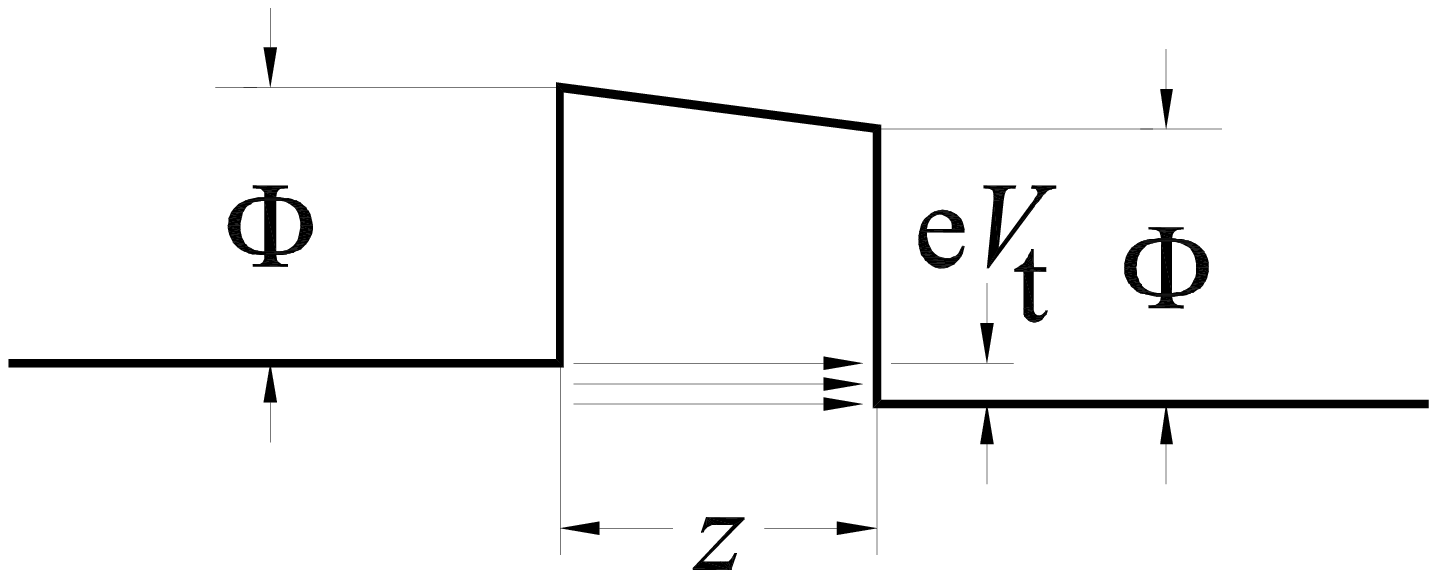}
  \caption{Energy diagram of an idealized tunneling gap. The image charge effect
  (see \citet{Chen:1993}) is not taken into account here.}\label{tunnelgap}
\end{figure}
When $\left|
V_{t}\right| $ is small compared to the workfunction $\Phi $, the
tunneling barrier is roughly rectangular (see Fig. \ref{tunnelgap})
with a width $z$ and a height
given by the workfunction $\Phi $. According to elementary quantum
mechanics, the tunneling current is given by:
\begin{equation}
I_{t}(z)=I_{0}e^{-2\kappa _{t}z}.  \label{It_z}
\end{equation}
$I_{0}$ is a function of the applied voltage and the density of states in
both tip and sample and
\begin{equation}
\kappa _{t}=\sqrt{2m\Phi }/\hbar
\end{equation}
where $m$ is the mass of the electron and $\hbar $ is Planck's constant. For
metals, $\Phi \approx 4$\thinspace eV, thus $\kappa _{t}\approx 1$ \AA$^{-1}$%
. When $z$ is increased by one \AA ngstr\o m, the current drops by an order of
magnitude. This strong distance dependence is pivotal for the
atomic resolution capability of the STM. Most of the tunneling current is
carried by the atom that is closest to the
sample (\lq\lq front atom\rq\rq). If the sample is very flat, this front atom remains
the atom that is closest to the sample during scanning in $x$ and $y$ and
even relatively blunt tips yield atomic resolution easily.

\subsubsection{Experimental measurement and noise}

\begin{figure}[h]
  \centering \includegraphics[width=6cm,clip=true]{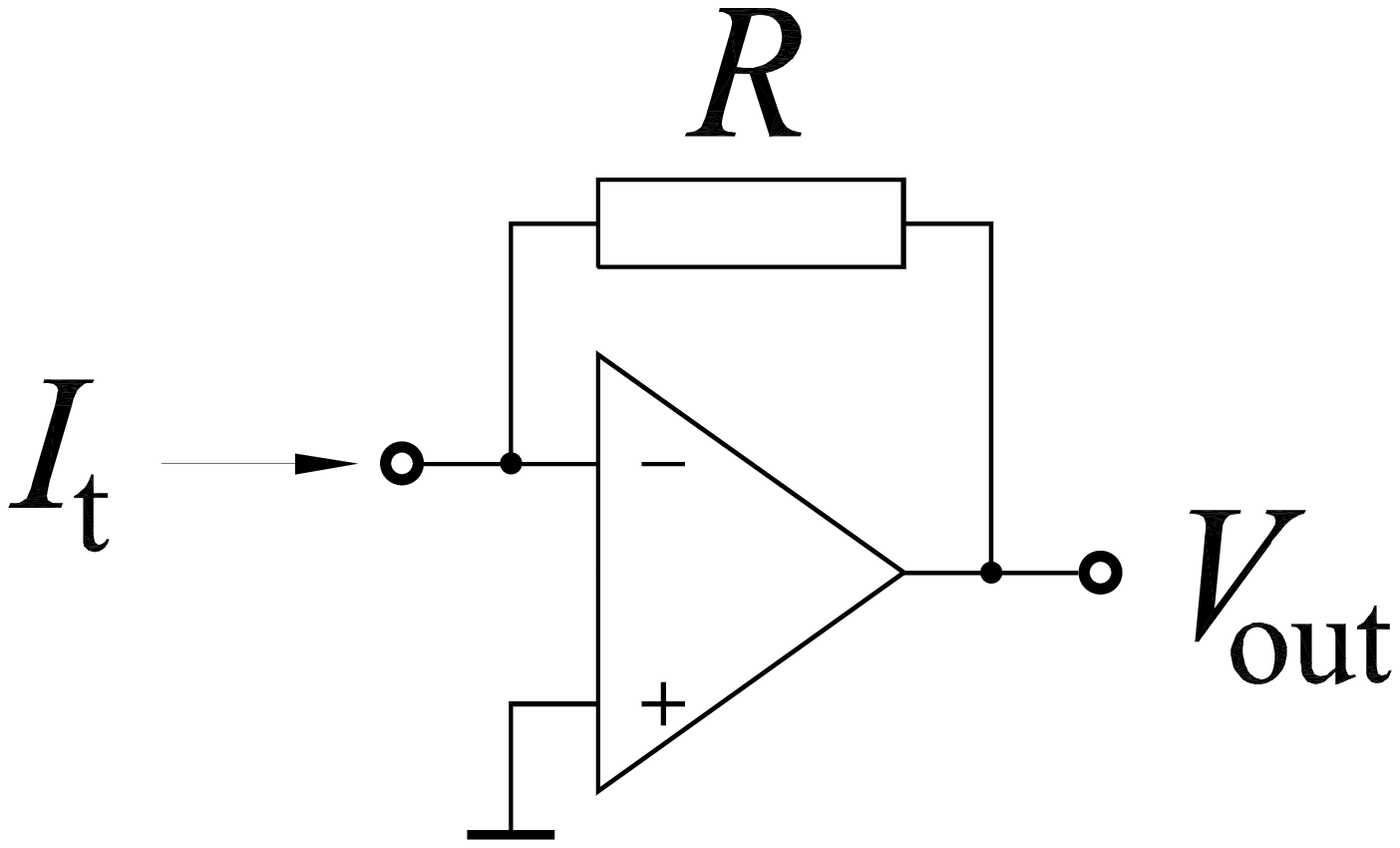}
  \caption{A simple current-to-voltage converter for an STM and for
  the qPlus sensor shown in Fig. \ref{qPlus}. It consists of an
  operational amplifier with high speed, low noise and low input bias
  current with a feedback resistor (typical impedance $R\approx
  10^8$\,$\Omega$) with low parasitic capacitance. The output voltage
  is given by $V_{out}=-R\times I_{t}$.}\label{cur_amp}
\end{figure}

\begin{figure}[h]
  \centering
  \includegraphics[width=8cm,clip=true]{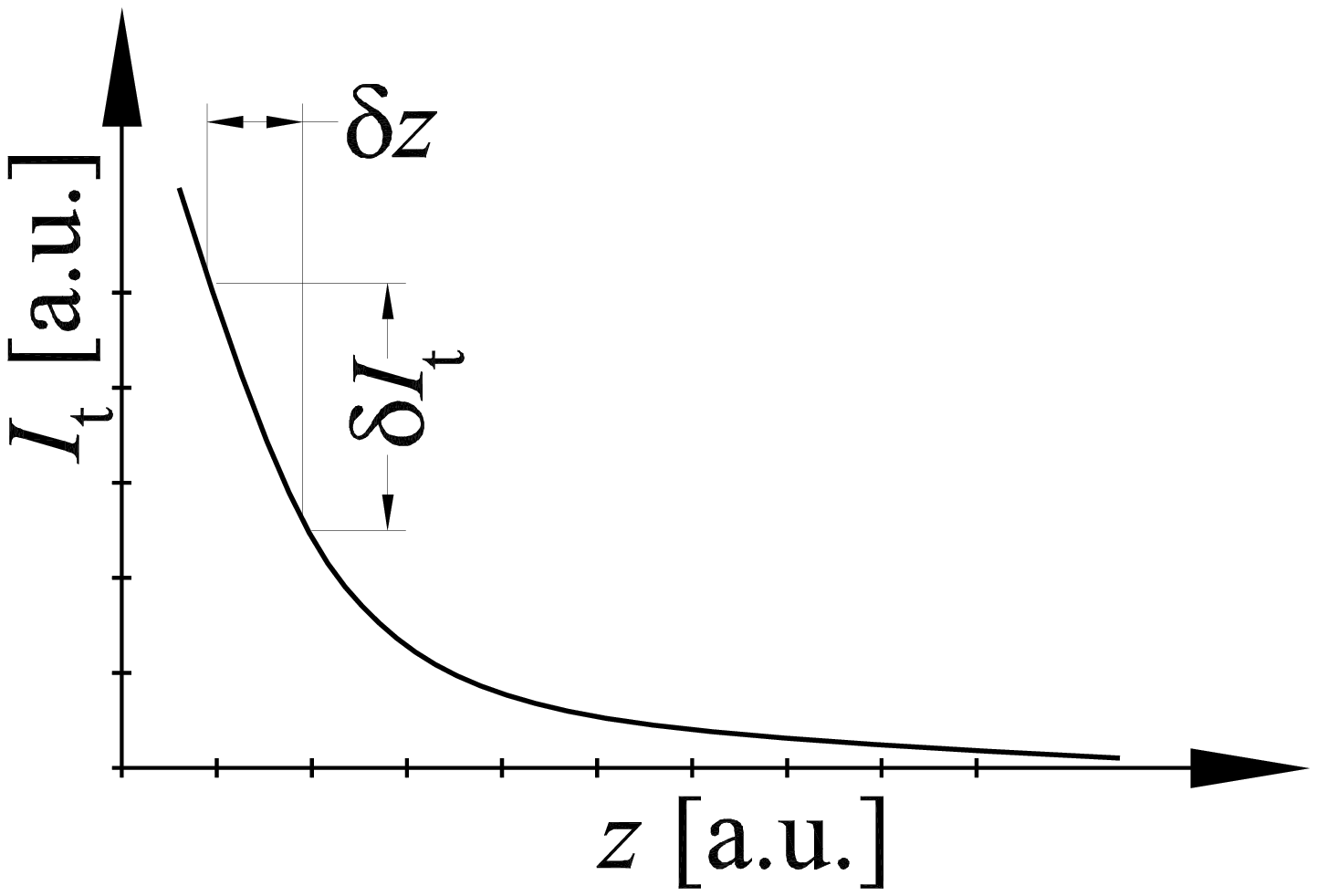}
  \caption{Tunneling current as a function of distance and relation
  between current noise $\delta I_{t}$ and vertical noise $\delta z$ (arbitrary units).}\label{current_z}
\end{figure}

The tunneling current is measured with a current-to-voltage
converter (see Fig. \ref{cur_amp}), a simple form of it consists
merely of a single operational amplifier (OPA) with low noise and
low input bias current, and a feedback resistor with a typical
impedance of $R=100$\,M$\Omega$ and small parasitic capacitance.
The tunneling current $I_{t}$ is used to measure the distance
between tip and sample. The noise in the imaging signal (tunneling
current in STM, force or some derived quantity in AFM) needs to be
small enough such that the corresponding vertical noise $\delta z$
is considerably smaller than the atomic corrugation of the sample.
In the following, the noise levels for imaging signals and
vertical positions are described by the root-mean-square (rms)
deviation of the mean value and indicated by the prefix $\delta$,
i.e.
\begin{equation}
\delta \xi \equiv \sqrt{<(\xi-<\xi>)^2>}.
\end{equation}
To achieve atomic resolution with an STM or AFM, a first
necessary condition is that the mechanical vibrations between tip
and sample are smaller than the atomic corrugations. This
condition is met by a microscope design emphasizing utmost
stability and establishing proper vibration isolation, as
described in Refs. \citet{Chen:1993, Park:1993, Kuk:1988}. In the
following, proper mechanical design and vibration isolation will
be presumed and are not discussed further. The inherent vertical
noise in an STM is connected to the noise in the current
measurement. Figure \ref{current_z} shows the qualitative
dependence of the tunneling current $I_{t}$ as a function of
vertical distance $z$.
Because the measurement of $I_{t}$ is
subject to noise, the vertical distance measurement is also subject to a
noise level $\delta z$:
\begin{equation}
\delta z_{I_{t}}=\frac{\delta I_{t}}{|\frac{\partial I_{t}}{\partial z}|}.
\label{z_noise_stm}
\end{equation}
It is shown below, that the noise in the current measurement $\delta I_{t}$ is small
and that $\frac{\partial I_{t}}{\partial z}$
is quite large, consequently the vertical noise in STM is very small.

The dominating noise source in the tunneling current is the
Johnson noise of both the feedback resistor $R$ in the current
amplifier, the Johnson noise in the tunneling junction, and the
input noise of the operational amplifier. The Johnson noise
density of a resistor $R$ at temperature $T$ is given by :
\begin{equation}
n_{R}=\sqrt{4k_{B}TR}
\end{equation}
\citep{Horowitz:1989} where $k_{B}$ is the Boltzmann constant. In
typical STMs, the tunneling current is of the order of
$I_{t}\approx 100$ pA and measured with an acquisition bandwidth
of $B \approx 1$ kHz, where $B$ is roughly determined by the spatial
frequency of features that are to be scanned times the scanning speed.
Thus, for a spatial frequency of 4 Atoms per nm and a scanning speed
of 250 nm/s, a bandwidth of $B=1$\,kHz is sufficient to map each atom
as a single sinusoidal wave.  With a
gain of $V/I=R=100$ M$\Omega$ and $T=300\,$K, the rms voltage
noise is $n_{i}\sqrt{B}$=$\sqrt{4k_{B}TRB}=40\mu V$ at room
temperature, corresponding to a current noise of $\delta I_t =0.4$
pA. With Eqs. \ref{It_z} and \ref{z_noise_stm}, the vertical noise
is
\begin{equation}
\delta z_{I_{t}} \approx \frac{\sqrt{4k_{B}TB/R}}{2\kappa_{t} |I_{t}| }
\label{z_noise_special}
\end{equation}
which amounts to a $z-$noise of $0.2$ pm in the present example.
Thus, in STM the thermal noise in the tunneling current is not
critical, because it is much smaller than the required resolution.
It is interesting to note that the noise in STM increases
proportional to the square root of the required bandwidth $B$, a
moderate rate compared to the $B^{1.5}$ dependence which holds for
frequency modulation AFM (see Eq. \ref{f_error_c}).

The spectacular spatial resolution and relative ease of obtaining atomic
resolution by STM rests on three properties of the tunneling current:
\begin{itemize}
  \item As a consequence of the strong distance dependence of the tunneling
  current, even with a relatively blunt tip the chance is high that a
  single atom protrudes far enough
  out of the tip such that it carries the main part of the
  tunneling current;

  \item Typical tunneling currents are in the nano-ampere range - measuring
  currents of this magnitude can be done with a very good signal to noise
  ratio even with a simple experimental setup;

  \item Because the tunneling current is a monotonic function of the tip-sample
   distance, it is easy to establish a feedback loop which controls the
   distance such that the current is constant.
\end{itemize}

It is shown in the next section that none of these conditions is
met in the case of the AFM and therefore, substantial hurdles had
to be overcome before atomic resolution by AFM became possible.

\subsection{Tip-sample forces $F_{ts}$}\label{subsection_tip_sample_forces}

The AFM is similar to an STM, except that the tunneling tip is replaced by a force sensor.
\begin{figure}[h]
  \centering
  \includegraphics[width=8cm,clip=true]{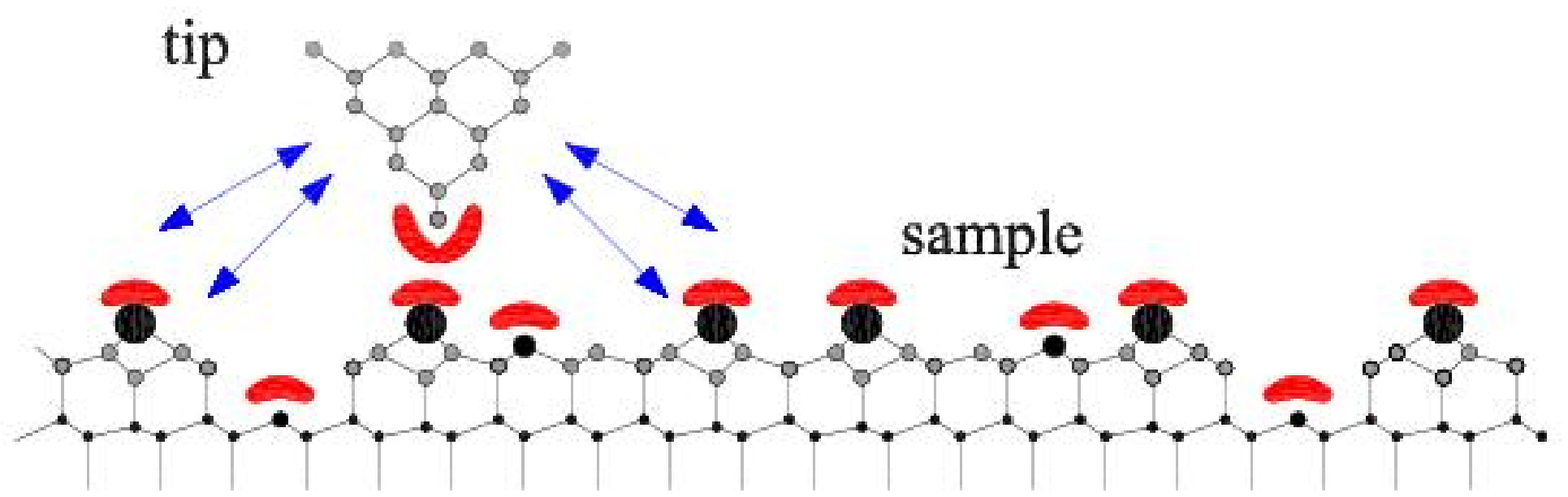}
  \caption{Schematic view of an AFM tip close to a sample. Chemical short range
  forces act when tip and sample orbitals (red crescents) overlap. Long range forces
  (indicated with blue arrows) originate in the full volume and surface of the tip and are a critical function of the mesoscopic
  tip shape.}\label{tip_sample_force}
\end{figure}
Figure \ref{tip_sample_force} shows a sharp tip close to a sample. The
potential energy between the tip and sample $V_{ts}$ causes a
$z$ component of the tip-sample force
$F_{ts}$=-$\frac{\partial V_{ts}}{\partial z}$ and a \lq\lq tip-sample
spring constant\rq\rq $k_{ts}$=-$\frac{\partial F_{ts}}{\partial z}$.
Depending on the mode of operation, the AFM uses $F_{ts}$ or some entity
derived from $F_{ts}$ as the imaging signal.

Unlike the tunneling current, which has a very short range, $F_{ts}$
has long- and short-range contributions. We can classify the
contributions by their range and strength. In vacuum, there are short
range chemical forces (fractions of nm) and van-der-Waals,
electrostatic and magnetic forces with a long range (up to 100 nm). In
ambient conditions, also meniscus forces formed by adhesion layers on
tip and sample (water or hydrocarbons) can be present.

A prototype of the chemical bond is treated in many textbooks on
quantum mechanics (see e.g. \citet{Baym:1969}): the H$_2^+$- ion as a
model for the covalent bond. This quantum mechanical problem can be
solved analytically and gains interesting insights into the character
of chemical bonds. The Morse Potential (see e.g.
\citet{Israelachvili:1991})
\begin{equation}
V_{Morse}=-E_{bond}(2e^{-\kappa (z-\sigma )}-e^{-2\kappa (z-\sigma )})
\end{equation}
describes a chemical bond with bonding
energy $E_{bond}$, equilibrium distance $\sigma$ and a decay length
$\kappa$. With a proper choice of $E_{bond},\sigma$ and $\kappa$, the
Morse potential is an excellent fit for the exact solution of the
H$_2^+$- problem.

The Lennard-Jones potential (see e.g.
\citet{Ashcroft:1981,Israelachvili:1991}):
\begin{equation}
V_{Lennard-Jones}=-E_{bond}( 2\frac{z^{6}}{\sigma^{6}} -
\frac{z^{12}}{\sigma^{12}} )
\end{equation}
has an attractive term $\propto r^{-6}$ originating from the
van-der-Waals interaction (see below) and a repulsive term $\propto
r^{-12}$.

While the Morse potential can be used for a qualitative description of
chemical forces, it lacks an important property of chemical bonds:
anisotropy. Chemical bonds, especially covalent bonds show an inherent
angular dependence of the bonding strength, see
\citet{Pauling:1957,Coulson:1991}. Empirical models which take the
directionality of covalent bonds into account are the Stillinger-Weber
potential \citep{Stillinger:1985}, the Tersoff potential and others.
For a review see \citet{Bazant:1997} and references therein.
The Stillinger-Weber potential appears to be a valid model for the
interaction of silicon tips with silicon samples in AFM:
\lq\lq \emph{Although the various terms }[of the Stillinger-Weber potential]
\emph{lose their physical significance for distortions of the diamond
lattice large enough to destroy sp$^3$ hybridization, the SW potential
seems to give a reasonable description of many states experimentally
relevant, such as point defects, certain surface structures, and the
liquid and amorphous states}\rq\rq \citep{Bazant:1997}.
Using the Stillinger-Weber potential, subatomic features in Si images
have been explained \citep{Giessibl:2000c}. Qualitatively,
these findings have been reproduced with \textit{ab initio}{} calculations
\citep{Huang:2003}.
The Stillinger-Weber potential
necessarily contains nearest and next nearest neighbor interactions.
Unlike solids with a face centered cubic or body centered cubic
lattice structure, solids which crystallize in the diamond structure
are unstable when only next-neighbour interactions are taken into
account. The nearest neighbor contribution of the Stillinger-Weber
potential is
\begin{equation}
V_{n}(r)=E_{bond}A\left[ B(\frac{r}{\sigma ^{\prime }})^{-p}-(\frac{r}{%
\sigma ^{\prime }})^{-q}\right] e^{\frac{1}{r/\sigma ^{\prime }-a}}\text{
for }r<a\sigma ^{\prime },\text{ else }V_{nn}(r)=0.
\end{equation}
\label{Vn}
The next nearest neighbor contribution is:
\begin{equation}
V_{nn}(\mathbf{r}_{i},\mathbf{r}_{j},\mathbf{r}_{k})=E_{bond}\left[
h(r_{ij},r_{ik},\theta _{jik})+h(r_{ji},r_{jk},\theta
_{ijk})+h(r_{ki},r_{kj},\theta _{ikj})\right]
\end{equation}
\label{Vnn} with
\begin{equation}
h(r_{ij},r_{ik},\theta _{jik})=\lambda e^{\gamma (\frac{1}{r_{ij}/\sigma
^{\prime }-a}+\frac{1}{r_{ik}/\sigma ^{\prime }-a})}(\cos \theta _{jik}+%
\frac{1}{3})^{2}\text{ for }r_{ij,ik}<a\sigma ^{\prime },\text{ else }0.
\end{equation}
Stillinger and Weber found optimal agreement with experimental data for the
following parameters:
\begin{center}
\begin{tabular}{ccc}
$A=7.049556277$ & $p=4$ & $\gamma =1.20$ \\
$B=0.6022245584$ & $q=0$ & $\lambda =21.0$ \\
$E_{bond}=3.4723$ aJ & $a=1.8$ & $\sigma^{\prime}=2.0951$ \AA
\end{tabular}
\end{center}
The equilibrium distance $\sigma$ is related to $\sigma^{\prime}$ by $%
\sigma=2^{1/6}\sigma^{\prime}$. The potential is constructed in such a way
to ensure that $V_{n}$ and $V_{nn}$ and all their derivatives with respect
to distance vanish for $r>a\sigma ^{\prime }=3.7718$\,\AA. The diamond
structure is favoured by the SW potential because of the factor $(\cos
\theta+\frac{1}{3})^{2}$ -- this factor is zero when $\theta$ equals the
tetraeder bond angle of $\theta=109.47^{\circ}$.

With increasing computer power, it becomes more and more feasible to
perform \textit{ab initio}{} calculations for tip-sample forces, see
e.g. \citet{Perez:1997,Perez:1998,Ke:2001,Tobik:2001,Huang:2003}.

The van-der-Waals interaction is caused by fluctuations in the electric dipole
moment of atoms and their mutual polarization. For two atoms at distance $z$,
the energy varies as $1/z^{6}$ (\citet{Baym:1969}). Assuming additivity and
disregarding the discrete nature of matter by replacing the sum over individual
 atoms by an integration over a volume with a fixed number density of atoms,
the van-der-Waals interaction between macroscopic bodies can be calculated
(\lq\lq \citet{Hamaker:1937} approach\rq\rq). This approach does
not account for retardation effects due to the finite speed of light and
is therefore only appropriate for distances up to several hundred \AA ngstr\o ms.
For a spherical tip with radius $R$ next to a flat surface ($z$ is the
distance between the plane connecting the centers of the surface atoms
and the center of the closest tip atom)
the van-der-Waals potential is given by \citet{Israelachvili:1991}:
\begin{equation}
V_{vdW}=-\frac{A_{H}R}{6z}.
\end{equation}
The van-der-Waals force for spherical tips is thus proportional to
$1/z^2$, while for pyramidal and conical tips, a $1/z$-force law holds
\citep{Giessibl:1997b}. The
\lq\lq Hamaker constant\rq\rq $A_{H}$ depends on the type of materials
(atomic polarizability and density) of the tip and sample. For most
solids and interactions across vacuum, $A_{H}$ is of the order of
1\,eV. For a list of $A_{H}$ for various materials, see
\citet{Krupp:1967,French:2000}. The van-der-Waals interaction can be quite large
-- the typical radius of an etched metal tip is 100\,nm and with
$z=0.5$\,nm, the van-der-Waals energy is $\approx -30$\,eV, and the
corresponding force is $\approx -10$\,nN. Because of their magnitude,
van-der-Waals forces are a major disturbance in force microscopy.
\citet{Ohnesorge:1993} have shown (see section \ref{early_AFM}) that
the large background vdW forces can be reduced dramatically by immersing the cantilever in
water.

A more modern approach to the calculation of van-der-Waals forces is
described in \citet{Hartmann:1991}.

When the tip and sample are both conductive and have an
electrostatic potential difference $U\neq0$, electrostatic forces
are important. For a spherical tip with radius $R$, the potential
energy is given by \citet{Sarid:1994}. If the distance between a
flat surface and a spherical tip with radius $R$ is small compared
to $R$, the force is approximately given by (see
\citet*{Olsson:1998,Law:2002}):
\begin{equation}
F_{electrostatic}(z)=-\frac{\pi \epsilon_{0}RU^{2}}{d}
\end{equation}
Like the van-der-Waals interaction, the electrostatic interaction can also
cause large forces -- for a tip radius of
100\,nm, $U=1$\,V and $z=0.5$\,nm, the electrostatic force is $\approx -5.5$\,nN.

It is interesting to
note that short-range van-der-Waals forces (energy $\propto 1/z^6$) add up to
long-range overall tip-sample forces because of their additivity. The opposite
effect can occur with electrostatic forces: in ionic crystals, where adjacent atoms
carry opposite charges, the
envelope of the electrostatic field
has a short-range exponential distance dependence \citep{Giessibl:1992a}.

More information about tip-sample forces can be found in
\citet{Garcia:2002,Drakova:2001,Ciraci:1990,Israelachvili:1991,Sarid:1994,Perez:1997,Shluger:1997,Perez:1998,
Shluger:1999,Ke:2001,Tobik:2001,Abdurixit:2000,Tsukada:2002bk,Ke:2002bk,Foster:2002bk}
and references therein.

\subsection{The force sensor (cantilever)}
\begin{figure}[h]
  \centering
  \includegraphics[width=8cm,clip=true]{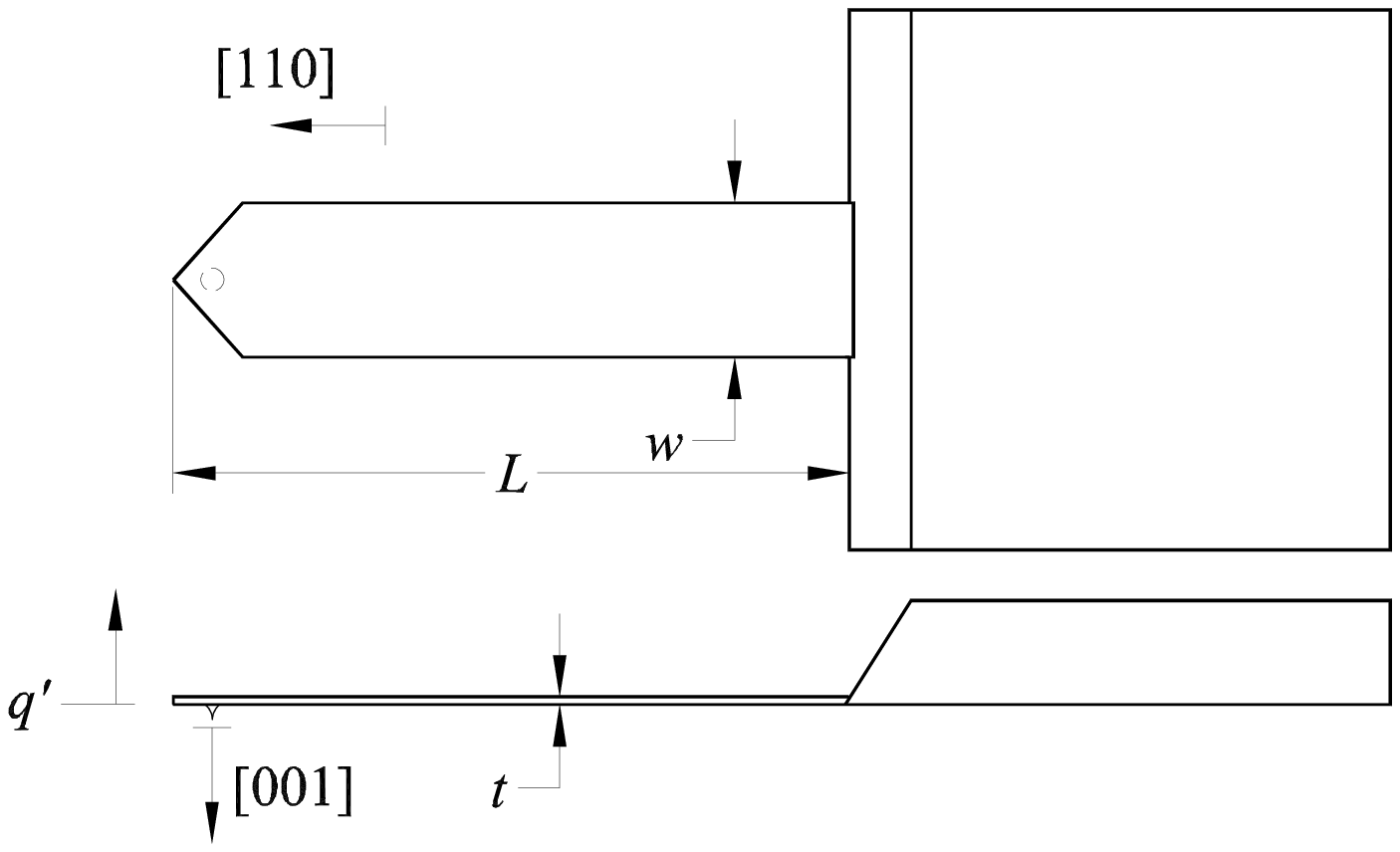}
  \caption{Top view and side view of a microfabricated cantilever (schematic). Most cantilevers have this diving board geometry.}\label{cl}
\end{figure}

Tip-sample forces can vary strongly on the atomic scale, and \citet{Pethica:1986} has proposed that they even
explain artifacts like giant corrugations apparent in STM experiments.
However, it is difficult to isolate force effects in STM, and a dedicated sensor for detecting forces is needed.
The central element of a force microscope and the major instrumental difference to the
scanning tunneling microscope is the spring which senses the force between
tip and sample. For sensing normal tip-sample forces,
the force sensor should be rigid in two axes and relatively soft in the third axis.
This property is fulfilled with a cantilever beam (\lq cantilever\rq{}), and therefore the cantilever geometry is typically used
for force detectors.
A generic cantilever is shown in Fig. \ref{cl}.
For a rectangular cantilever with dimensions $w,t$ and $L$ (see Fig. \ref{cl}),
the spring constant $k$ is given by \citep{Chen:1993}:
\begin{equation}
k=\frac{Ywt^{3}}{4L^{3}}.\label{k_cl}
\end{equation}
where $Y$ is Young's modulus. The fundamental eigenfrequency
$f_{0}$ is given by \citep{Chen:1993}:
\begin{equation}
f_{0}=0.162\frac{t}{L^{2}}\sqrt{ \frac{Y}{\rho} }\label{f0_cl}
\end{equation}
where $\rho$ is the mass density of the cantilever material.

The properties of interest are the stiffness $k$, the
eigenfrequency $f_0$, the quality factor $Q$, the variation of the
eigenfrequency with temperature $\partial f_{0}/\partial T$ and of
course the chemical and structural composition of the tip. The
first AFMs were mostly operated in the static contact mode (see
below), and for this mode the stiffness of the cantilever should
be less than the interatomic spring constants of atoms in a solid
\citep{Rugar:1990}, which amounts to $k \leq 10$\,N/m. This
constraint on $k$ was assumed to hold for dynamic AFM as well.
It turned out later that in dynamic AFM,
$k-$values exceeding hundreds of N/m help to reduce noise and
increase stability \citep{Giessibl:1999a}. The $Q$-factor depends
on the damping mechanisms present in the cantilever. For
micromachined cantilevers operated in air, $Q$ is mainly limited
by viscous drag and typically amounts to a few hundred while in
vacuum, internal and surface effects in the cantilever material
are responsible for damping and $Q$ reaches hundreds of thousands.

The first cantilevers were made from a gold foil with a small
diamond tip attached to it \citep{Binnig:1986a}. Simple
cantilevers can even be cut from household aluminum foil
\citep{Rugar:1990} and etched tungsten wires
\citep{McClelland:1987}. Later, silicon micromachining technology
was employed to build cantilevers in parallel production with well
defined mechanical properties. The first micromachined cantilevers
were built at Stanford in the group of Calvin F. Quate. Initially,
mass produced cantilevers were built from SiO$_2$ and
Si$_3$N$_4$ \citep{Albrecht:1990}. Later, cantilevers with
integrated tips were machined from silicon-on-insulator wafers
\citep{Akamine:1990}. The most common cantilevers in use today are
built from all-silicon with integrated tips pointing in a [001]
crystal direction and go back to \citet*{Wolter:1991} at IBM
Sindelfingen, Germany.
\begin{figure}[h]
  \centering
  \includegraphics[width=8cm,clip=true]{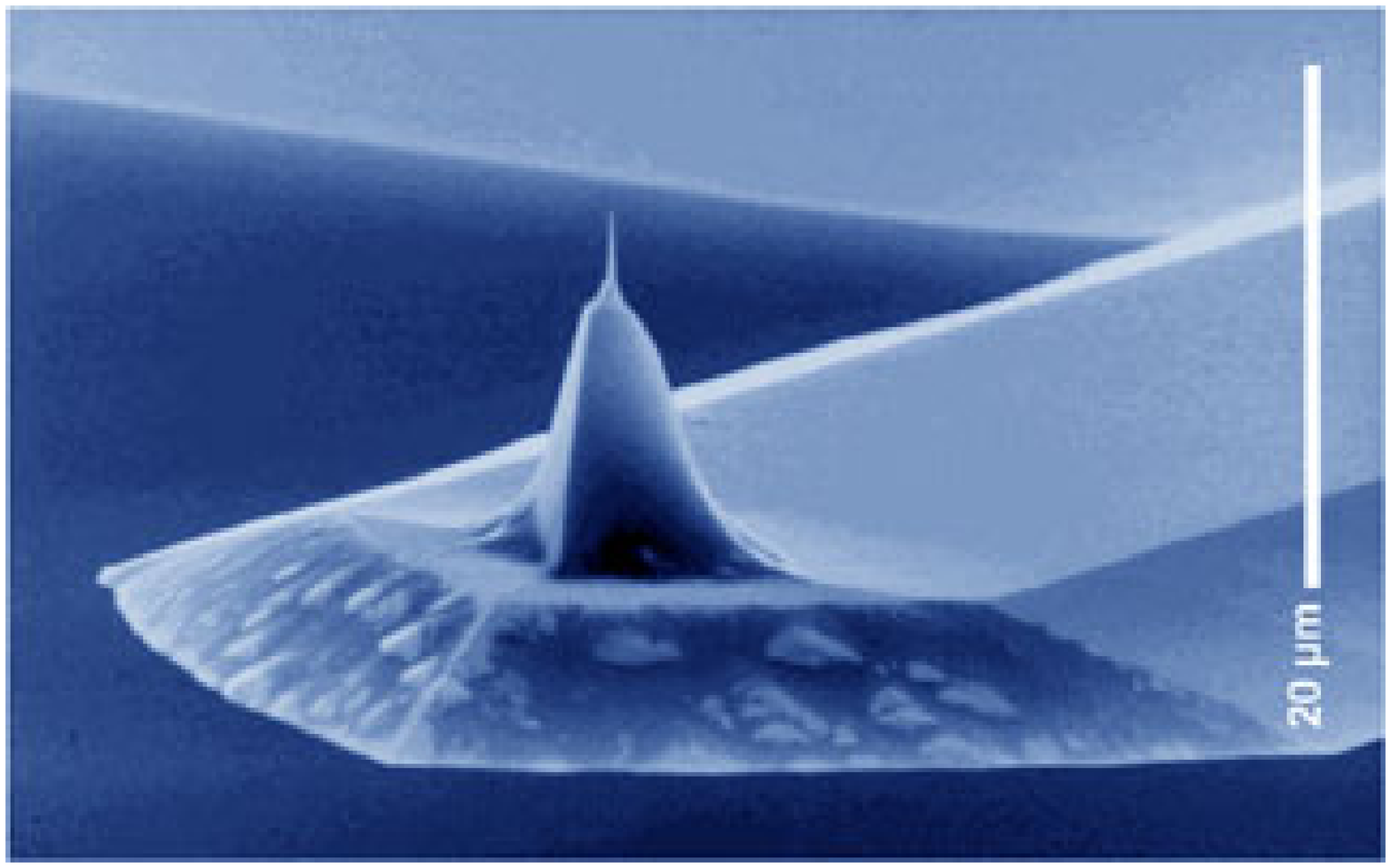}
  \caption{Scanning electron micrograph of a micromachined silicon cantilever with an integrated
  tip pointing in the [001] crystal direction. Source: \citet{Nanosensors}, see \cite{Wolter:1991}.}\label{wolterlever}
\end{figure}
Figures \ref{wolterlever} and \ref{olympuslever} show the type of
cantilevers which are mainly used today: micromachined silicon
cantilevers with integrated tips.
\begin{figure}[h]
  \centering
  \includegraphics[width=8cm,clip=true]{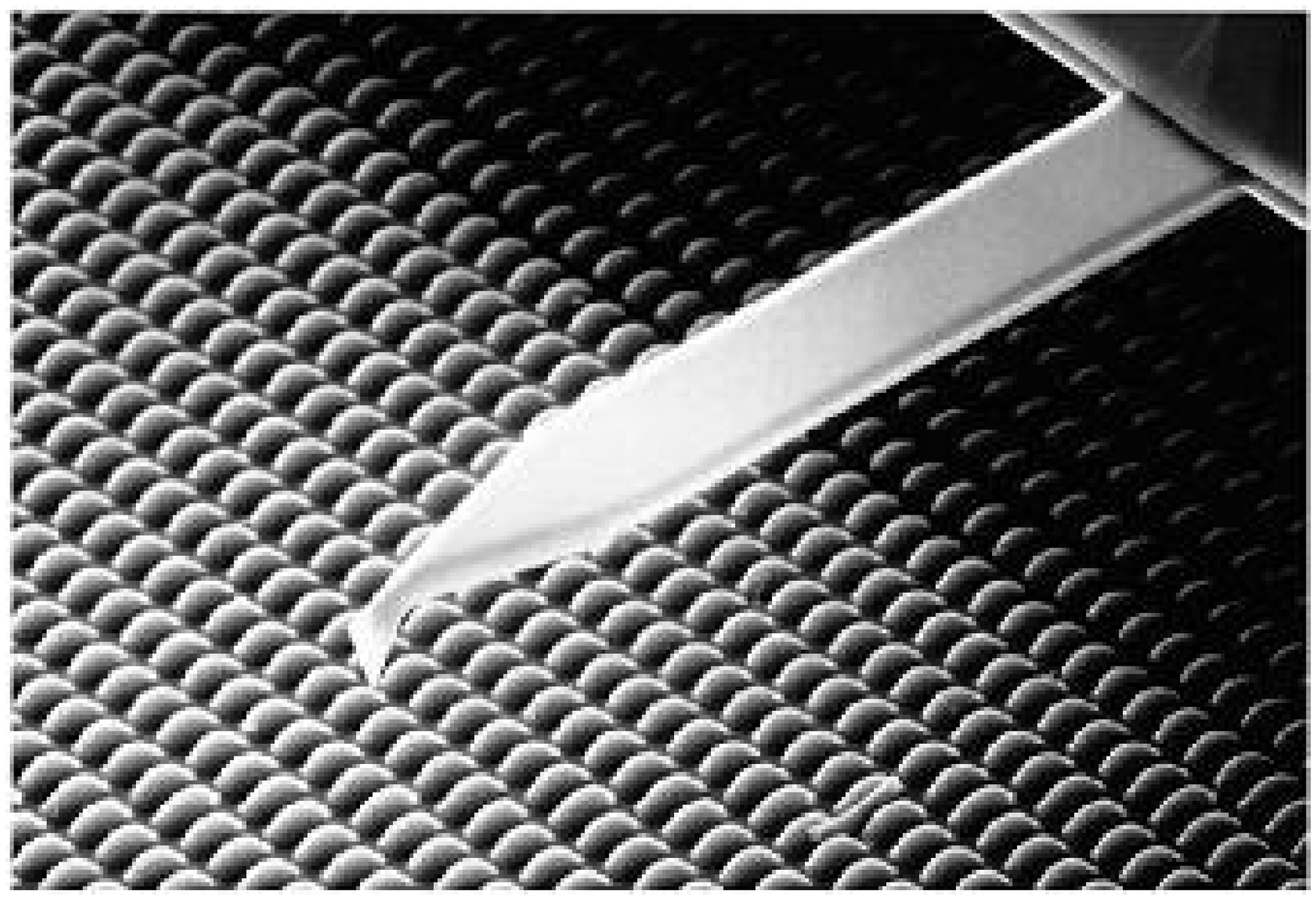}
  \caption{Scanning electron micrograph of a micromachined silicon cantilever with an integrated
  tip pointing in the [001] crystal direction. In this type, the tip is etched free such that the sample area which is
  adjacent to the tip is visible in an optical microscope. Length: 120\,$\mu$m, width:
  30\,$\mu$m, thickness: 2.8\,$\mu$m, $k=15$\,N/m, $f_0=300$\,kHz. Source: \citet{Olympus}.}\label{olympuslever}
\end{figure}
\citet*{Tortonese:1993} have built a self-sensing cantilevers with integrated tips and a built-in
deflection measuring scheme utilizing the piezoresistive
effect in silicon (see Fig. \ref{pl}).

\begin{figure}[h]
  \centering
  \includegraphics[width=8cm,clip=true]{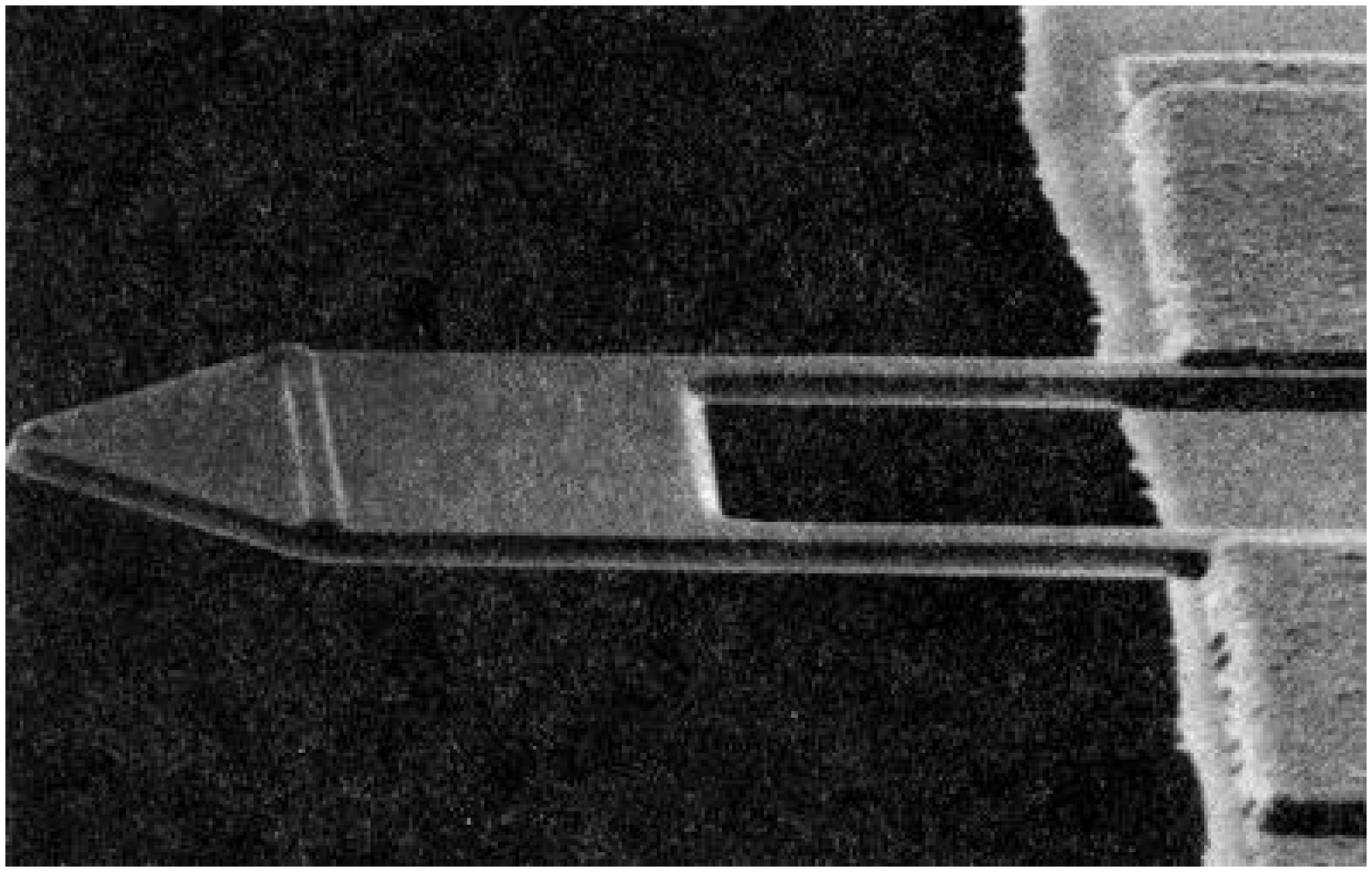}
  \caption{Scanning electron micrograph of a piezoresistive cantilever built from silicon.
  Length: 250\,$\mu$m, full width: 80\,$\mu$m, thickness: 2\,$\mu$m. Source: \citet{Tortonese:1993}.}\label{pl}
\end{figure}
\begin{figure}[h]
  \centering \includegraphics[width=8cm,clip=true]{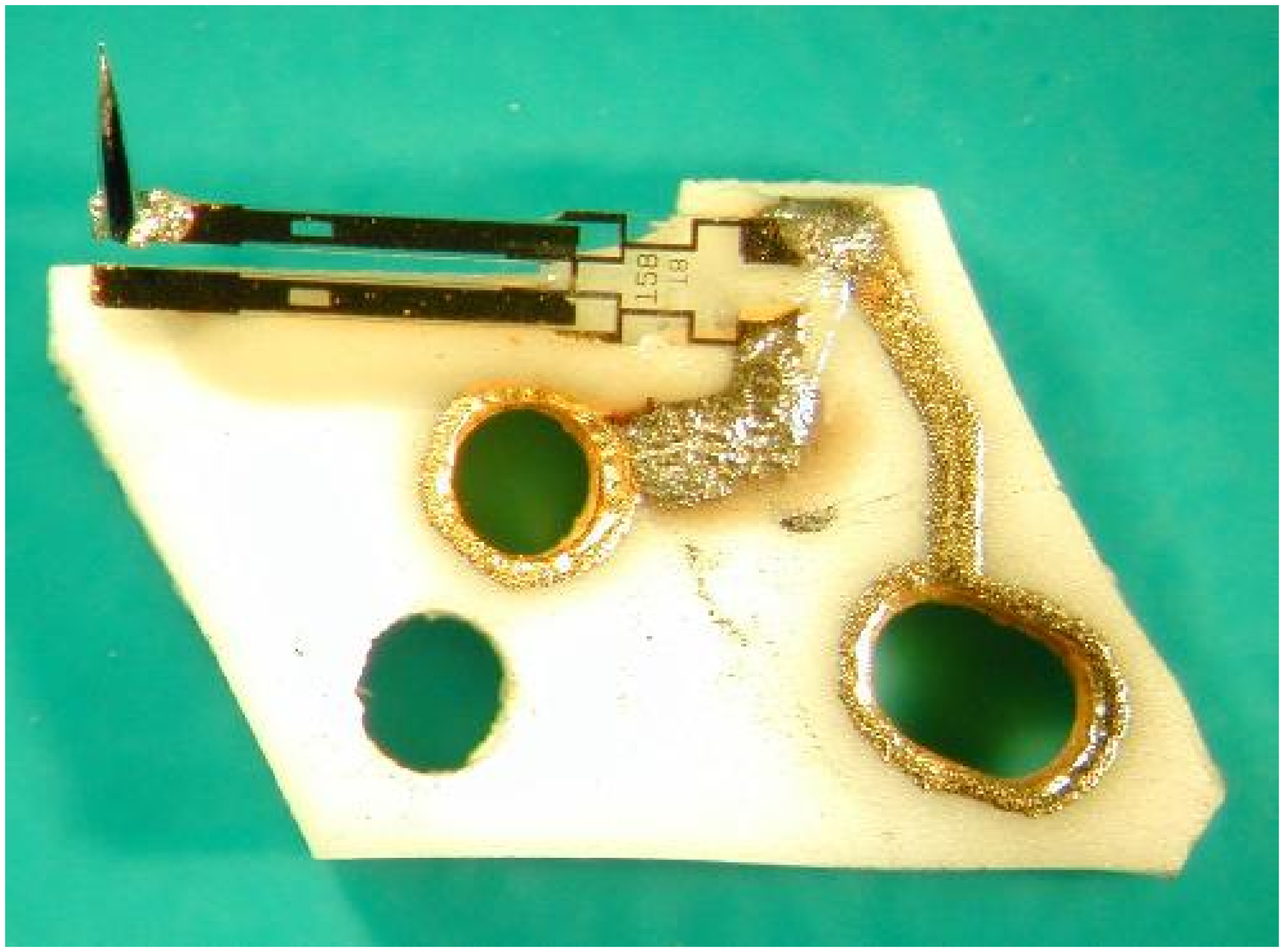}
  \caption{Micrograph of a \lq qPlus\rq{} sensor - a cantilever made
  from a quartz tuning fork. One of the prongs is fixed to a large
  substrate and a tip is mounted to the free prong. Because the fixed
  prong is attached to a heavy mass, the device is mechanically
  equivalent to a traditional cantilever. The dimensions of the free
  prong are: Length: 2\,400\,$\mu$m, width: 130\,$\mu$m, thickness:
  214\,$\mu$m.}\label{qPlus}
\end{figure}
\begin{figure}[h]
  \centering \includegraphics[width=5cm,clip=true]{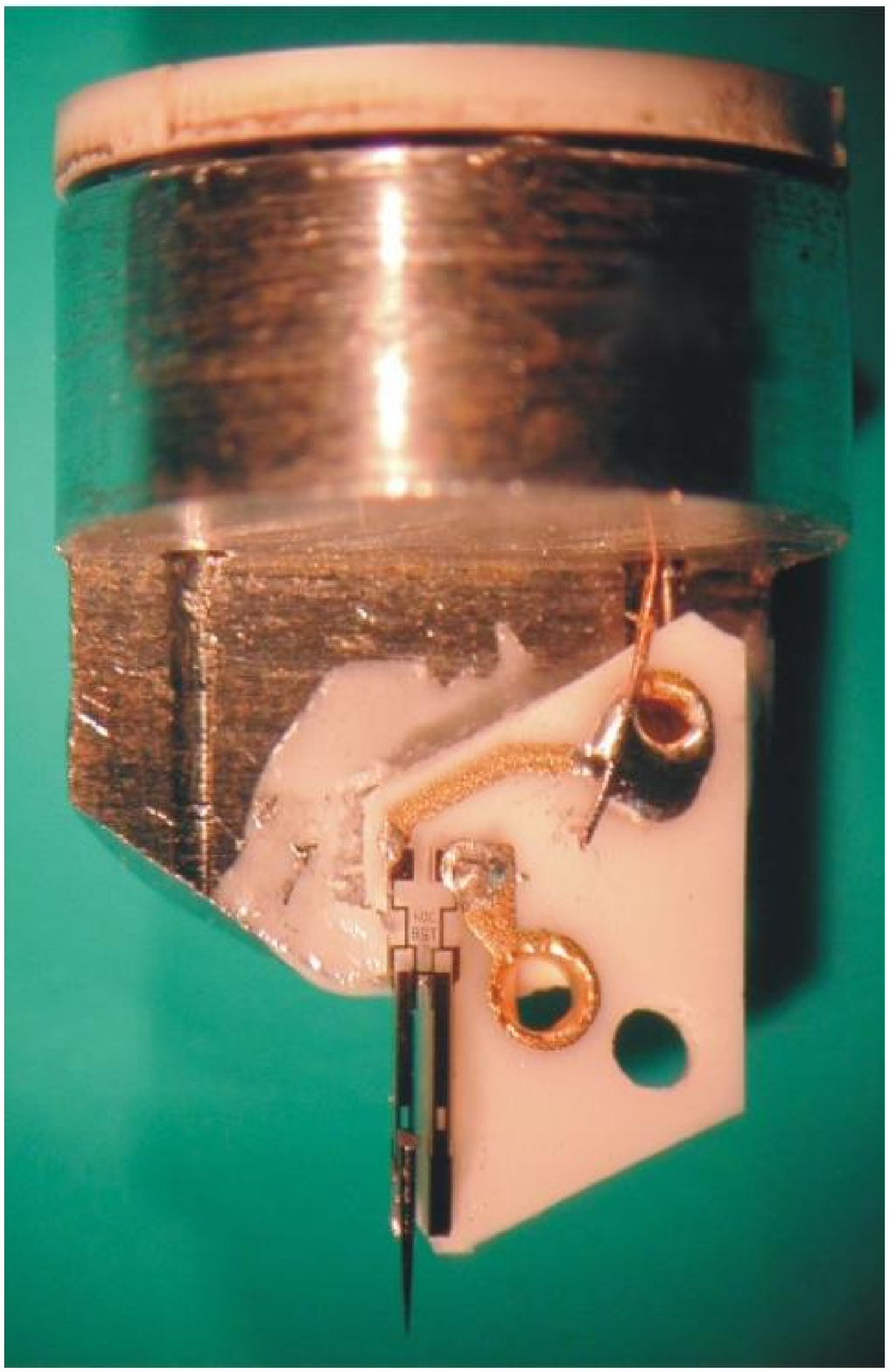}
  \caption{Micrograph of a \lq qPlus\rq{} lateral force sensor. The
  lateral force sensor is similar to the normal force sensor in Fig.
  \ref{qPlus}. It is rotated 90$^{\circ}$ with respect to the normal
  force sensor and its tip is aligned parallel to the free
  prong.}\label{qPluslateral}
\end{figure}

In dynamic AFM, some requirements for the force sensor are similar
to the desired properties of the time keeping element in a watch:
utmost frequency stability over time and temperature changes and little
energy consumption. Around 1970, the watch industry was
revolutionized with the introduction of quartz tuning forks as
frequency standards in clocks \citep{Momosaki:1997,Walls:1985}.
Billions of these devices are now manufactured annually, and the
deviations of even low cost watches are no more than a few seconds
a week. Experimental studies of using quartz based force sensors
were done soon after the invention of the AFM.
\citet{Guethner:1989} and \citet{Guethner:1992} have used tuning forks as
force sensors in acoustic near field microscopy and
\citet{Karrai:1995} have used a tuning fork to control the
distance between the optical near field probe and the surface in a
scanning near-field-optical microscope. \citet{Bartzke:1993} has
proposed the 'needle sensor', a force sensor based on a quartz bar
oscillator. \citet{Rychen:1999} and \citet{Hembacher:2002} have
demonstrated the use of quartz tuning forks at low temperature and
other applications of quartz tuning forks as force sensors can be
found in
\citet{Edwards:1997,Rensen:1999,Ruiter:1997,Todorovic:1998,Tsai:1998,Wang:1998}.
Quartz tuning forks have many attractive properties, but their
geometry is a decisive disadvantage for using them as force
sensors. The great benefit of the fork geometry is the high
$Q$-factor which is a consequence of the presence of an
oscillation mode where both prongs oscillate opposite to each
other. The dynamic forces necessary to keep the two prongs
oscillating cancel in this case exactly. However, this only works
if the eigenfrequency of both prongs matches precisely. The mass
of the tip mounted on one prong and the interaction of this tip
with a sample breaks the symmetry of tuning fork geometry. This
problem can be avoided by fixing one of the two beams and turning
the fork symmetry into a cantilever symmetry, where the cantilever
is attached to a high-mass substrate with a low-loss material.
Figure \ref{qPlus} shows a quartz cantilever based on a quartz
tuning fork \citep{Giessibl:1996,Giessibl:1998,Giessibl:2000b}.
Quartz tuning forks are available in several sizes. We found
optimal performance with the type of tuning forks used in Swatch
wristwatches. In contrast to micromachined silicon cantilevers,
the quartz forks are large. Therefore, a wide selection of tips
can be mounted onto a tuning fork with the mere help of tweezers
and a stereoscopic microscope - sophisticated micromachining
equipment is not needed. Tips made from tungsten, diamond, silicon,
iron, cobalt, samarium, CoSm permanent magnets and iridium have
been built in our laboratory for various purposes.
Figure \ref{qPluslateral} shows a quartz cantilever oriented for lateral
force detection (see section \ref{DLFM}) \citep{Giessibl:2002a}.
Piezoelectric sensors
based on thin films of materials with much higher piezoelectric constants than quartz
\citep{Itoh:1996} are also available. However, these devices lack the very low internal
dissipation and high frequency stability of quartz.
The general advantage of piezoelectric sensors
versus piezoresistive sensors is that the latter dissipate power in the mW range,
while electric dissipation is negligible in piezoelectric sensors. Therefore,
piezoelectric sensors are preferred over piezoresistive schemes for low temperature applications.

\subsubsection{Cantilever tips}

\begin{figure}[h]
  \centering \includegraphics[width=6cm,clip=true]{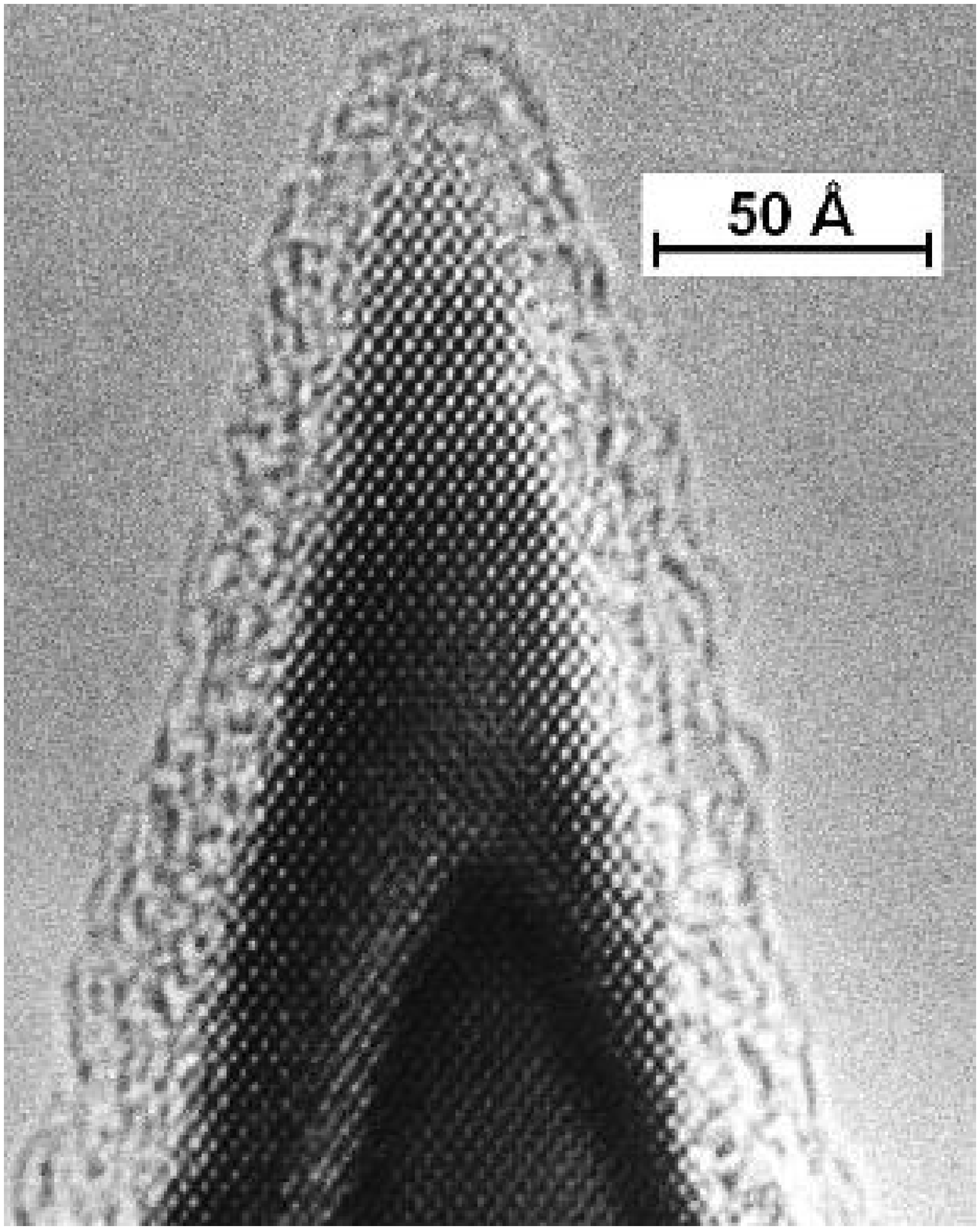}
  \caption{Transmission Electron Micrograph of an extremely sharp
  silicon tip. The native oxide has been etched away with hydrofluoric
  acid before imaging. The 15 -- 20 \AA{} thick coating of the tip is
  mostly due to hydrocarbons which have been polymerized by the
  electron beam. Interestingly, the crystal structure appears to
  remain bulk-like up to the apex of the tip. Source:
  \citet{Marcus:1990}.}\label{marcustip}
\end{figure}

\begin{figure}[h]
  \centering
  \includegraphics[width=8cm,clip=true]{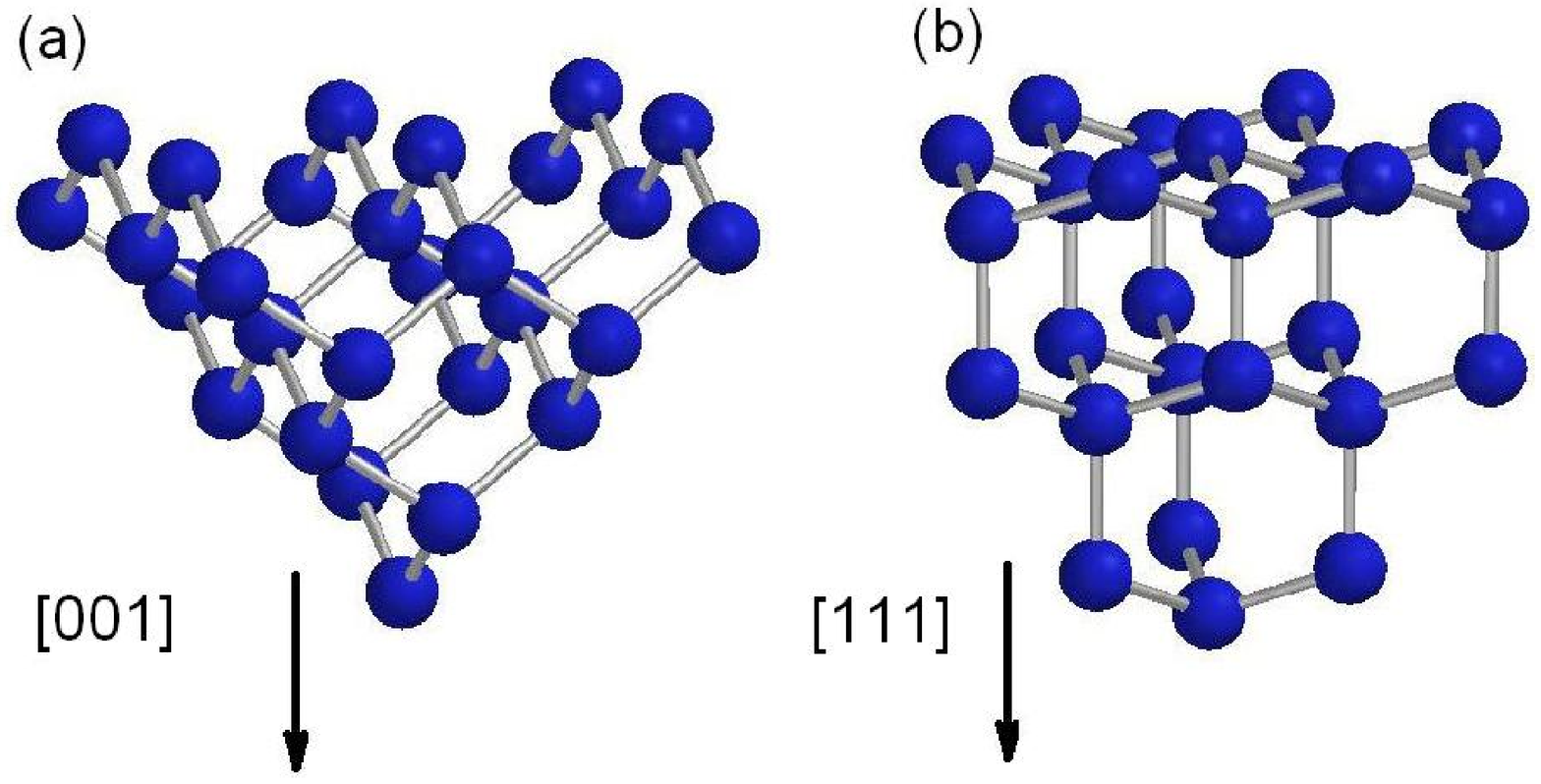}
  \caption{Model of atomic arrangements for bulk-like terminated silicon tips, pointing in a [001] direction (a) and
  in a [111] direction (b).}\label{tipmodels}
\end{figure}

\begin{figure}[h]
  \centering
  \includegraphics[width=6cm,clip=true]{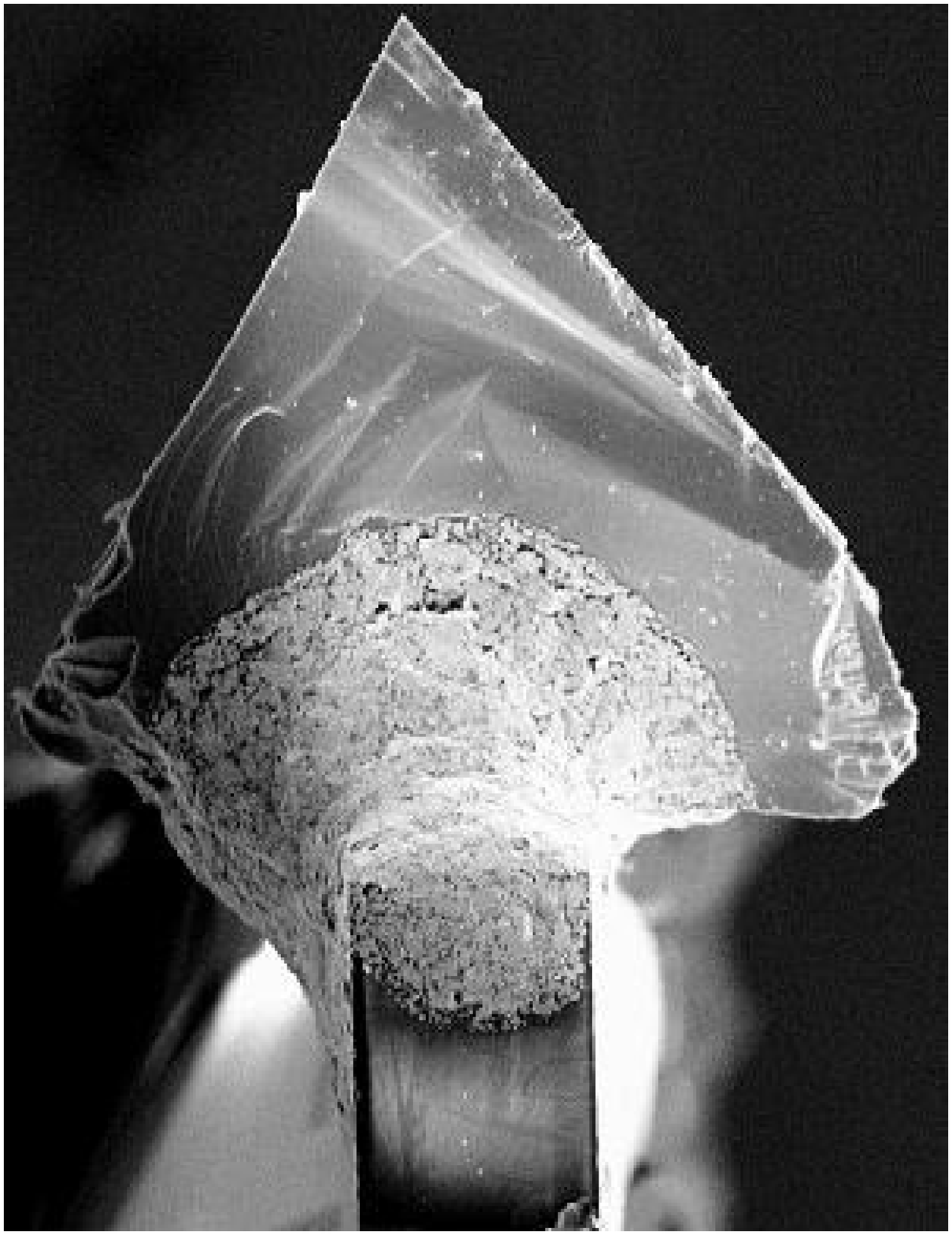}
  \caption{Scanning Electron Micrograph of a cleaved single crystal silicon
  tip attached to the free prong of a qPlus sensor. The rectangular section
  is the end of the free prong with a width of 130\,$\mu$m and a thickness of 214\,$\mu$m.
  The tip is pointed in the [111] direction and bounded by ($\bar{1}$$\bar{1}$$\bar{1}$), (1$\bar{1}$$\bar{1}$) and ($\bar{1}$1$\bar{1}$) planes
  after \citet{Giessibl:2001a}.
  Source: \citet{Schiller:2003}.}\label{qPlus111tip}
\end{figure}

For atomic-resolution AFM, the front atom of the tip should
ideally be the only atom which interacts strongly with the sample.
In order to reduce the forces caused by the shaft of the tip, the
tip radius should be as small as possible, see section
\ref{subsection_tip_sample_forces}. Cantilevers made of silicon
with integrated tips are typically oriented such that the tip
points in the [001] crystal direction. Due to the anisotropic
etching rates of Si and SiO$_2$, these tips can be etched such that
they develop a very
sharp apex \citep{Marcus:1990}, as shown in Fig. \ref{marcustip}.
Recently, it has turned out that not only the sharpness of a tip
is important for AFM, but also the coordination of the front atom.
Tip and sample can be viewed as two giant molecules
\citep{Chen:1993}. In chemical reactions between two atoms or
molecules, the chemical identity and the spatial arrangement of
both partners plays a crucial role. For AFM with true atomic
resolution, the chemical identity and bonding configuration of the
front atom is therefore critical. In [001] oriented silicon tips,
the front atom exposes two dangling bonds (if bulk termination is
assumed), and the front atom has only two connecting bonds to the
rest of the tip. If we assume bulk termination, it is immediately
evident that tips pointing in the [111] direction are more stable,
because then the front atom has \emph{three} bonds to the rest of
the tip, see Figs. \ref{tipmodels}. In a simple picture where only
nearest-neighbor interactions are contributing significantly to
the bonding energy, the front atom of a [111] oriented silicon tip
has 3/4 of the bulk atomic bonding energy. For a [111] oriented
metal tip with fcc bulk structure, the bonding energy of the front
atom has only 3/12 of the bulk value. This trivial picture might
explain, why silicon can be imaged with atomic resolution using
positive frequency shifts (i.e. repulsive forces) with a [111]
silicon tip (to be discussed below).
Even if the (111) sidewalls of these tips reconstruct to e.g. Si
7$\times$7, the front atom is fixed by three bonds and a very
stable tip should emerge. Figure \ref{qPlus111tip} shows a tip
with [111] orientation. The tip is cleaved from a silicon wafer.
Experiments show, that these tips can come very close to a surface
without getting damaged \citep{Giessibl:2001a}.

\subsubsection{Measurement of cantilever deflection and noise}

In the first AFM, the deflection of the cantilever was measured
with an STM - the backside of the cantilever was metalized, and a
tunneling tip was brought close to it to measure the deflection
\citep{Binnig:1986b}. While the tunneling effect is very sensitive
to distance variations, this method has a number of drawbacks:
\begin{itemize}
\item It is difficult to position a tunneling tip such that it
aligns with the very small area at the end of the cantilever.

\item The tunneling tip exerts forces on the cantilever and it is
impossible to distinguish between forces caused by
cantilever-sample and cantilever-tunneling tip interactions.

\item When the cantilever is deflected, the lateral position of
the tip on the backside of the cantilever is shifted. The atomic
roughness of the cantilever backside along with the lateral motion
results in a nonlinear deflection signal.

\end{itemize}
 Subsequent designs used optical
(interferometer, beam-bounce) or electrical methods
(piezoresistive, piezoelectric) for measuring the cantilever
deflection. The deflection of silicon cantilevers is most commonly
measured by optical detection through an interferometer or by
bouncing a light beam of the cantilever and measuring its
deflection (\lq\lq beam bounce method\rq\rq). For detailed
descriptions of these techniques, see \citet{Sarid:1994}, optical
detection techniques are discussed extensively in
\citet{Howald:1994a}. The deflection of piezoresistive cantilevers
is usually measured by making them part of a Wheatstone bridge,
see \citet{Tortonese:1993}.

\begin{figure}[h]
  \centering
  \includegraphics[width=8cm,clip=true]{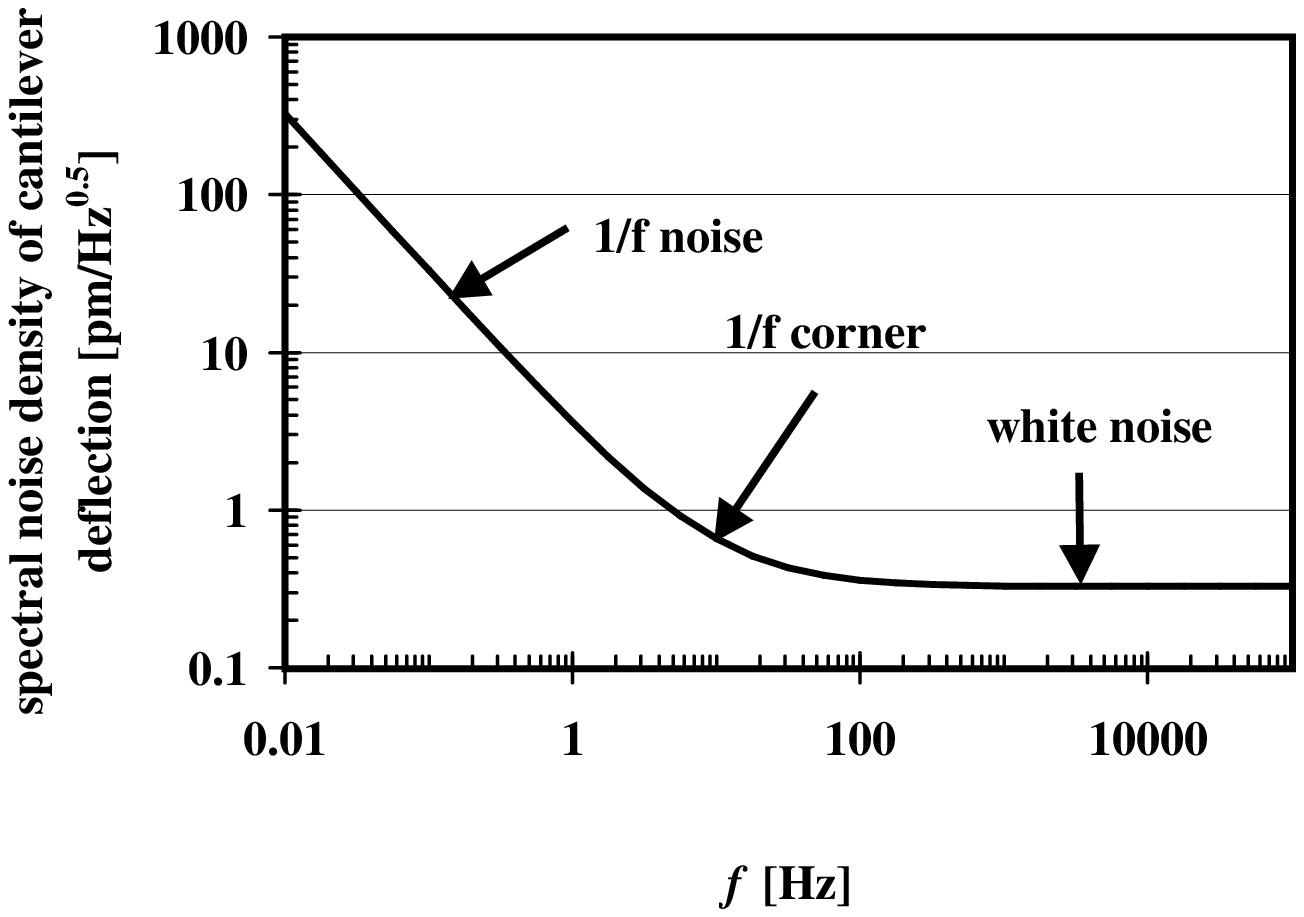}
  \caption{Noise spectrum of a typical cantilever deflection detector (schematic),
  characterized by $1/f$ noise for low frequencies and white noise for intermediate frequencies.
  For very high frequencies, the deflection noise density of typical cantilever deflection sensors
  goes up again (\lq blue noise\rq, not shown here).}\label{pink_noise.eps}
\end{figure}

The deflection of the cantilever is subject to thermal drift and other noise
factors. This can be expressed in a plot of the deflection noise density
versus frequency. A typical noise density is plotted in Fig. \ref{pink_noise.eps},
showing a $1/f$ dependence for low frequency that merges into a
constant noise density (\lq\lq white noise\rq\rq) above the \lq\lq
$1/f$ corner frequency\rq\rq. This $1/f$ noise is also apparent in
macroscopic force sensing devices, such as scales. Typically, scales
have a reset or zero button, which allows the user to reset the
effects of long-term drift. Machining AFMs from materials with low
thermal expansion coefficients like Invar or operation at low
temperatures helps to minimize $1/f$ noise.

\subsubsection{Thermal stability}
\begin{figure}[h]
  \centering \includegraphics[width=8cm,clip=true]{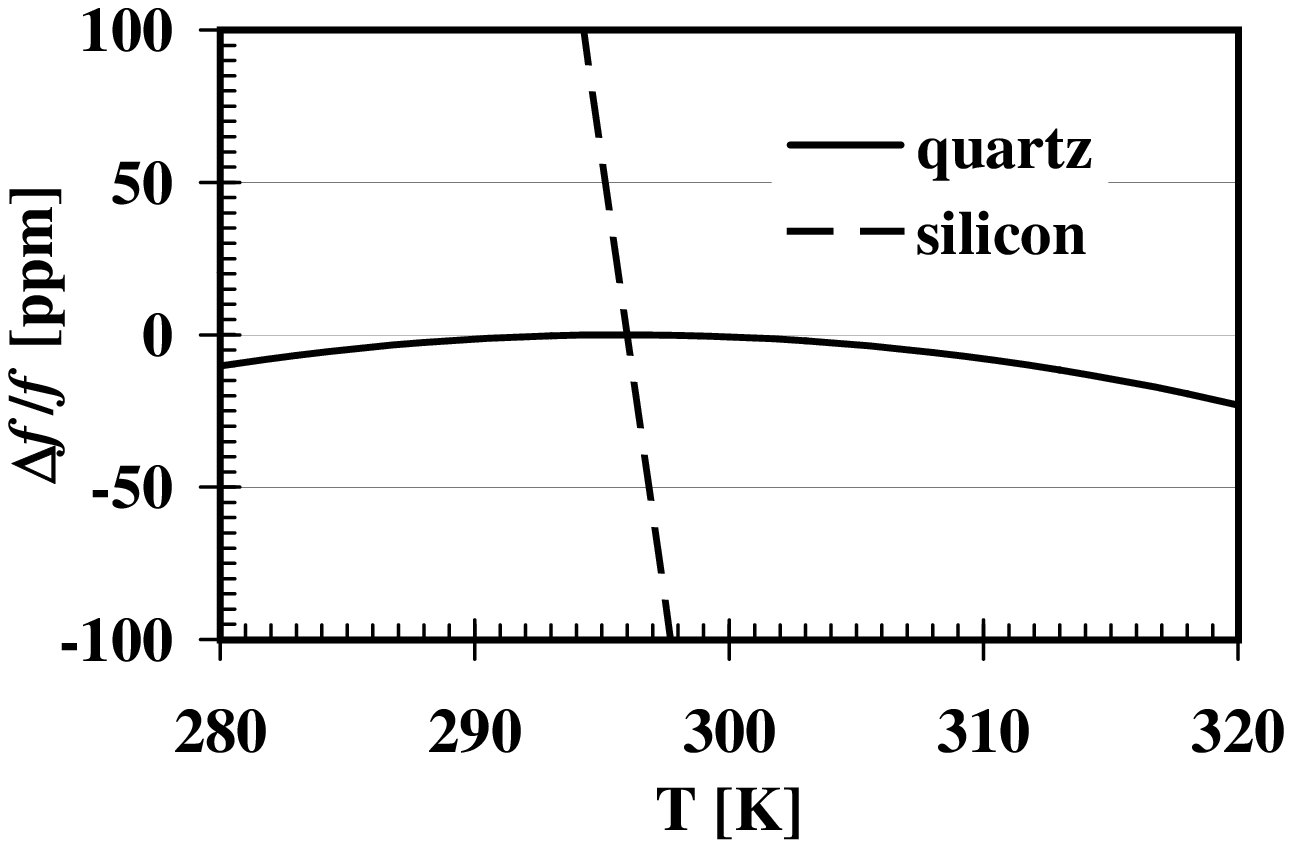}
  \caption{Frequency variation as a function of temperature for
  silicon [110] oriented cantilevers and quartz tuning forks in X+5
  $^\circ$ -cut (see \citet{Momosaki:1997}).}\label{f_T_quartz_si}
\end{figure}

A change in temperature can cause bending of the
cantilever and a change in its eigenfrequency. In this respect,
quartz is clearly superior to silicon as a cantilever material,
as quartz can be cut along
specific crystal orientations such that the variation of
oscillation frequency of a tuning fork or cantilever is zero for a
certain temperature $T_{0}$. For quartz cut in the
X+5$^{\circ}$ direction, $T_{0}\approx 300$\,K, see e.g.
\citet{Momosaki:1997}. This cannot be accomplished with silicon
cantilevers. \label{drifts_in_eigenfrequency} In the dynamic
operating modes (see section \ref{section_FM-AFM}), drifts in
$f_{0}$, caused by variations in temperature, add to the vertical
noise. The eigenfrequency (see Eq. \ref{f0}) is determined by the
spring constant and the effective mass of the cantilever. The
spring constant changes with temperature, due to thermal expansion
and the change of Young's modulus $Y$ with temperature. Changes of
the effective mass due to picking up a few atoms from the sample
or transferring some atoms from the tip to the sample are
insignificant, because a typical cantilever contains at least
$10^{14}$ atoms. The resonance frequency of a cantilever is given
in Eq. \ref{f0_cl}. With the velocity of sound in the cantilever
material $v_{s}=\sqrt{Y/\rho}$, Eq. \ref{f0_cl} can be expressed
as \citep{Chen:1993}:
\begin{equation}
f_{0}=0.162\,v_{s}\,\frac{t}{L^{2}}  .\label{f0_cl_vs}
\end{equation}
The temperature dependence of the eigenfrequency
is then given by
\begin{equation}
\frac{1}{f_{0}}\frac{\partial f_{0}}{\partial T}=
\frac{1}{v_{s}}\frac{\partial v_{s}}{\partial T}-\alpha   \label{df0_dT}
\end{equation}
where $\alpha$ is the thermal expansion coefficient. For silicon
oriented along the [110]-crystal direction (see Fig. \ref{cl}),
$\frac{1}{v_{s}}\frac{\partial v_{s}}{\partial T}=
-5.5\times 10^{-5}K^{-1}$ and $\alpha = 2.55\times 10^{-6}K^{-1}$
at $T=290$\,K \citep{Kuchling:1982,Landolt:1982}. The resulting
relative frequency shift for (rectangular) silicon cantilevers is
then $-5.8\times 10^{-5}K^{-1}$. This is is a large noise source
in classical FM-AFM, where relative frequency shifts can be as
small as $-6\,$Hz$/151\textrm{kHz}=-4\times 10^{-5}$ (see row 5 in
Table \ref{table1}) and a temperature variation of $\Delta
T=+0.69$\,K causes an equal shift in resonance frequency. The
drift of $f_{0}$ with temperature is much smaller for cantilevers
made of quartz. Figure \ref{f_T_quartz_si} shows a comparison of typical frequency
variations as a function of temperature for silicon and quartz.
The data for silicon is calculated with Eq. \ref{df0_dT}, the
quartz data is taken from \citet{Momosaki:1997}. As can be seen,
quartz is remarkably stable at room temperature compared to
silicon.
Less significant noise sources, like the thermal fluctuation of
$A$, are discussed in \citet{Giessibl:1999a}.
\citet{Hembacher:2002} have measured the frequency variations of a
quartz tuning fork sensor from room temperature to 5\,K.

\subsection{Operating Modes of AFMs}
\subsubsection{Static AFM}
In AFM, the force $F_{ts}$ which acts between the tip and sample
is used as the imaging signal. In the static mode of operation,
the force translates into a deflection $q^{\prime }=F_{ts}/k$ of
the cantilever. Because the deflection of the cantilever should be
significantly larger than the deformation of the tip and sample,
restrictions on the useful range of $k$ apply. In the static mode,
the cantilever should be much softer than the bonds between the
bulk atoms in tip and sample. Interatomic force constants in
solids are in a range from 10 N/m to about 100 N/m - in biological
samples, they can be as small as 0.1 N/m. Thus, typical values for
$k$ in the static mode are $0.01-5$\,N/m.
The eigenfrequency $f_{0}$ should be significantly
higher than the desired detection bandwidth, i.e. if 10 lines per second are
recorded during imaging a width of say 100 atoms, $f_{0}$ should be at least
$10\times 2\times 100$ s$^{-1}=2$ kHz in order to prevent resonant
excitation of the cantilever.

Even though it has been demonstrated that
atomic resolution is possible with static AFM \citep{Giessibl:1992b,Ohnesorge:1993,Schimmel:1999}, the method can only be applied
in certain cases. The magnitude of $1/f$-noise can be reduced by
low temperature operation (\citet{Giessibl:1992a}), where the coefficients of thermal expansion are
very small or by building the AFM of a material with a low thermal expansion
coefficient. The long-range attractive forces have to be
cancelled by immersing tip and sample in a liquid (\citet{Ohnesorge:1993}) or by
partly compensating the attractive force by pulling at the cantilever after
jump-to-contact has occurred (\citet{Giessibl:1991b,Giessibl:1992a,Giessibl:1992b}).
\citet{Jarvis:1996,Jarvis:1997} have introduced a method to cancel the long-range attractive force with an electromagnetic force applied
to the cantilever.

While the experimental realization of static AFM is difficult, the
physical interpretation of static AFM images is simple: The image
is a map $z(x,y,F_{ts}=const.)$.

\subsubsection{Dynamic AFM}

In the dynamic operation modes, the cantilever is deliberately
vibrated. The cantilever is mounted onto an actuator to allow an
external excitation of an oscillation. There are two basic methods
of dynamic operation: amplitude modulation (AM) - and frequency
modulation (FM) operation. In AM-AFM \citep*{Martin:1987}, the
actuator is driven by a fixed amplitude $A_{drive}$
at a fixed frequency $f_{drive}$ where $f_{drive}$ is close to but different from $%
f_{0}$. When the tip approaches the sample, elastic and inelastic
interactions cause a change in both the amplitude and the phase
(relative to the driving signal) of the cantilever. These changes
are used as the feedback signal. The change in amplitude in AM
mode does not occur instantaneously with a change in the
tip-sample interaction, but on a timescale of $\tau_{AM} \approx
2Q/f_{0}$. With $Q$-factors reaching 100000 in vacuum, the AM mode
is very slow. \citet*{Albrecht:1991} solved this problem by
introducing the frequency modulation (FM) mode, where the change
in the eigenfrequency occurs within a single oscillation cycle on
a timescale of $\tau_{FM} \approx 1/f_{0}$.

Both AM and FM modes were initially meant to be ``non-contact"
modes, i.e. the cantilever was far away from the surface and the
net force between the front atom of the tip and the sample was
clearly attractive. The AM mode was later used very successfully
at a closer distance range in ambient conditions involving
repulsive tip-sample interactions (\lq\lq Tapping Mode\rq\rq{}
\citet{Zhong:1993}) and \citet{Erlandsson:1996} obtained atomic
resolution on Si in vacuum with an etched tungsten cantilever
operated in AM mode in 1996. Using the FM mode in vacuum, the resolution was
improved dramatically (\citet{Giessibl:1994a,Giessibl:1994b}) and
finally atomic resolution (\citet{Giessibl:1995}) was obtained. A detailed
description of the FM-mode is given in section
\ref{section_FM-AFM}.

\section{CHALLENGES FACED BY AFM WITH RESPECT TO STM} \label{section_AFM_challenges}

\begin{figure}[h]
  \centering
  \includegraphics[width=8cm,clip=true]{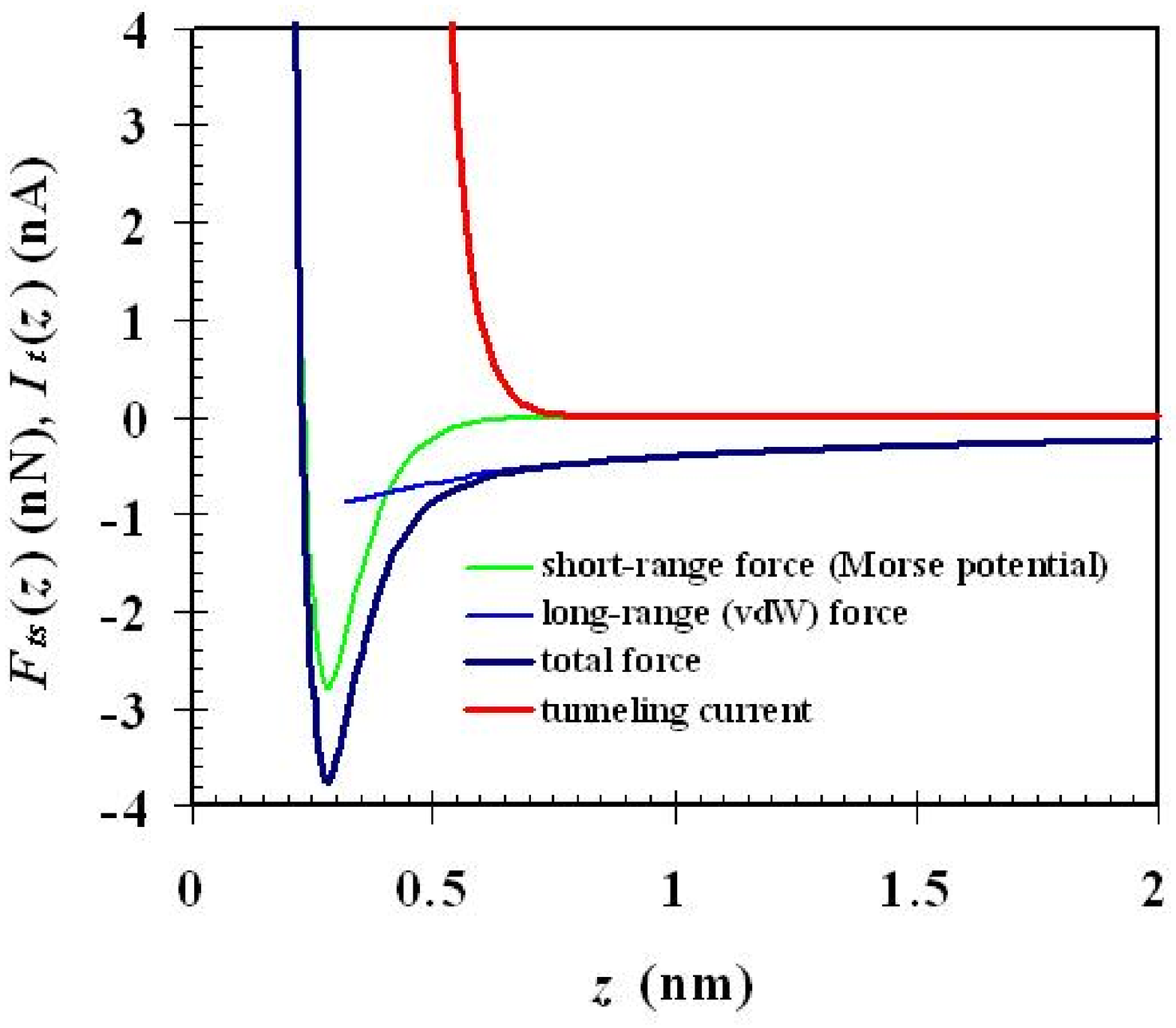}
  \caption{Plot of tunneling current $I_{t}$ and force $F_{ts}$
  (typical values) as a function of distance $z$ between center of
  front atom and plane defined by the centers of surface atom layer.}\label{cur&force}
\end{figure}

In a scanning tunneling microscope, a tip has to be scanned across a
surface with a precision of pico-meters while a feedback mechanism
adjusts the $z-$ position such that the tunneling current is constant.
This task seems daunting and the successful realization of STM is an
amazing accomplishment. Yet, implementing an AFM capable of atomic
resolution poses even more obstacles. Some of these challenges become
apparent when comparing the characteristics of the physical
observables used in the two types of microscopes. Figure
\ref{cur&force} is a plot of tunneling current and tip sample force as
a function of distance. For experimental measurements of force and
tunneling current, see e.g. \citet{Schirmeisen:2000}.
The tunneling current is a monotonic function of the tip-sample
distance and increases sharply with decreasing distance. In contrast,
the tip-sample force has long- and short- range components and is not
monotonic.

\subsection{Stability} \label{section_JTC}

Van-der-Waals forces in vacuum are always attractive, and if chemical
bonding between tip and sample can occur the chemical forces are also
attractive for distances greater than the equilibrium distance. Because the tip
is mounted on a spring, approaching the tip can cause a sudden
\lq\lq jump-to-contact\rq\rq{} when the stiffness of the cantilever is smaller than a
certain value.

This instability occurs in the quasistatic mode if
\begin{equation}
k<\max (-\frac{\partial ^{2}V_{ts}}{\partial z^{2}})=k_{ts}^{\max
} \label{JTCstatic}
\end{equation}
\citep {Tabor:1969,McClelland:1987,Burnham:1989}. The jump to
contact can be avoided even for soft cantilevers by oscillating it
at a large enough amplitude $A$:
\begin{equation}
kA>\max (-F_{ts})=F_{ts}^{\max}
\end{equation}
\citep{Giessibl:1997b}. If hysteresis occurs in the $F_{ts}(z)$-relation, the energy $\Delta E_{ts}$
needs to be supplied to the cantilever for each oscillation cycle. If
this energy loss is large compared to the intrinsic energy loss of
the cantilever, amplitude control can become difficult (see the discussion
after Eq. \ref{Adrive}). An new conjecture
regarding $k$ and $A$ is then
\begin{equation}
\frac{k}{2}A^2 \geq  \Delta E_{ts}\frac{Q}{2\pi}.
\end{equation}
The validity of these criteria is supported by an
analysis of the values of $k$ and $A$ for many NC-AFM experiments
with atomic resolution in table \ref{table1}.

Fulfilment of the stability criteria thus requires either the use of large amplitudes,
cantilevers with large spring constants, or both. However, using large amplitudes has critical disadvantages, which are discussed
in chapter \ref{chapter_noise_in_fmafm}.

\subsection{Non-monotonic imaging signal}

The magnitude of the tunneling current increases continuously as the
tip-sample distance decreases, i.e. the tunneling current is a strictly
monotic decreasing function of
the distance (see Fig. \ref{current_z} on page \pageref{current_z}).
This property allows a simple implementation of a feedback
loop: the tunneling current is fed into a logarithmic amplifier to
produce an error signal that is linear with the tip-sample distance.

In contrast, the tip-sample force is not monotonic. In general, the force is
attractive for large distances and upon decreasing the distance between tip and sample,
the force turns repulsive (see Fig. \ref{cur&force}). Stable feedback is only
possible on a branch of the force curve, where it is monotonic.

Because the tunneling current is monotonic for the whole distance range and
the tip-sample force is not monotonic, it is much easier to establish a $z-$
distance feedback loop for STMs than for AFMs.

\subsection{Contribution of long-range forces}

The force between tip and sample is composed of many
contributions: electrostatic-, magnetic-, van-der-Waals- and
chemical forces in vacuum. In ambient conditions there are also
meniscus forces. While electrostatic-, magnetic- and meniscus
forces can be eliminated by equalizing the electrostatic potential
between tip and sample, using nonmagnetic tips and vacuum
operation, the van-der-Waals forces cannot be switched off. For
imaging by AFM with atomic resolution, it is desirable to filter
out the long-range force contributions and only measure the force
components which vary at the atomic scale. In STM, the rapid decay
of the tunneling current with distance naturally blocks
contributions of tip atoms that are further distant to the sample,
even for fairly blunt tips. In contrast, in \emph{static} AFM,
long- and short-range forces add up to the imaging signal. In
\emph{dynamic} AFM, attenuation of the long-range contributions is
achieved by proper choice of the cantilever's oscillation
amplitude $A$, see section \ref{subsection_frequency_shift_for}.

\subsection{Noise in the imaging signal}
Forces can be measured by the deflection of a spring. However, measuring the
deflection is not a trivial task and is subject to noise, especially at low
frequencies ($1/f$ noise). In static AFM, the imaging signal is given by
the dc deflection of the cantilever, which is subject to $1/f$ noise. In
dynamic AFM, the low-frequency noise is discriminated if the eigenfrequency $%
f_{0}$ is larger than the $1/f$ corner frequency. With a bandpass filter with a
center frequency around $f_{0}$ only the white noise density is integrated
across the bandwidth $B$ of the bandpass filter.
\newline
\newline
Frequency modulation AFM, described in detail in chapter \ref{section_FM-AFM},
helps to overcome three of these four challenges. The non-monotonic
force vs. distance relation is a remaining complication for AFM.

\section{EARLY AFM EXPERIMENTS} \label{early_AFM}
\label{sec:early experiments} The first description of the AFM by
\citet{Binnig:1986b} already lists several possible ways to
operate the microscope: contact and non-contact, static and
dynamic modes. Initially,  AFMs were mainly operated in the static
contact mode. However, soon after the invention of the AFM,
\citet*{Duerig:1986} have measured the forces acting during
tunneling in STM in UHV with a dynamic technique. In these
experiments, the interaction between a tungsten STM tip and a thin
film of Ag condensed on a metal cantilever was studied. The
thermally excited oscillation of the metal cantilever was observed
in the spectrum of the tunneling current, and the force gradient
between tip and sample caused a shift in the resonance frequency
of the cantilever. In a later experiment, \citet*{Duerig:1990}
used Ir tips and an Ir sample. While variations of the force on
atomic scale were not reported in these experiments, it was shown
that both repulsive (W tip, Ag sample) and attractive forces (Ir
tip, Ir sample) of the order of a few nN can act during STM
operation.

G. Binnig, Ch. Gerber, and others started the IBM Physics Group at
the Ludwig-Maximilian-Universit\ät in Munich. The author joined
this group in May 1988 and helped to build a low-temperature UHV
AFM to probe the resolution limits of AFM. If atomic resolution
was possible, we thought that the best bet would be to try it at
low temperatures in order to minimize the detrimental effects of
thermal noise. The microscope was fitted to a quite complex vacuum
system which was designed by G. Binnig, Ch. Gerber and T. Heppell
with colleagues (VG Special Systems Hastings, England). Because it
was anticipated that the design of the instrument had to go
through many iterations which involves the breaking of the vacuum,
the vacuum system was designed
 in an effort to keep the bake-out time
short and to allow rapid cooling to 4\,K, see
\citet*{Giessibl:1991a}. Our instrument could resolve atoms in STM
mode on graphite at $T=4$\,K in 1989, but AFM operation with
atomic resolution was not possible yet. As AFM test samples, we
used ionic crystals and in particular alkali halides. Alkali
halides can be viewed as consisting of hard spheres which are
charged by plus/minus one elementary charge
(\citet{Ashcroft:1981}). These materials are easily prepared by
cleaving in vacuum, where large (001) planes with fairly low step
densities develop.

In late 1989, E. \citet{MeyerE:1990a} has shown quasiatomic
resolution on LiF(001) in ambient conditions. The AFM images were
explained with the \lq contact-hard-spheres-model\rq{} by
\citet{MeyerE:1990b}, which assumes that the front atom of the tip
and the sample atoms are hard spheres. Also in 1990, G.
\citet{MeyerG:1990} published a paper about the successful imaging
of NaCl in UHV at room temperature with quasiatomic resolution.
Quasiatomic resolution means that the images reflect atomic
periodicities, but no atomic defects. The images appear to arise
from a tip which has several or possibly many atomic contacts
(minitips) spaced by one or several surface lattice vectors. This
hypothesis is supported by the common observation in contact AFM
that the resolution appears to improve after the tip is scanned
for a while. Wear can cause the tip to develop a set of minitips
which are spaced by multiple sample surface lattice vectors,
yielding a quasi-atomic resolution. This mechanism is not observed
and not expected to occur in todays non-contact AFM experiments.

\begin{figure}[h]
  \centering \includegraphics[width=8cm,clip=true]{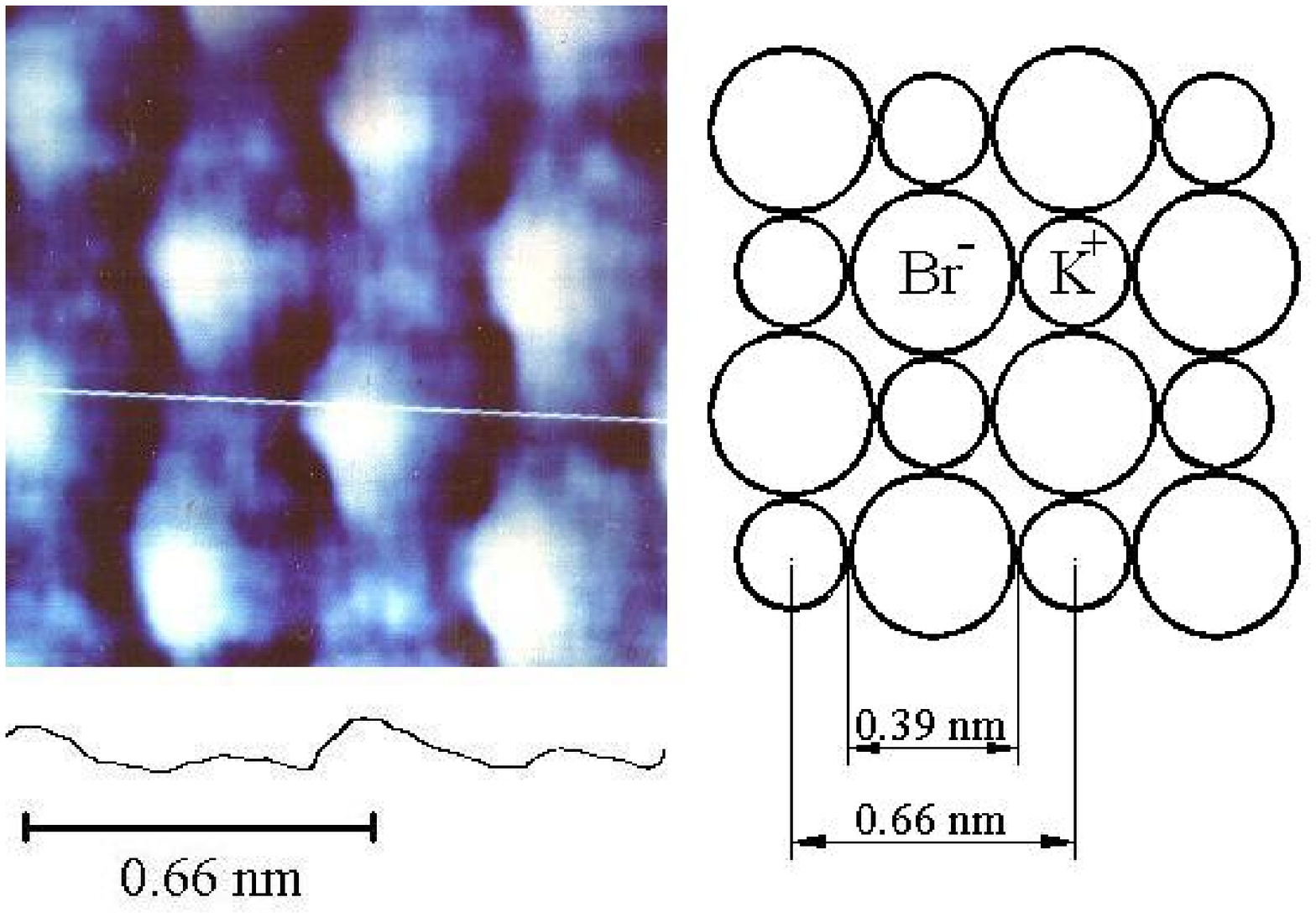}
  \caption{Atomically resolved image of KBr (001) in contact AFM mode.
  The small and large protrusions are attributed to K$^+$- and
  Br$^-$-ions, respectively. Source:
  \citet{Giessibl:1992b}.}\label{KBr_uc}
\end{figure}

\begin{figure}[h]
  \centering \includegraphics[width=8cm,clip=true]{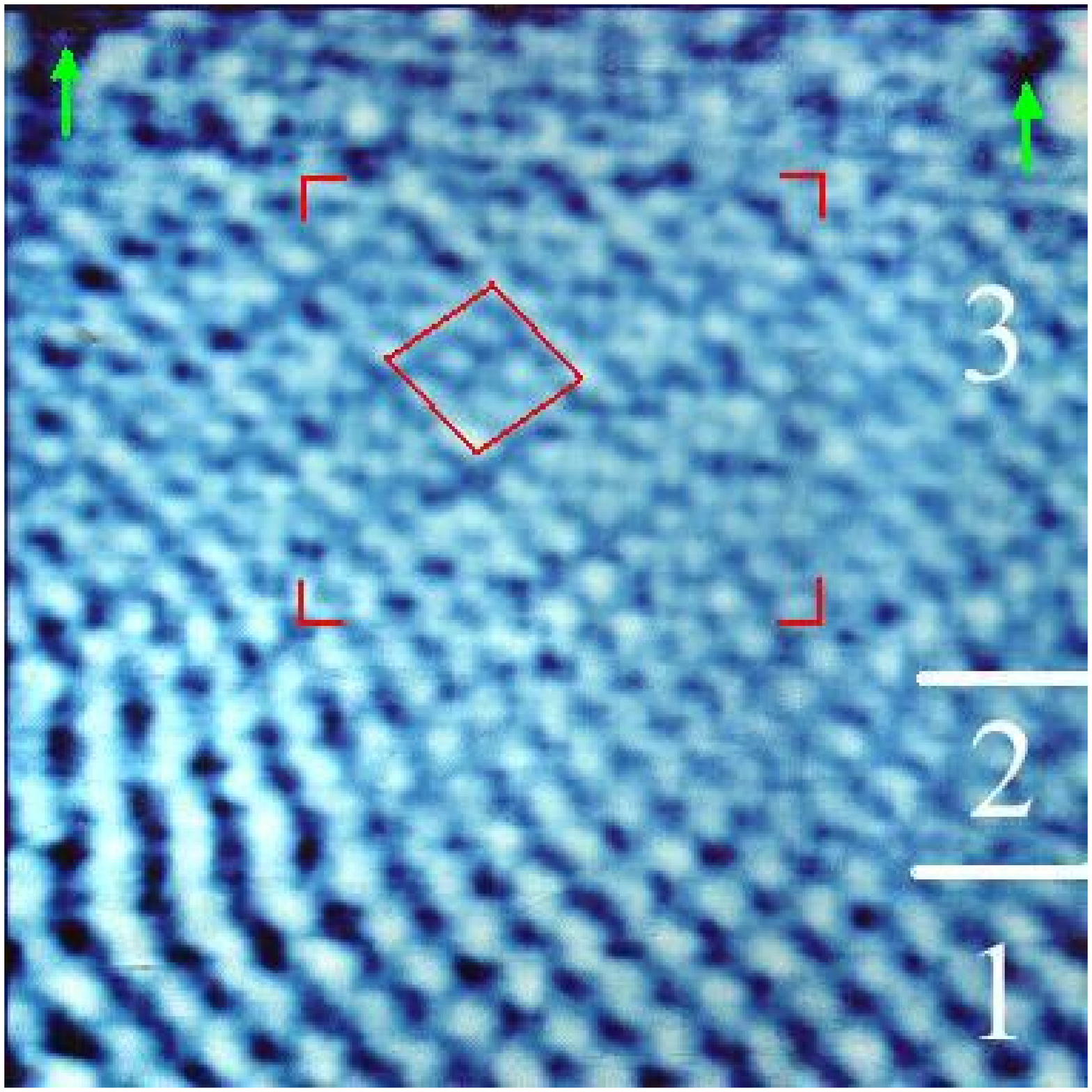}
  \caption{Atomically resolved image of KBr (001) in contact AFM mode.
Scan width 5\,nm in region 1, continuously increased from 5\,nm to
10\,nm in region 2 and 10\,nm in region 3, see text. The
$\sqrt{3}\times\sqrt{3}$\,R\,$45^{\circ}$ superstructure and the slight
depression in the central 5\,nm $\times$ 5\,nm area (enclosed by red
angles) is probably caused by the repulsive force of 1\,nN between the
front atom of the tip and the sample. The red square shows the
unit-cell of the $\sqrt{3}\times\sqrt{3}$\,R\,$45^{\circ}$ reconstruction,
the green arrows indicate atomic-size defects. Source:
\citet{Giessibl:1992c}.} \label{KBr_def}
\end{figure}

In both contact- AFM experiments (E. Meyer et al. and G. Meyer et
al.), only one type of ion was apparent in the force microscope
images. In 1990, we improved our 4\,K UHV instrument by mounting
the whole vacuum system on air legs and adding a vibration
insulation stage directly at the microscope. The major
experimental challenge was the detection of the cantilever
deflection. Like in the first AFM by \citet{Binnig:1986a},
tunneling detection was used to measure the deflection of a
micromachined \lq V\rq{}-shaped cantilever with a spring constant
of $k=0.37$\,N/m after \citet{Albrecht:1990}. The cantilever was
made from SiO$_2$ and plated with a thin gold film for electrical
conductance. The tunneling tip had to be adjusted to an area of a
few $\mu$m$^2$ before the microscope was inserted into the low
temperature vacuum system. As it turned out later, successful
tunneling between the platinum coated tungsten tip and the gold
plated cantilever was only possible if the tip had drifted towards
the fixed end of the cantilever beam during cooling the instrument
from room temperature to 4\,K. When the tunneling tip was adjacent
to the free end of the cantilever jump-to-contact between
tunneling tip and cantilever occured and stable tunneling
conditions were hard to achieve. However, if the tunneling tip
meets the cantilever at a distance $L_{tunneling}$ from the fixed
end of the cantilever with total length $L$, the effective
stiffness of the cantilever increases by a factor of
$(L/L_{tunneling})^3$ (see Eq. \ref{k_cl}) and jump-to-contact is
less likely to occur. Endurance was critical, because only in one
of about ten cooling cycles all parts of the complicated
microscope worked. KBr (cleaved in situ) was used as a sample.
After the sample was approached to the cantilever, jump-to-contact
occured and the sample area where the cantilever had landed was
destroyed. After jump-to-contact occured, the pressure on the tip
region was released by pulling back the sample such that
cantilever still stayed in contact with the sample, however the
repulsive force between the front atom of the cantilever and the
sample was reduced to $\approx$ 1\,nN. With the reduced tip-sample
force, the sample was moved laterally to an undisturbed area, and
atomic resolution was immediately obtained. In summer 1991, we
finally succeeded in obtaining true atomic resolution on KBr.
Figure \ref{KBr_uc} shows the KBr (001) surface imaged in contact
mode. Both ionic species are visible, because repulsive forces are used
for imaging. The small bumps are
attributed as K$^+$ ions and the large bumps as Br$^-$ ions. Today
even with refined non-contact AFM, only one atomic species appears
as a protrusion in images of ionic crystals. Most likely, the dominant
interaction between front atom and sample in non-contact AFM is
electrostatic, so the charge of the front atom determines if cations or
anions appear as protrusions, see e.g. \citet{Livshits:1999a,Shluger:1999,Livshits:1999}.
Figure \ref{KBr_def} shows atomic resolution on KBr with linear
singularities and atomic defects. This image was obtained by scanning an area of 5\,nm
$\times$ 5\,nm for a while and then doubling the scan size to 10\,nm
$\times$ 10\,nm. The fast scanning direction was horizontal, the slow
scanning direction vertical from bottom to top. In the lower section
in Fig. \ref{KBr_def} (region 1) the scan size was 5\,nm $\times$
5\,nm. Region 2 is a transition area, where the $x-$ and $y-$scan widths
were continuously increased to 10\,nm (an analog scan electronics was
used where the widths of both scanning axes are independently
controlled by a potentiometer in real time). In region 3, the scan
size was set at 10\,nm $\times$ 10\,nm. Initially, we interpreted the
singularity as a monoatomic step (\citet{Binnig:1992}). However, the
height difference between the central area in Fig. \ref{KBr_def} and
the surrounding area is much smaller than a single step (3.3\,\AA).
Therefore, today it appears that the central area is a \lq scan
window\rq{}, i.e. a region slightly damaged by the pressure of the
scanning cantilever. Even scanning at very small loads of a nN or so
has disturbed the surface slightly and created a depressed area with a
$\sqrt{3}\times\sqrt{3}$\,R\,$45^{\circ}$ superstructure on the KBr
surface. The presence of atomic defects (green arrows in Fig.
\ref{KBr_def}), linear defects and superstructures strengthened our
confidence in true atomic resolution capability of the AFM. However,
the experimental difficulties with low-temperature operation, sample
preparation, tunneling detection etc. were quite impractical for
routine measurements.

In 1993,
\citet*{Ohnesorge:1993} pursued a different method to cancel the damaging long-range forces.
The long-range
attractive forces which cause jump-to-contact were reduced by
immersing the cantilever and sample
into a liquid, as explained by \citet{Israelachvili:1991}.
\citet*{Ohnesorge:1993} achieved true
atomic resolution by AFM across steps on Calcite in repulsive \emph{and} attractive mode.
True atomic resolution of inert surfaces by AFM had thus been clearly established.
However, the enigmatic icon of atomic resolution microscopy, Si(111)-(7$\times$7)
remained an unsolved challenge. Even experts in
experimental AFM were convinced that this goal is impossible to reach because of
silicons high reactivity and the strong bonds that are formed between cantilever tips
and the Si surface.

\section{THE RUSH FOR SILICON}
\label{sec:THE RUSH FOR SILICON} 
\begin{figure}[h]
  \centering \includegraphics[width=8cm,clip=true]{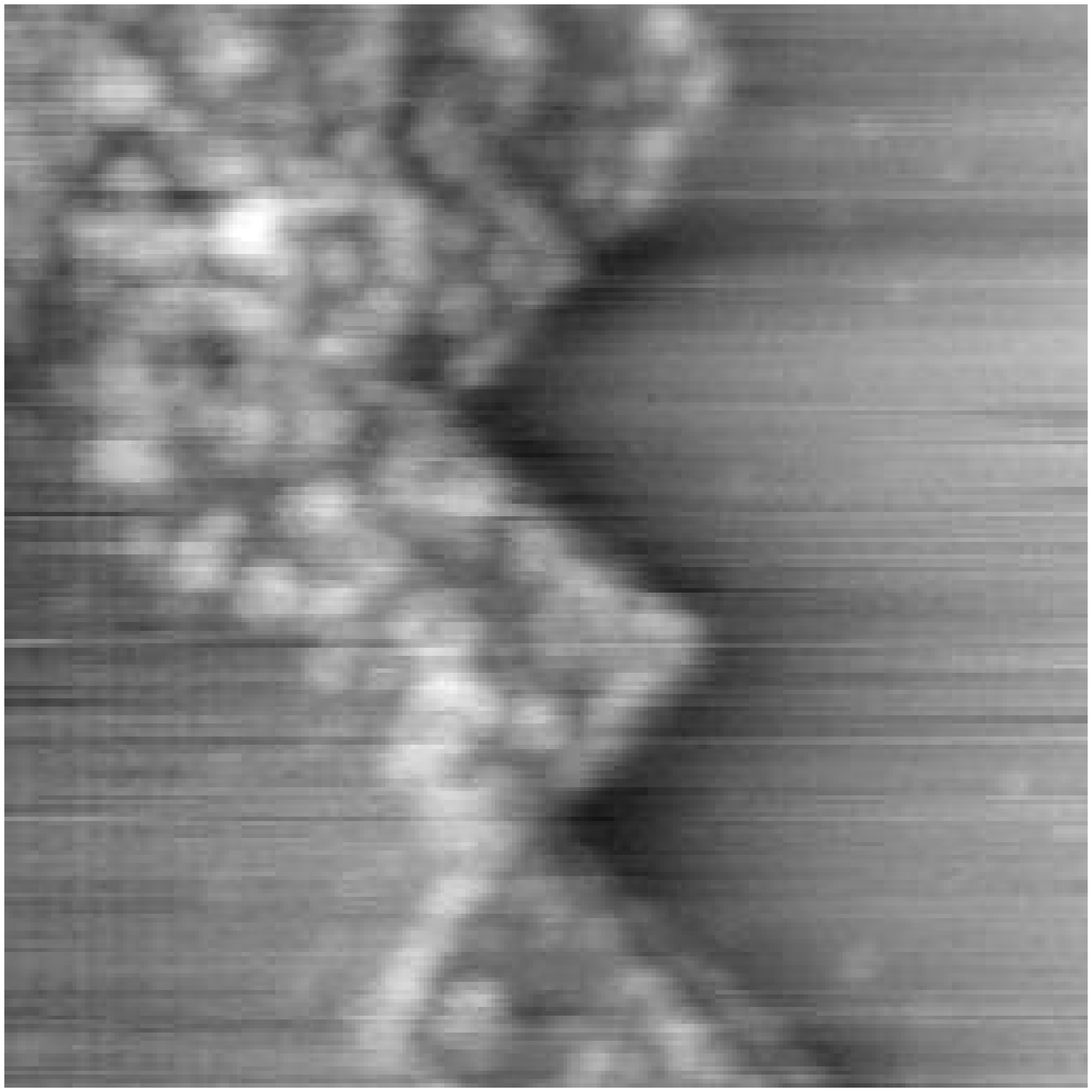}
  \caption{non-contact AFM image of a cleavage face of KCl (001) with
  mono- and double steps of a height of 3.1 and 6.2\,\AA{}
  respectively. Image size 120\,nm $\times$ 120\,nm. Source:
  \citet{Giessibl:1994a}}\label{KBr1993}
\end{figure}

\begin{figure}[h]
  \centering
  \includegraphics[width=8cm,clip=true]{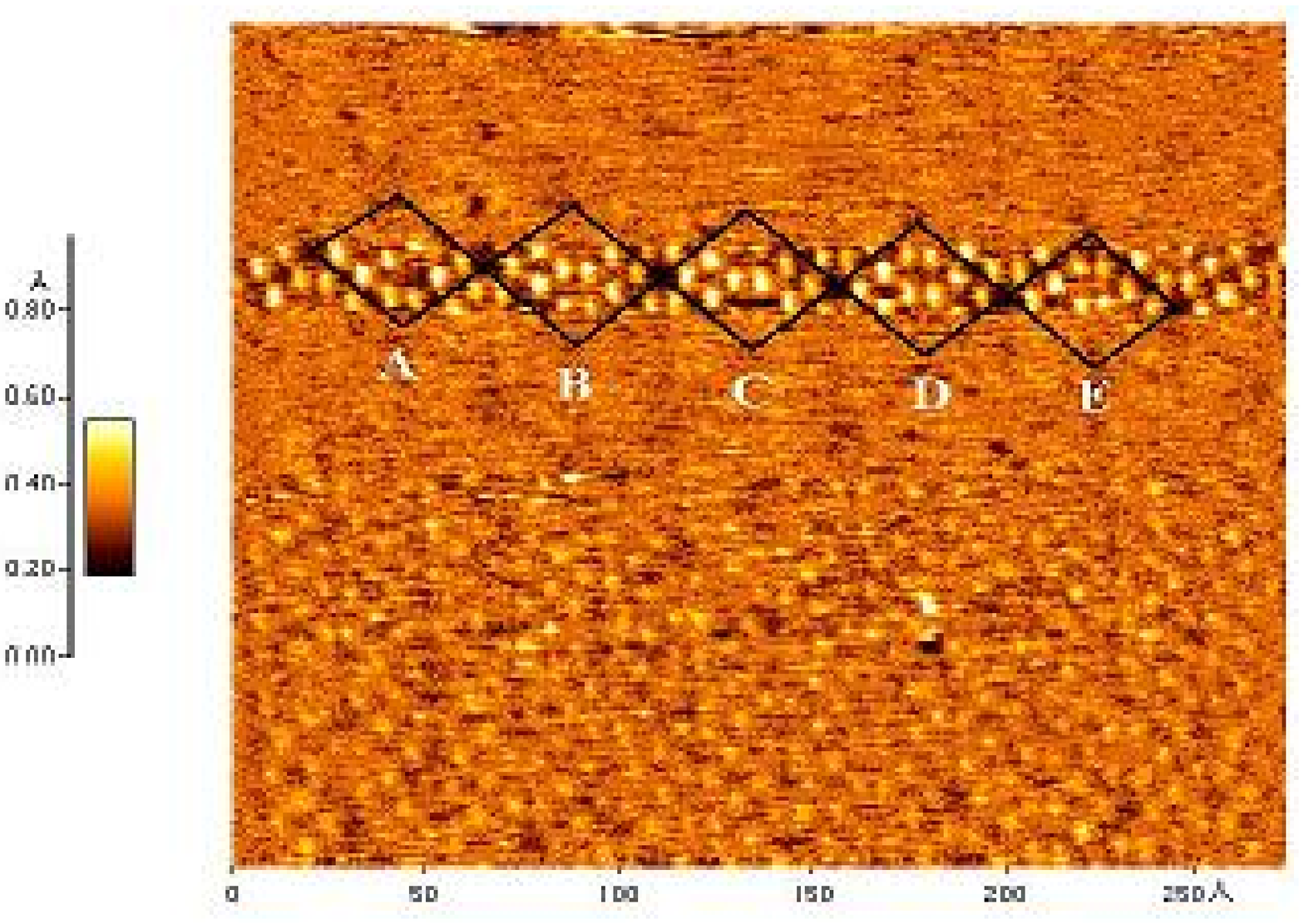}
  \caption{First AFM image of the silicon 7$\times$7 reconstruction with true atomic resolution. Parameters:
  $k=17$\,N/m, $f_0=114$\,kHz, $A=34$\,nm, $\Delta f=-70$\,Hz and $Q=28\,000$. Source: \citet{Giessibl:1995}}\label{7x7}
\end{figure}

Imaging the Si (111)-(7$\times$7)
reconstruction has been crucial for the success of the STM, and
therefore imaging silicon by AFM with atomic resolution has been a
goal for many AFM researchers. However, so far atomic resolution
had not been obtained on reactive surfaces. The ions in alkali-halides
form more or less closed noble gas shells and are
therefore inert. In contrast, Silicon is known to form strong
covalent bonds with a cohesive energy of roughly 2\,eV per bond.
The jump-to-contact problem outlined in section \ref{section_JTC}
is even more severe for silicon, and using silicon cantilevers on
silicon samples in contact mode in vacuum has proven not to work.
At Park Scientific Instruments, the frequency modulation technique
pioneered by \citet{Albrecht:1991} was used at that time in
ambient conditions, and it was tempting to incorporate
the technique into our newly designed UHV microscope (\lq
AutoProbe VP\rq). Marco Tortonese had developed piezoresistive
cantilevers during his time as a graduate student in Cal Quate's
group in Stanford and made them available to us. In vacuum, the
piezoresistive cantilevers have excellent $Q-$values and were thus
predestined for using them in the FM mode. In late 1993,
\citet{Giessibl:1994a} observed single steps and kinks on KBr
using the FM method (see Fig. \ref{KBr1993}).
Also in 1993, \citet{Giessibl:1994b} observed atomic rows on Si (111)-(7$\times$7)
and in May 1994 \citep{Giessibl:1995}, the first clear images of the 7$\times$7 pattern
appeared (see Fig. \ref{7x7}).

\begin{figure}[h]
  \centering
  \includegraphics[width=8cm,clip=true]{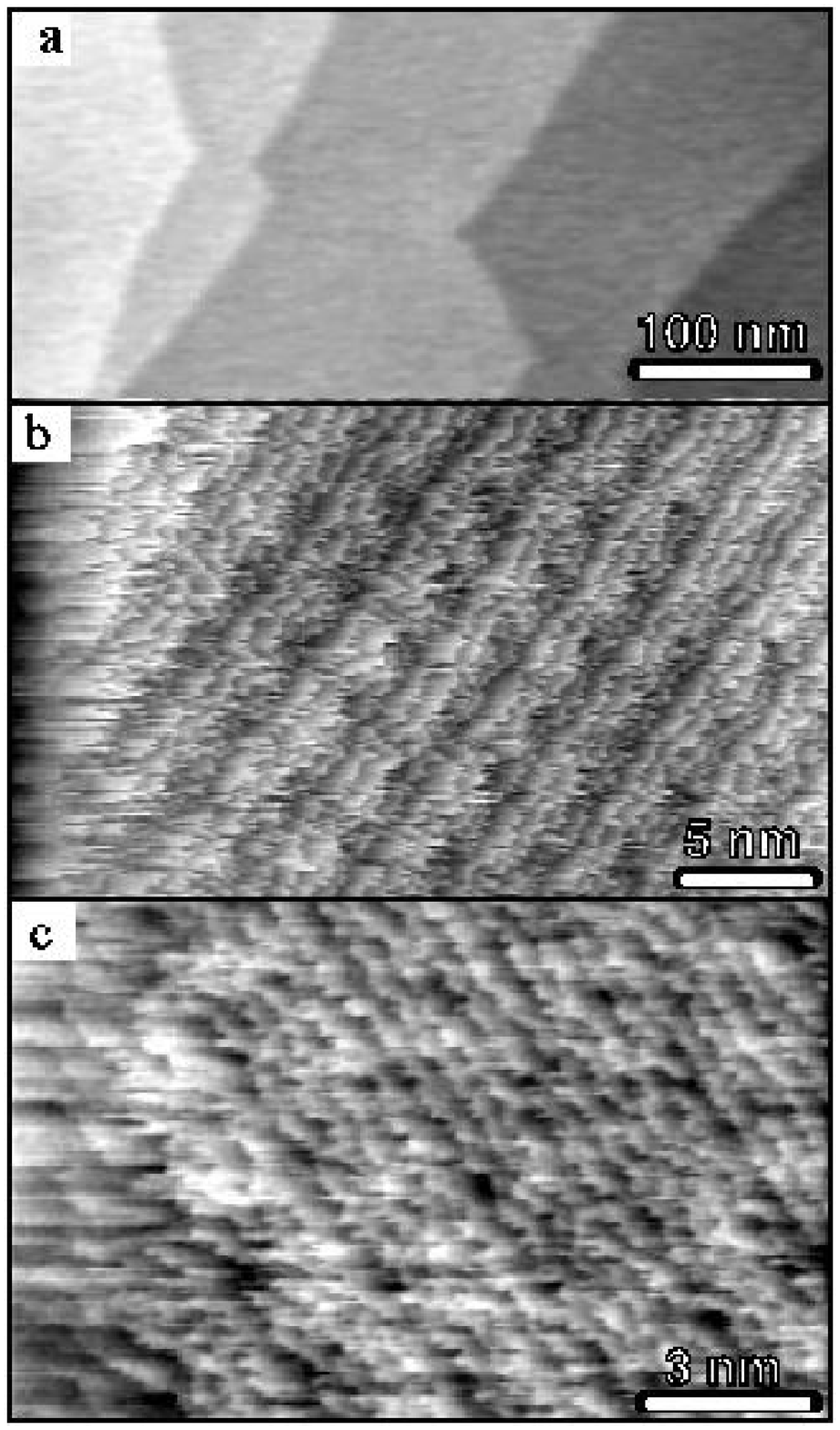}
  \caption{(a) Normal-force image on the surface of Si(111)7$\times$7. The step heights are
  3 and 6 \AA. (b),(c) Lateral-force image on Si(111)7$\times$7. A repulsive force of
  10$^{-9}$N is applied between the probing tip and sample. Variations of the lateral force
  are digitized while keeping the normal force constant. Source: \citet{Howald:1994}.}\label{lukas}
\end{figure}

\begin{figure}[h]
  \centering
  \includegraphics[width=8cm,clip=true]{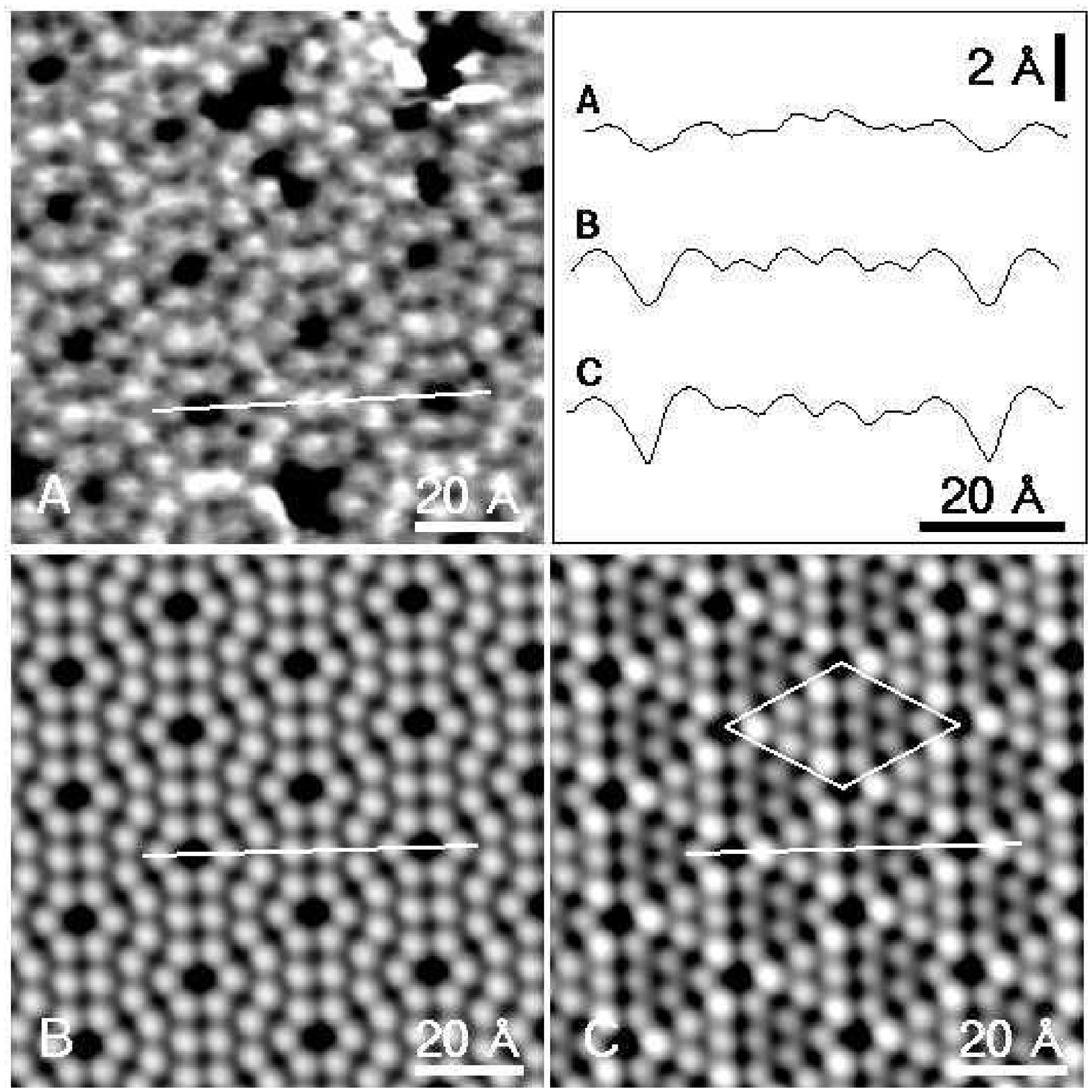}
  \caption{AFM image of the silicon 7$\times$7 reconstruction with AM mode. Image size 100\,\AA $\times$ 100\,\AA.
  A comparison between (A) an AFM image and (B) empty- and (C) filled-state STM images. The grey scales in the images correspond to a heigth difference of 1 \AA.
  The STM images were recorded with tip voltages of -2 and +2.2\,V,
  respectively, and a constant current of 0.1\,nA. The AFM image
  has been low-pass filtered using a $3\times 3$ convolution
  filter while the STM images show unfiltered data. The cross
  sections through the four inequivalent adatoms are obtained from
  raw data. The $7\times 7$ unit cell is outlined in the
  filled-state STM image. The faulted and unfaulted halves
  correspond to the left-hand and right-hand side, respectively.
  Source: \citet{Erlandsson:1996}}\label{erlandsson}
\end{figure}

A different route had been pursued in the group of H.-J.
G\üntherodt: \citet*{Howald:1994} have coated the tip of the
cantilever with PTFE (poly-tetra-fluoro-ethylen) and found that
atomic steps and even the Si(111) $7\times 7$ unit cell
periodicity could be imaged in contact mode in vacuum (see Fig.
\ref{lukas}).
\citet{Erlandsson:1996} could show that atomic resolution on Si(111)7$\times$7 is also possible
with the amplitude modulation technique (see Fig. \ref{erlandsson}).

\section{FREQUENCY MODULATION AFM (FM-AFM)}                              \label{section_FM-AFM}
\subsection{Experimental setup}

\begin{figure}[h]
  \centering
  \includegraphics[width=8cm,clip=true]{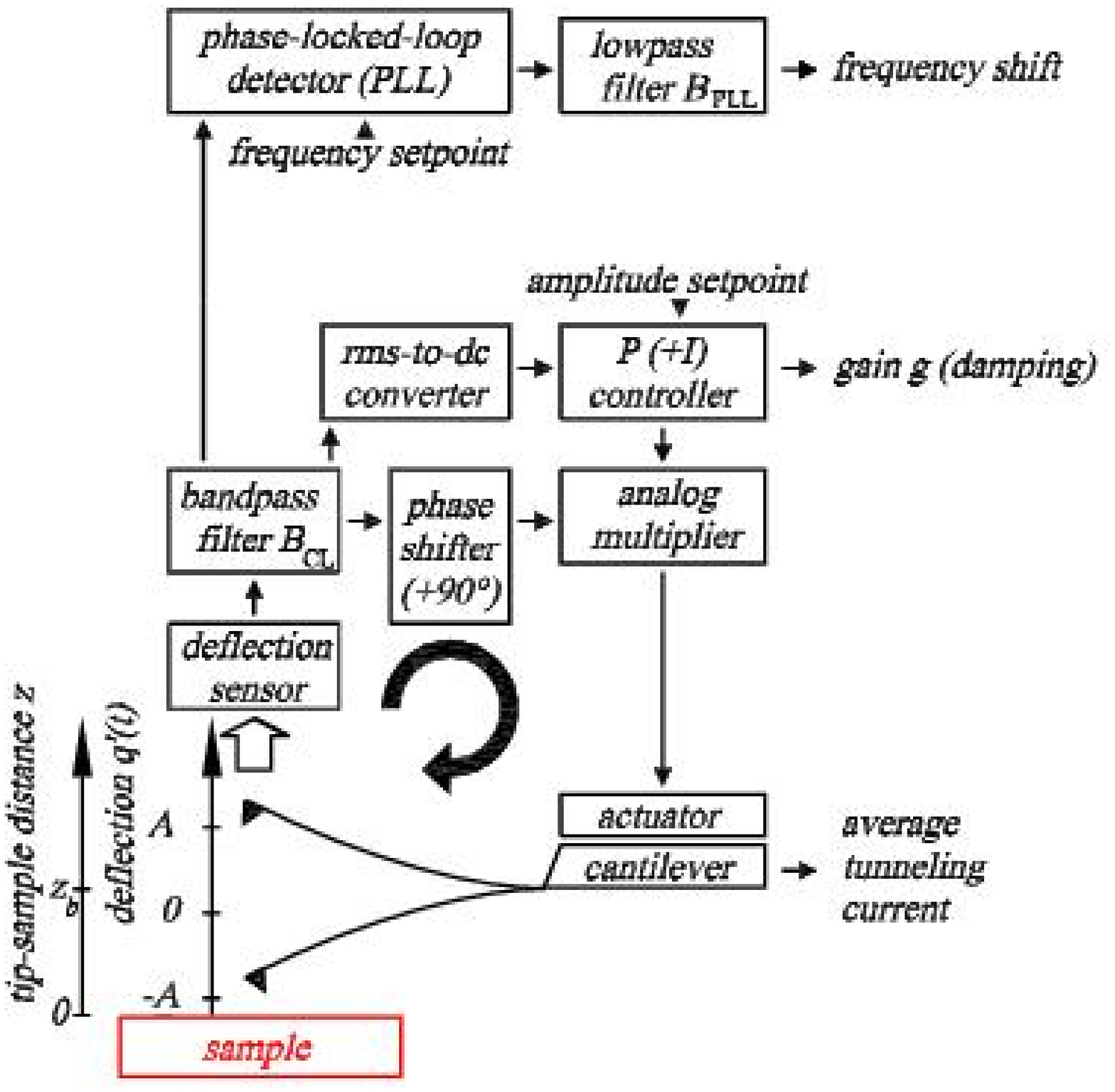}
  \caption{Block diagram of the frequency-modulation AFM feedback loop for constant amplitude control
  and frequency shift measurement. Three physical observables are available: frequency shift, damping signal
  and (average) tunneling current.}\label{fmprinciple}
\end{figure}

In FM-AFM, a cantilever with eigenfrequency $f_{0}$ and spring
constant $k$ is subject to controlled positive feedback such that
it oscillates with a constant amplitude $A$
\citep{Albrecht:1991,Duerig:1992} as shown in Fig.
\ref{fmprinciple}. The deflection signal first enters a bandpass
filter. Then the signal splits in three branches: one branch is
phase shifted, routed through an analog multiplier and fed back to
the cantilever via an actuator; one branch is used to compute the
actual oscillation amplitude, this signal is used to calculate a
gain input $g$ for the analog multiplier and one branch is used to
feed a frequency detector. The frequency $f$ is determined by the
eigenfrequency $f_{0}$ of the cantilever and the phase shift
$\varphi$ between the mechanical excitation generated at the
actuator and the deflection of the cantilever. If $\varphi=\pi/2$,
the loop oscillates at $f=f_{0}$.

Forces between tip and sample cause a change in $f=f_{0} + \Delta
f$. The eigenfrequency of a harmonic oscillator is given by
$(k^*/m^*)^{0.5}/(2\pi)$, where $k^*$ is the effective spring
constant and $m^*$ is the effective mass. If the second derivative
of the tip-sample potential $k_{ts}=\frac{\partial
^{2}V_{ts}}{\partial z^{2}}$ is constant for the whole range
covered by the oscillating cantilever, $k^*=k+k_{ts}$. If $k_{ts}
<< k$, the square root can be expanded as a Taylor series and the
shift in eigenfrequency is approximately given by:
\begin{equation}
\Delta f=\frac{k_{ts}}{2k} f_{0}.
\end{equation}
The case where $k_{ts}$ is not constant is treated in the next
chapter. By measuring the frequency shift $\Delta f$, the
tip-sample force gradient can be determined.

The oscillator circuit is a critical component in FM-AFM.
The function of this device is understood best by analyzing the cantilever motion.
The cantilever can be treated as a damped harmonic oscillator that is
externally driven. For sinusoidal excitations $A_{drive}e^{i2\pi f_{drive} t}$ and a quality factor $Q\gg1$,
the response of the oscillation amplitude of the cantilever is given by
\begin{equation}
\frac{A}{A_{drive}}=\frac{1}
{1-f_{drive}^{2}/f_{0}^{2}+i f_{drive}/(f_{0}Q) }.
\end{equation}
The absolute value of the amplitude is given by
\begin{equation}
|A|=\frac{|A_{drive}|}
{ \sqrt{(1-f_{drive}^{2}/f_{0}^{2})^{2}+f_{drive}^{2}/(f_{0}^{2}Q^{2})} } \label{absA_f}
\end{equation}
and the phase angle between the driving and resulting signals is
\begin{equation}
\varphi=\arctan[\frac{f_{drive}} {Qf_{0}(1-f_{drive}^{2}/f_{0}^{2})}]
\end{equation}

In the case of a closed feedback loop as shown in Fig. \ref{fmprinciple}, the driving frequency
cannot be choosen freely anymore but is determined by $f_{0}$ of the cantilever, the phase shift $\varphi$
and the tip-sample forces. The purpose of the oscillator circuit is to provide controlled positive
feedback (with a phase angle of $\varphi=\pi/2$) such that the cantilever oscillates at a constant
amplitude. This requirement is fulfilled with the setup shown in Fig. \ref{fmprinciple}.

The cantilever deflection signal is first routed through a
bandpass filter which cuts off the noise from unwanted frequency
bands. The filtered deflection signal branches into an rms-to-dc
converter and a phase shifter (see \citet{Horowitz:1989}). The
rms-to-dc chip computes a dc signal which corresponds to the
rms-value of the amplitude. This signal is added to the inverted
setpoint rms amplitude, yielding the amplitude error signal. The
amplitude error enters a proportional (P) and optional integral
(I) controller and the resulting signal $g$ is multiplied with the
phase shifted cantilever deflection signal $q''$ with an analog
multiplier chip. This signal drives the actuator. The phase
shifter is adjusted so that the driving signal required for
establishing the desired oscillation amplitude is minimal;
$\varphi$ is exactly $\pi/2$ in this case. \citet{Duerig:1992} and
\citet{Gauthier:2001,Gauthier:2002} have analyzed the stability
issues related to this forced motion.

The filtered cantilever deflection signal is fed into a
frequency-to-voltage converter. Initially, analog circuits were
used as frequency to voltage converters \citep{Albrecht:1991}.
Recently, commercial digital phase-locked-loop (PLL) detectors
\citep{Nanosurf} and analog quartz-stabilized PLLs
\citep{Kobayashi:2001} became available which are more precise and
more convenient. The PLL allows to set a reference frequency
$f_{ref}$ and outputs a signal which is proportional to the
difference between the input frequency $f$ and the reference
frequency $f_{ref}$. This signal $\Delta f=f-f_{ref}$ is used as
the imaging signal in FM-AFM. Some researchers use the PLLs
oscillator signal to drive the cantilever. The advantage is the
greater spectral cleanliness of the PLL oscillator signal. A
disadvantage is that the cantilever drive loop becomes more
convoluted, and once the PLL is out of lock, the oscillation of
the cantilever stops.

\subsection{Experimental parameters}

While it was believed initially that the net force between the
front atom of the tip and the sample has to be attractive when
atomic resolution is desired, theoretical
\citep{Jarvis:2001,Sokolov:1999} and experimental evidence
\citep{Giessibl:2001a,Giessibl:2001d} suggests that atomic
resolution even on highly reactive samples is possible with
repulsive forces. Nevertheless, the dynamic modes are commonly
still called \lq\lq non-contact\rq\rq{} modes. For atomic studies
in vacuum, the FM-mode is now the preferred AFM technique.

FM-AFM was introduced by \citet*{Albrecht:1991} in magnetic force microscopy.
In these experiments, Albrecht \textit{et al.} imaged a thin film CoPtCr magnetic recording disk
(Fig. 7a in \citet{Albrecht:1991}) with a cantilever with a spring constant $k\approx10$\,N/m,
eigenfrequency $f_{0}=68\,485$\,Hz, amplitude $A=5$\,nm, a $Q$ value
of 40000 (\citet{Albrecht:2000}) and a tip with a thin magnetic film coverage. The noise level and
imaging speed was enhanced significantly
compared to amplitude modulation techniques. In 1993, the frequency modulation method was implemented in the
prototype of a commercial STM/AFM for ultra-high vacuum \citep{Giessibl:1994a}. Initial experiments on
KCl yielded excellent resolution
and soon after, the Si (111)-(7$\times$7) surface was imaged with true atomic resolution
for the first time \citep{Giessibl:1995}.
FM-AFM has five operating parameters:
\begin{enumerate}
\item The spring constant of the cantilever $k$.
\item The eigenfrequency of the cantilever $f_{0}$.
\item The $Q$-value of the cantilever $Q$.
\item The oscillation amplitude $A$.
\item The frequency shift of the cantilever $\Delta f$.
\end{enumerate}
The first three parameters are determined by the type of
cantilever that is used, while the latter two parameters can be
freely adjusted. The initial parameters which provided true atomic
resolution ($k=17$\,N/m, $f_{0}=114$\,kHz, $Q=28\,000$,
$A=34$\,nm, $\Delta f=-70$\,Hz) were found empirically.
Surprisingly, the amplitude necessary for obtaining good results
was very large compared to atomic dimensions. The necessity of
using large amplitudes for obtaining good results seems
counterintuitive, because the tip of the cantilever spends only a
small fraction during an oscillation cycle in close vicinity to
the sample. In hindsight, it appears that the large amplitudes
were required to prevent instabilities of the cantilever
oscillation (see section \ref{section_JTC}). Apparently, the
product between spring constant and amplitude (column \lq\lq
$kA\textrm{[nN]}$\rq\rq \,in Table \ref{table1}) has to be larger
than $\approx 100$\,nN to provide a sufficiently strong
withdrawing force. In the experiments conducted in 1994 (see rows
1 and 2 in Table \ref{table1}), this condition was not met, and
correspondingly, the resolution was not quite atomic yet. An
additional lower-threshold condition for $A$ is proposed:
$E=\frac{1}{2}kA^2$ (column \lq\lq $E\textrm{[keV]}$\rq\rq \,in
Table \ref{table1}) should be large compared to $\Delta E_{ts}$
defined in Eq. \ref{deltaEts}. This condition is required for
maintaining stable oscillation amplitudes as exemplified below. As
shown in Table \ref{table1}, atomic resolution on silicon and
other samples was reproduced by other groups with similar
operating parameters $\Delta f\approx -100$\,Hz, $k\approx
20$\,N/m, $f_{0}\approx 200$\,kHz and $A\approx 10$\,nm. Several
commercial vendors now offer FM-AFMs that operate with these
parameters \citep{Jeol,Omicron}.
\begin{table} [h]
\tiny
\begin{tabular}{|l||c|c|c|c|c|c|c|c|c|c|}
\hline
year\hspace{1.2cm} & $k$  & $f_{0}$&$\Delta f$&$A$&$\gamma$& $kA$ &$E$ &$\Delta E_{CL}$ &  sample &  \,\,Ref.\,\,   \\
 &  \,\,N/m\,\, & \, \,kHz\, \, & \,\,Hz\,\, &\,\,nm\,\,&\,\,fN$\sqrt{\textrm{m}}\,\,$& \,\,nN \,\,& \,\,keV\, & \,\,eV***\, & &\\ \hline\hline
1994*  &  2.5 & 60.0  &  -16   & 15.0 &     -1.26 & 37.5 &   1.8 &  0.06 & KCl(001) & \citet{Giessibl:1994a} \\ \hline
1994*  &  2.5 & 60.0  &  -32   &  3.3 &     -0.29 & 8.25 &   0.1 &  0.4  & Si(111) & \citet{Giessibl:1994b} \\ \hline \hline
1995   & 17.0 & 114.0 &  -70   & 34.0 &     -66.3 & 544  &  61   & 14    & Si(111) & \citet{Giessibl:1995} \\ \hline
1995   & 43.0 & 276.0 &  -60   & 40.0 &     -75.6 & 1720 & 215   & 27    & Si(111) & \citet{Kitamura:1995} \\ \hline
1995   & 34.0 & 151.0 &   -6   & 20.0 &     -3.91 & 680  &  42   &  5    & InP(110) & \citet{Sugawara:1995} \\ \hline
1996   & 23.5 & 153.0 &  -70   & 19.0 &     -28.8 & 447  &  27   &  3.3  & Si(111) & \citet{Luethi:1996} \\ \hline
1996   & 33.0 & 264.0 & -670   &  4.0 &     -23.6 & 132  &  12   &  1.45 & Si(001) & \citet{Kitamura:1996} \\ \hline
1996   & 10.0 & 290.0 &  -95   & 10.0 &     -3.42 & 100  &   3.1 &  0.4  & Si(111) & \citet{Guethner:1996} \\ \hline
1997   & 30.0 & 168.0 &  -80   & 13.0 &     -21.9 & 390  &  16   &  2    & NaCl(001) & \citet{Bammerlin:1997} \\ \hline
1997   & 28.0 & 270.0 &  -80   & 15.0 &     -15.7 & 420  &  20   &  2.5  & TiO$_{2}$(110) & \citet{Fukui:1997} \\ \hline
1997   & 41.0 & 172.0 &  -10   & 16.0 &     -4.96 & 654  &  33   &  4    & Si(111) & \citet{Sugawara:1997} \\ \hline
1999   & 35.0 & 160.0 &  -63   &  8.8 &     -10.1 & 338  &  10   &  1.4  & \,\,HOPG(0001)\,\, & \citet{Allers:1999a} \\ \hline
1999   & 36.0 & 160.0 &  -60.5 & 12.7 &     -18.1 & 457  &  18   &  2.3  & InAs(110) & \citet{Schwarz:1999} \\ \hline
1999   & 36.0 & 160.0 &  -92   &  9.4 &     -19.8 & 338  &  10   &  1.2  & Xe(111) & \citet{Allers:1999b} \\ \hline
1999   & 27.4 & 152.3 &  -10   & 11.4 &      -2.2 & 312  &  11   &  1.4  & Ag(111) & \citet{Orisaka:1999} \\ \hline
2000   & 28.6 & 155.7 &  -31   &  5.0 &     -4.1  & 143  &   2.2 &  0.04 & Si(111) & \citet{Lantz:2000} \\ \hline
2000   & 30.0 & 168.0 &  -70   &  6.5 &     -6.6  & 195  &   4.0 &  0.5  & Cu(111) & \citet{Loppacher:2000a} \\ \hline
2001   &  3.0 &  75.0 &  -56   & 76   &     -46.9 & 228  &  54.1 &  7    & Al$_{2}$O$_{3}$(0001) & \citet{Barth:2001} \\ \hline
2002   & 24.0 & 164.7 &   -8   & 12.0 &     -1.5  & 288  &   2.2 &  1.4  & KCl$_{0.6}$Br$_{0.4}$ & \citet{Bennewitz:2002} \\ \hline
2002   & 46.0 & 298.0 &   -20  &  2.8 &     -0.46 & 129  &   1.1 &  0.13 & Si(111) & \citet{Eguchi:2002} \\ \hline \hline
2000** & 1800 & 16.86 & -160   &  0.8 &     -387  &1440  &   3.6 & 11    & Si(111) & \citet{Giessibl:2000c} \\ \hline
2001** & 1800 & 20.53 &   85   &  0.25&     +29.5 & 450  &   0.4 &  1    & Si(111) & \citet{Giessibl:2001d} \\ \hline
\end{tabular}
\caption{Operating parameters of various FM-AFM experiments: *early experiments with
nearly atomic resolution, experiments with standard parameters (classic NC-AFM) on semiconductors, metals and
insulators and **small amplitude experiments.
*** Internal cantilever damping calculated from $\Delta E=2\pi E/Q$. When $Q$ is
not quoted in the original publication, a $Q-$value of $50\,000$
is used as an estimate.} \label{table1}
\end{table}

 The use of high-$Q$ cantilevers with a stiffness of $k\approx
20$\,N/m oscillating with an amplitude of $A\approx 10$\,nm has
enabled many groups to routinely achieve atomic resolution by
FM-AFM. As shown in table \ref{table1}, this mode is used in many
laboratories now and we therefore call it the \lq\lq classic\rq\rq
FM-AFM mode. While the operating parameters of the classic FM-AFM
mode provide good results routinely, it was not proved initially
that these parameters yield optimal resolution. The search space
for finding the optimal parameters was not completely open,
because micromachined cantilevers were only available with a
limited selection of spring constants. A theoretical study has
shown later \citep{Giessibl:1999a}, that the optimal amplitudes
are in the \AA-range, requiring spring constants of the order of a
few hundred N/m, much stiffer than the spring constant of
commercially available cantilevers. This result has been verified
experimentally by achieving unprecedented resolution with
a cantilever with $k=1800$\,N/m and sub-nm oscillation amplitudes \citep{Giessibl:2000c,Giessibl:2001d}.
\citet{Eguchi:2002} have also achieved extremely high resolution with
a silicon cantilever with a stiffness of 46.0\,N/m and an oscillation
amplitude of only 2.8\,nm.

\section{PHYSICAL OBSERVABLES IN FM-AFM}

\subsection{Frequency shift and conservative forces} \label{delta_f_calculation}

\subsubsection{Generic calculation}
The oscillation frequency is the main observable in FM-AFM and it is important
to establish a connection between frequency shift and the forces acting between
tip and sample. While the frequency can be calculated numerically \citep{Anczykowski:1996},
an analytic calculation is important for finding the functional relationships between
operational parameters and the physical tip-sample forces.
\begin{figure}[h]
  \centering
  \includegraphics[width=8cm,clip=true]{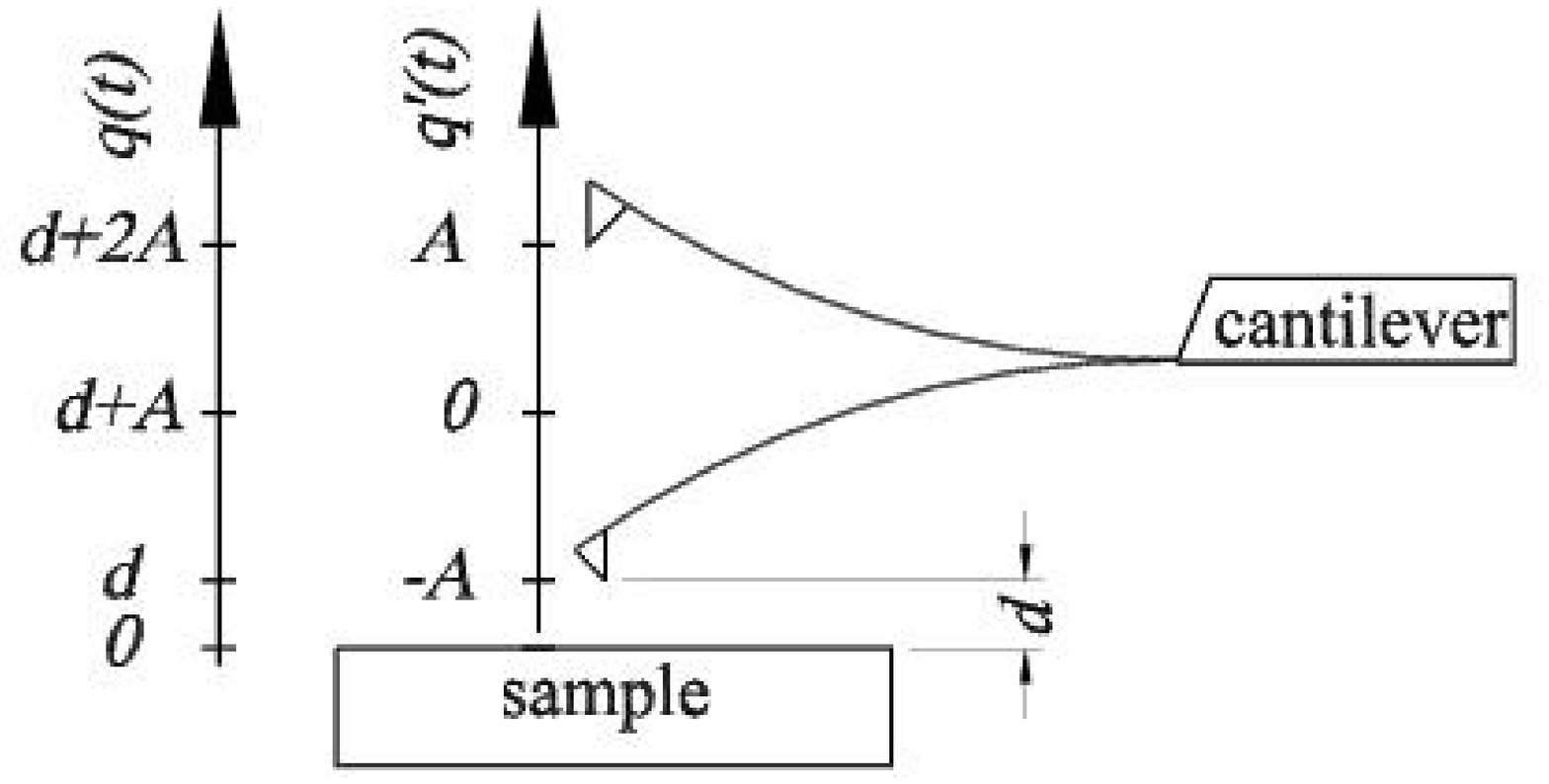}
  \caption{Schematic view of an oscillating cantilever at its upper and lower turnaround points.
  The minimum tip-sample distance is $d$ and the amplitude is $A$.}\label{oscillating_cl}
\end{figure}
The motion of the cantilever (spring constant $k$, effective mass $m^{*}$) can
be described by a weakly disturbed harmonic
oscillator. Figure \ref{oscillating_cl} shows the deflection $q^{\prime }(t)$ of the tip of the
cantilever: it oscillates with an amplitude $A$ at a distance $q(t)$ to a
sample. The closest point to the sample is $q=d$ and $q(t)=q^{\prime
}(t)+d+A $. The Hamiltonian of the cantilever is:
\begin{equation}
H=\frac{p^{2}}{2m^{*}}+\frac{kq^{\prime 2}}{2}+V_{ts}(q)  \label{H0}
\end{equation}
where $p=m^{*}dq^{\prime }/dt$. The unperturbed motion is given by:
\begin{equation}
q^{\prime }(t)=A\cos (2\pi f_{0}t)  \label{q0(t)}
\end{equation}
and the frequency is:
\begin{equation}
f_{0}=\frac{1} {2\pi} \sqrt{\frac{k} {m^{*}} }.  \label{f0}
\end{equation}
If the force
gradient $k_{ts}=-\frac{\partial F_{ts}}{\partial z}$ is constant during the
oscillation cycle, the calculation of the frequency shift is trivial:
\begin{equation}
\Delta f=f_{0}\frac{k_{ts}}{2k}.\label{df_Aklein}
\end{equation}
However, in classic FM-AFM $k_{ts}$ varies orders of magnitude during one
oscillation cycle and a perturbation approach as shown below has to be employed
for the calculation of the frequency shift.

The first derivation of the frequency shift in FM-AFM
\citep{Giessibl:1997b} utilized canonical perturbation theory (see
e.g. \citet{Goldstein:1980}). The result of this calculation is:
\begin{equation}
\Delta f=-\frac{f_{0}}{kA^{2}}<F_{ts}q^{\prime }>.  \label{df_HJ}
\end{equation}
where the pointed brackets indicate averaging across one oscillation cycle.

The applicability of first-order perturbation theory depends on the
magnitude of the perturbation, i.e. on the ratio between $V_{ts}$ and the
energy of the oscillating cantilever $E=H_{0}$. In FM-AFM, $E$ is typically
in the range of several keVs (see table \ref{table1}), while $V_{ts}$ is only
a few electron volts
and first order perturbation theory yields results for $\Delta f$ with
excellent precision.

An alternate approach to the calculation of $\Delta f$ has been
followed by \citet{Baratoff:1997},
\citet{Duerig:1999a,Duerig:1999b} and \citet{Livshits:1999}. This
approach also derives the magnitude of the higher harmonics and
the constant deflection of the cantilever.

This method involves solving Newton's equation of
motion for the cantilever (effective mass $\mu ^{*}$, spring constant $k$):
\begin{equation}
\mu ^{*}\frac{d^{2}q^{\prime }}{dt^{2}}=-kq^{\prime }+F_{ts}(q^{\prime }).
\label{DGL}
\end{equation}
The cantilever motion is assumed to be periodic, therefore it is expressed
as a Fourier series with fundamental frequency $f$:
\begin{equation}
q^{\prime }(t)=\sum_{m=0}^{\infty }a_{m}\cos (m2\pi ft).
\end{equation}
Insertion into Eq. \ref{DGL} yields:
\begin{equation}
\sum_{m=0}^{\infty }a_{m}\left[ -(m2\pi f)^{2}\mu ^{*}+k\right] \cos (m2\pi
ft)=F_{ts}(q^{\prime }).
\end{equation}
Multiplication by $\cos (l2\pi ft)$ and integration from $t=0$ to $t=1/f$
yields:
\begin{equation}
a_{m}\left[ -(m2\pi f)^{2}\mu ^{*}+k\right] \pi (1+\delta _{m0})=2\pi
f\int_{0}^{1/f}F_{ts}(q^{\prime })\cos (m2\pi ft)dt
\end{equation}
by making use of the orthogonality of the angular functions
\begin{equation}
\int_{0}^{2\pi }\cos (mx)\cos (lx)dx=\pi \delta _{ml}(1+\delta _{m0}).
\end{equation}
If the perturbation is weak, $q^{\prime }(t)\approx A\cos (2\pi ft)$
with $f=f_{0}+\Delta f$, $f_{0}=\frac{1}{2\pi} \sqrt{\frac{k}{\mu ^{*}}}$ and $\left|
\Delta f\right| \ll f_{0}.$ To first order, the frequency shift is given by:
\begin{equation}
\Delta f=-\frac{f_{0}^{2}}{kA}\int_{0}^{1/f_{0}}F_{ts}(q^{\prime
})\cos (2\pi f_{0}t)dt=-\frac{f_{0}}{kA^{2}}<F_{ts}q^{\prime }>
\label{df_F}
\end{equation}
which of course equals the result of the Hamilton-Jacobi method.

The results of these calculations are also applicable for
amplitude modulation AFM (\citet{Bielefeldt:1999}).
\citet{Hoelscher:1999} have also used a canonical perturbation
theory approach and extended it to show that the frequency shift
as a function of amplitude for inverse power forces can be
expressed as a rational function for \emph{all} amplitudes, not
just in the large amplitude limes. Sasaki and Tsukada have
obtained a similar result to Eq. \ref{df_HJ} with a different type
of perturbation theory
\citep{Sasaki:1998,Sasaki:1999,Tsukada:2002bk}.

\subsubsection{An intuitive expression for frequency shifts as a function of amplitude}
\begin{figure}[h]
  \centering
  \includegraphics[width=8cm,clip=true]{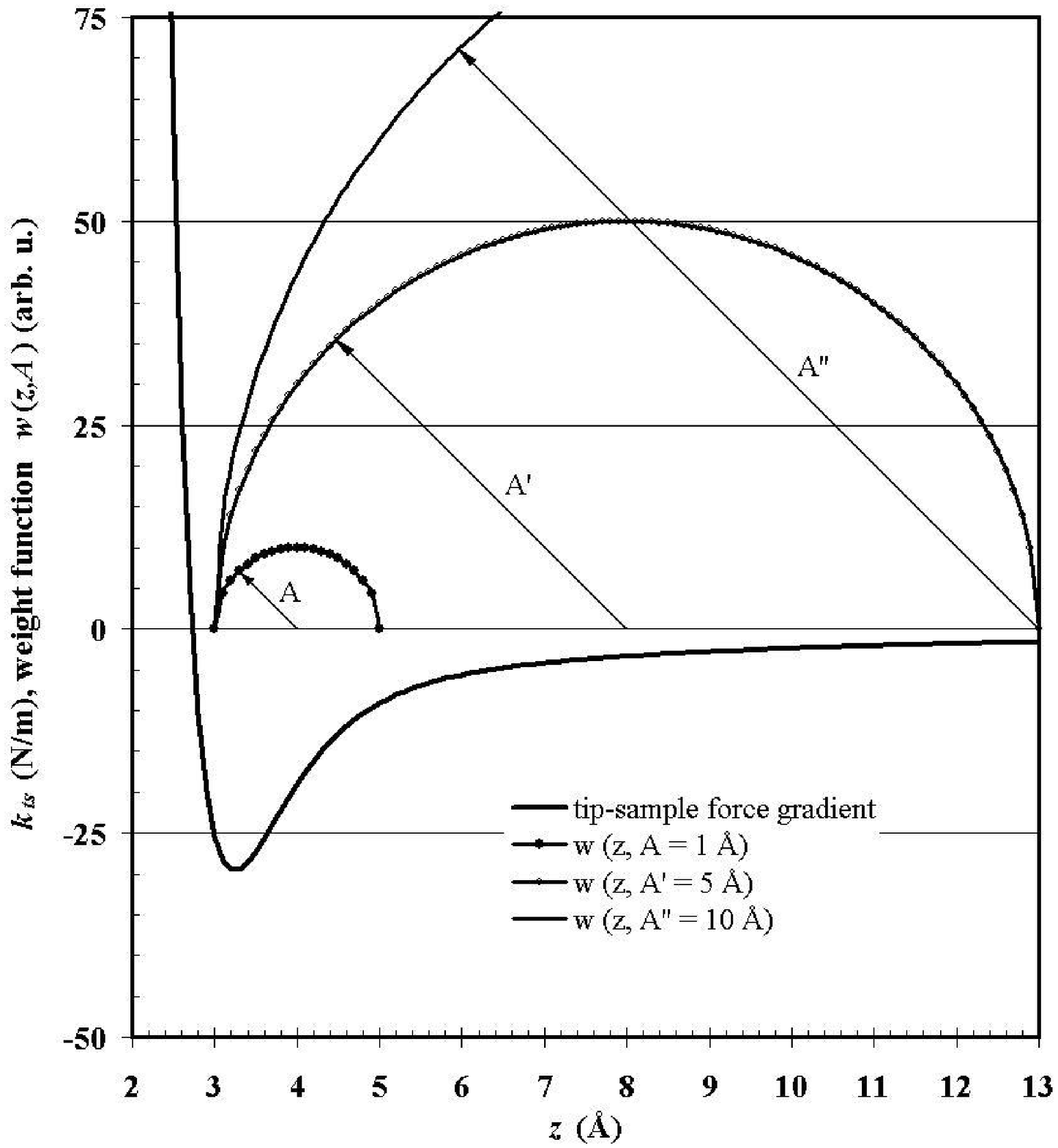}
  \caption{Calculation of the frequency shift $\Delta f$: $\Delta f$ is a convolution of a semi-spherical
  weight function with the tip-sample force gradient. The radius $A$ of the weight function is equal to the
  oscillation amplitude of the cantilever. The weight function $w$ is plotted in arbitrary units in this scheme -- $w$ has
  to be divided by $\pi A^2/2$ for normalization (see Fig. \ref{convolutionb}).}\label{convolutiona}
\end{figure}

\begin{figure}[h]
  \centering
  \includegraphics[width=8cm,clip=true]{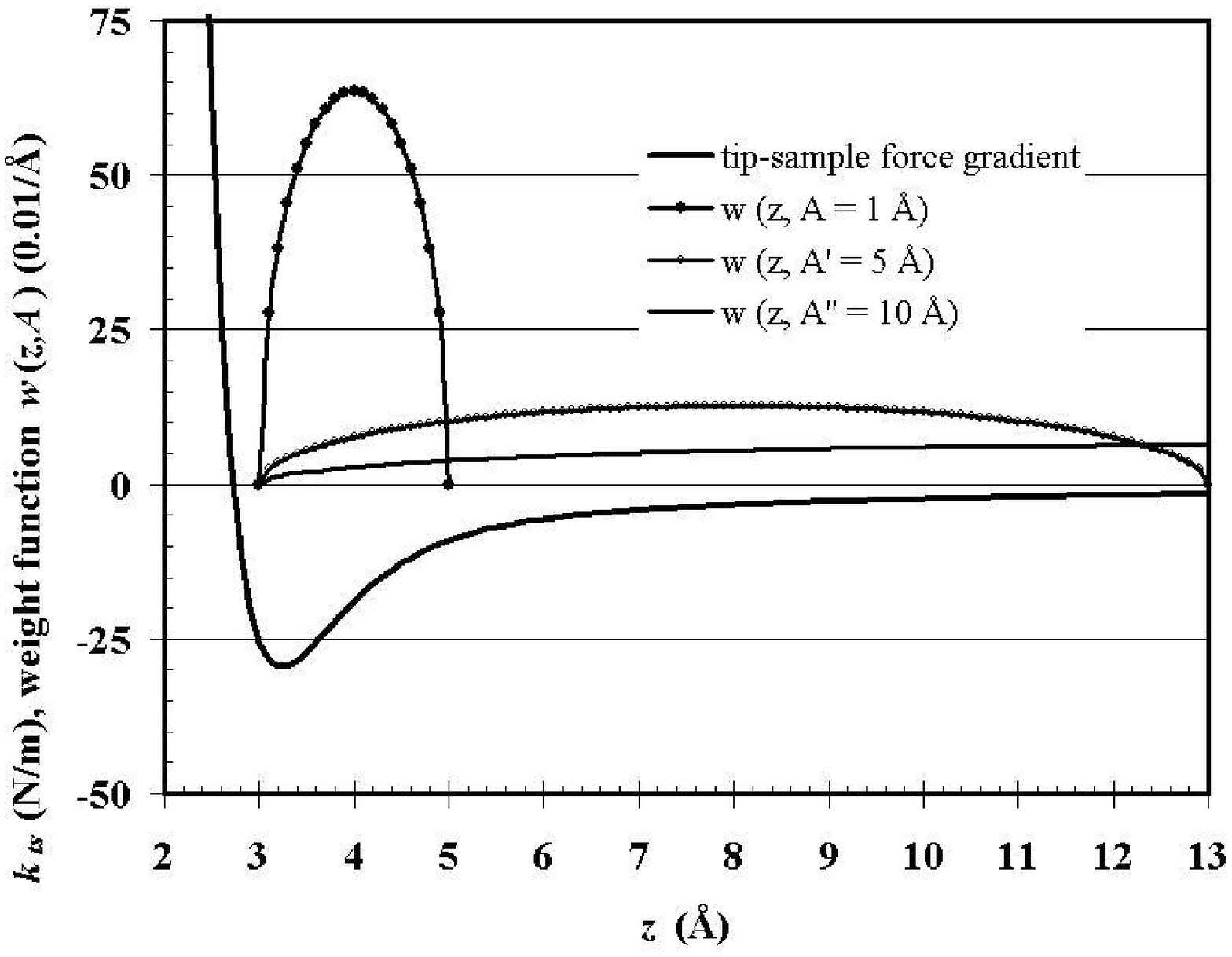}
  \caption{Calculation of the frequency shift $\Delta f$: $\Delta f$ is a convolution of a
  weight function $w$ with the tip-sample force gradient. For small amplitudes, short-range interactions
  contribute heavily to the frequency shift, while long-range interactions are attenuated. However,
  for large amplitudes the long-range interactions cause the main part of the frequency shift, and short-range
  interactions play a minor role.}\label{convolutionb}
\end{figure}

For small amplitudes, the frequency shift is a very simple
function of the tip-sample forces -- it is proportional to the
tip-sample force gradient $k_{ts}$. For large amplitudes, the
frequency shift is given by the rather complicated expressions Eq.
\ref{df_HJ} and Eq. \ref{df_F}. With integration by parts, these
complicated formulas transform into a very simple expression that
resembles Eq. \ref{df_Aklein} \citep{Giessibl:2001e}.
\begin{equation}
\Delta f(z)=f_{0}\frac{\langle k_{ts}(z) \rangle}{2k}
\label{deltaf_intuitive}
\end{equation}
with
\begin{equation}
\langle k_{ts}(z)\rangle=\frac{1}{\frac{\pi }{2}A^{2}}%
\int_{-A}^{A}k_{ts}(z-q^{\prime })\sqrt{A^{2}-q^{\prime 2}}dq^{\prime }.
\label{keff}
\end{equation}
This expression is closely related to Eq. \ref{df_Aklein}: the constant $%
k_{ts}$ of Eq. \ref{df_Aklein} is replaced by a weighted average $\langle k_{ts}\rangle$,
where the weight function $w(q^{\prime },A)$ is a semi circle with radius $A$
divided by the area of the semicircle $\Gamma =\pi A^{2}/2$ (see Fig. \ref{convolutiona}).
For $A \rightarrow 0$, the semicircular weight function with its
normalization factor $2/\pi A^2$ is a representation of Dirac's
Delta function. Figure \ref{convolutionb} shows the convolution
with the proper normalization factor, and it is immediately
apparent from this figure how the use of small amplitudes
increases the weight of the short-range atomic forces over the
unwanted long-range forces. The amplitude in FM-AFM allows to tune
the sensitivity of the AFM to forces of various ranges.

\subsubsection{Frequency shift for a typical tip-sample force} \label{subsection_frequency_shift_for}

The interaction of a macroscopic tip of an AFM with a sample is a
complicated many-body problem and $F_{ts}$ cannot be described by
a simple function. However, quite realistic model forces can be
constructed from linear combinations of the following basic types:
a) inverse-power forces, b) power forces and c) exponential
forces. Analytic expressions for the frequency shift as a function
of tip-sample distance $z$ and amplitude $A$ are listed in
\citet{Giessibl:2000a}. A typical tip-sample force is composed of
long range contributions and short range contributions. This force
can be approximated by a long-range van-der-Waals component and a
short-range Morse type interaction:
\begin{equation}
F_{ts}(z)=\frac{C}{z+\sigma}+2\kappa E_{bond}(-e^{-\kappa (z-\sigma )}+e^{-2\kappa (z-\sigma )}).
\label{specific_force}
\end{equation}
$C$ depends on the tip angle and the Hamaker constant of tip and sample,
and $E_{bond}, \sigma$ and $\kappa$ are the bonding energy, equilibrium
distance and decay length of the Morse potential respectively.
With the results derived in \citet{Giessibl:2000a}, the resulting frequency shift is:

\begin{eqnarray}
\Delta f(z,A)&=&
\frac{f_{0}} {k A} \frac{C}{z+\sigma}
\left[ F_{1}^{1,1/2}(\frac{-2A}{z+\sigma})-F_{2}^{1,3/2}(\frac{-2A}{z+\sigma})\right] \nonumber\\
&&-  f_{0}\frac{2\kappa E_{bond}} {k A} \left\{e^{-\kappa z}\left[ M_{1}^{1/2}(-2\kappa
A)-M_{2}^{3/2}(-2\kappa A)\right]\right. \nonumber\\
&&+  \left. e^{-2\kappa z}\left[ M_{1}^{1/2}(-4\kappa
A)-M_{2}^{3/2}(-4\kappa A)\right]\right\}.                    \label{df_specific}
\end{eqnarray}

where $F_{c}^{a,b}(z)$ is the Hypergeometric Function and
$M_{b}^{a}(z)$ is Kummer's Function \citep{Abramowitz:1970}.

\begin{figure}[h]
  \centering
  \includegraphics[width=8cm,clip=true]{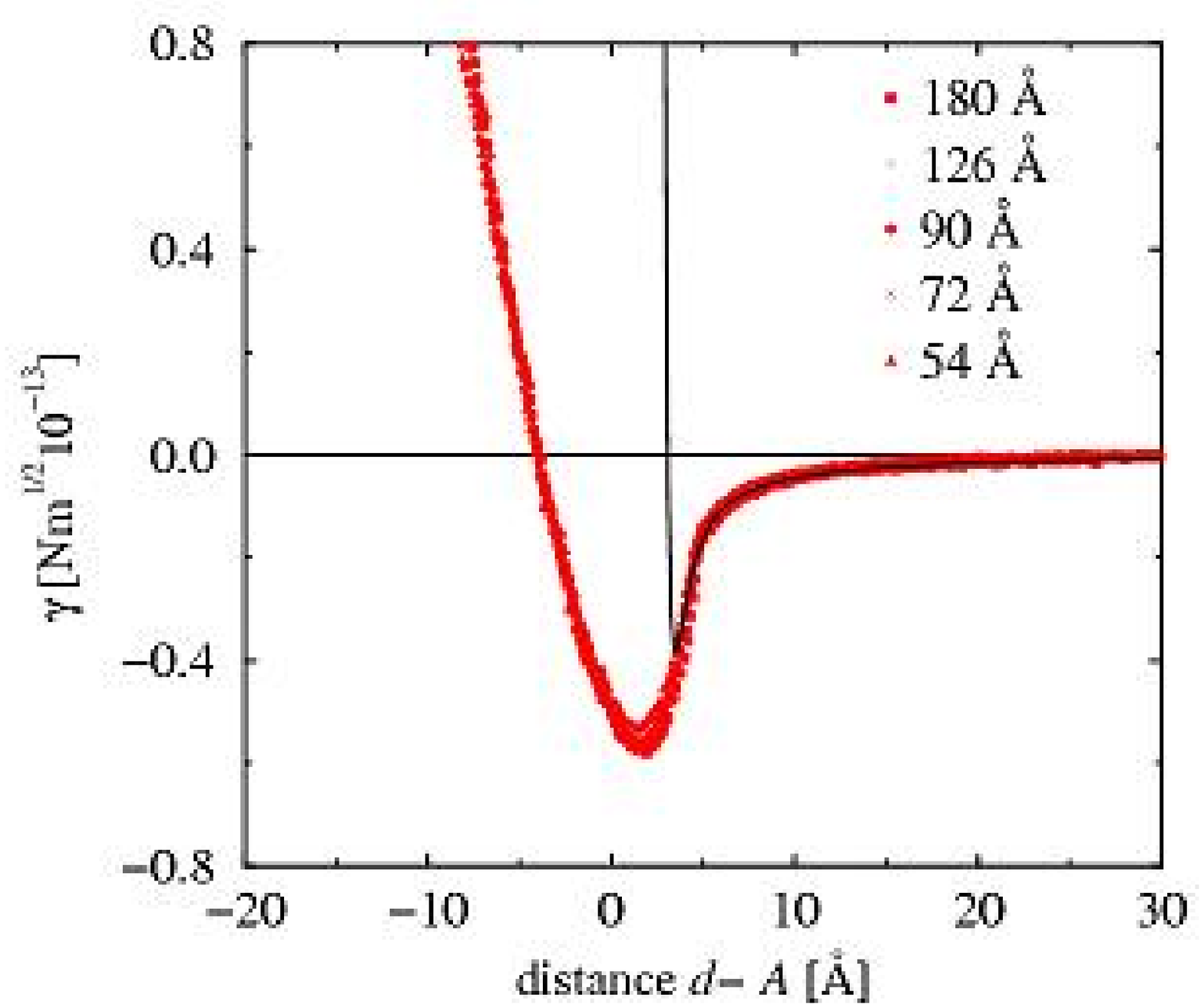}
  \caption{Experimental normalized frequency-shift versus
  distance data acquired with a low-temperature UHV AFM with a
  graphite sample surface and a silicon cantilever. The average
  distance between the center of the tip's front atom and the plane
  defined by the centers of the surface atom layer is $d$, thus the
  minimal tip-sample distance is $d-A$. Five experimental frequency
  shift versus distance data sets with amplitudes from
  54\,\AA{} to 180\,\AA{} are expressed in normalized frequency shift
  $\gamma=kA^{3/2}\Delta f/f_0$. The five experimental data sets
  match exactly, proving the validity of the concept of a normalized
  frequency shift.
  The black curve is a simulated $\gamma(d-A)$-curve using a Lennard-Jones
  short-range force and a $1/(d-A)$ long-range force.
  In the repulsive regime, the deviation between the experimental dots
  and the simulated curve is substantial, because the
  $1/(d-A)^{12}$-dependence of the repulsive Lennard-Jones potential
  only describes the interaction of the tip- and sample atom.
  The sample atom is embedded in a lattice with finite stiffness, and
  in particular graphite is a very soft material. Hertzian
  contact theory (see e.g. \citet{Chen:1993}) is an appropriate model
  when the repulsive forces are large enough to cause overall sample
  deformations. Source: \citet{Hoelscher:2000}.}\label{gamma_graphite}
\end{figure}

\begin{figure}[h]
  \centering \includegraphics[width=8cm,clip=true]{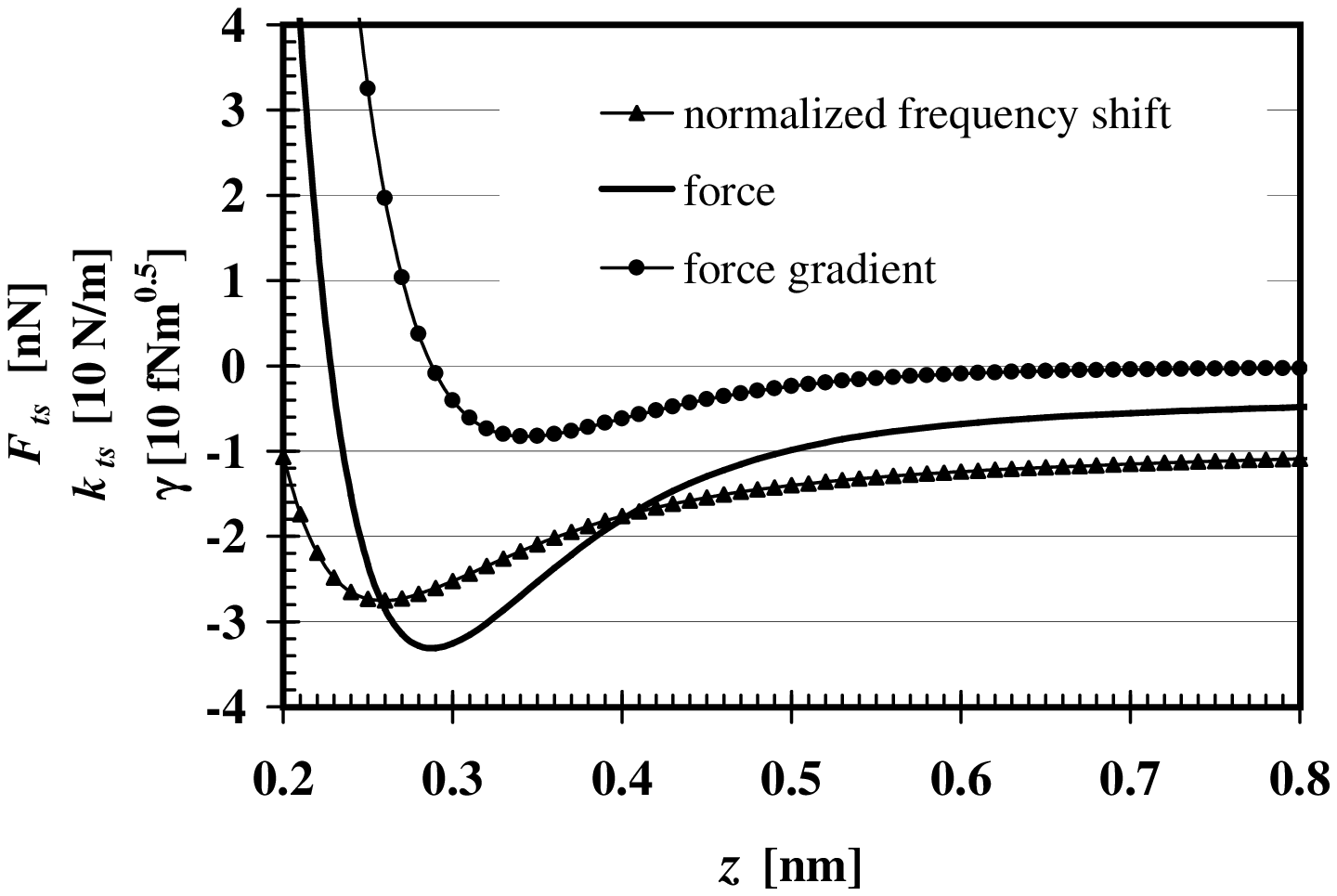}
  \caption{Force $F_{ts}(z)$, force gradient $k_{ts}(z)$ and
  large-amplitude normalized frequency shift $\gamma_{lA}(z)$ for the
  tip-sample force defined in Eq. \ref{specific_force}. When the
  cantilevers oscillation amplitude $A$ is small compared to the range
  of the tip-sample forces $\lambda$, the frequency shift is
  proportional to the force gradient, while for $A>>\lambda$, the
  frequency shift is proportional to $\gamma_{lA}(z)$.}\label{deltafA}
\end{figure}

Equation \ref{df_specific} describes the frequency shift as a
function of amplitude and tip-sample distance. For small
amplitudes, the frequency shift is independent of the amplitude
and proportional to the tip-sample force gradient $k_{ts}$ (Eq.
\ref{df_Aklein}). For amplitudes that are large compared to the
range of the tip-sample force, the frequency shift is a function
of the amplitude $\Delta f \propto A^{-1.5}$.
If amplitudes larger than the range of the relevant forces
are used, it is helpful to introduce
a \lq\lq normalized frequency shift\rq\rq $\gamma$ defined by:
\begin{equation}
\gamma (z,A):=\frac{kA^{3/2}}{f_{0}}\Delta f(z,A).
\label{def_gamma}
\end{equation}
For large amplitudes, $\gamma (z,A)$ asymptotically approaches a
constant value (see Fig. 2 in Ref. \citet{Giessibl:2000a} ), i.e.
$\lim_{A\to\infty}\gamma (z,A) \equiv \gamma_{lA} (z)$. The
normalized frequency shift is calculated from the tip-sample force
with
\begin{equation}
\gamma _{lA}(z)=\frac{1}{\sqrt{2}\pi }\int_{0}^{\infty }\frac{F_{ts}(z+z')}{%
\sqrt{z'}}dz'.  \label{def_gamma_la}
\end{equation}
The normalized frequency shift helps to characterize AFM experiments
and has a similar role as the tunneling impedance in STM on metals.
\citet{Hoelscher:2000} have performed frequency shift versus
distance measurements with a silicon cantilever on a graphite
surface with amplitudes ranging from 54\,\AA{} to 180\,\AA{} and
verified the concept of the normalized frequency shift $\gamma =
\Delta f \times k \times A^{3/2}/f_0$ as the pertinent imaging
parameter in classic FM-AFM (see Fig. \ref{gamma_graphite}). Five
frequency shift curves taken with amplitudes ranging from
54\,\AA{} to 180\,\AA{} match precisely when rescaled using the
normalized frequency shift.
Thus, for small amplitudes the frequency shift is very sensitive to short-range forces, because
short-range forces have a very strong force gradient, while for large amplitudes, long-range
forces contribute heavily to the frequency shift.
Figure \ref{deltafA} shows the tip-sample force defined in Eq. \ref{specific_force} and the
corresponding force gradient and normalized frequency shift $\gamma_{lA}$. The parameters for the short-range
interaction are adopted from \citet{Perez:1998}: $\kappa=12.76$ nm$^{-1}$,
$E_{bond}=2.273$\,eV
and $\sigma=2.357$\,\AA.
The force gradient is vanishing for $z>6$\,\AA, while the
normalized frequency shift for large amplitudes reaches almost
half its maximum at this distance. The dependence of the frequency
shift with amplitude shows that \emph{small amplitudes increase
the sensitivity to short-range forces!} The possibility of
adjusting the amplitude in FM-AFM compares to tuning an optical
spectrometer to a passing wavelength. When short-range
interactions are to be probed, the amplitude should be in the
range of the short-range forces. While using amplitudes in the \AA
-range has been elusive with conventional cantilevers because of
the instability problem described in subsection \ref{section_JTC},
stiff sensors such as the qPlus sensor displayed in Fig.
\ref{qPlus} are suited well for small-amplitude operation.

\subsubsection{Deconvolution of forces from frequency shifts}

Frequency shifts can be measured with high accuracy and low noise,
while the measurement of dc-forces is subject to large noise.
However, forces and not frequency shifts are of primary physical
interest. A number of methods have been proposed to derive forces
from the frequency shift curves.

The first type of methods requires the relation of frequency shift
versus distance $\Delta f (z)$ over the region of interest.
Because force and frequency shift are connected through a
convolution, a deconvolution scheme is needed to connect forces
(or force gradients) to the frequency shift and vice versa.
\citet{Giessibl:1997b} has proposed to build a model force
composed from basic functions (inverse power- and exponential
forces) and fit the parameters (range and strength) of the model
force such that its corresponding frequency shift matches the
experimental frequency shift. \citet{Gotsmann:1999} have proposed
a numerical algorithm and \citet{Duerig:1999b} has invented an
iterative scheme for force deconvolution. \citet{Giessibl:2001e}
has proposed a simple and intuitive matrix method to deconvolute
forces from frequency shifts.

The second type of spectroscopy methods requires to know the
frequency as a function of cantilever amplitude $\Delta f (A)$.
\citet{Hoelscher:1999a} and \citet{Hoelscher:2002bk} have modified
the method elucidated in $\S 12$ of Landau's textbook on classical
mechanics \citep{Landau:1990} to recover the interaction potential
from the dependence of the oscillation period $T=1/f$ from energy
$E=kA^2/2$.

For the third type, invented by \citet{Duerig:2000}, the full
tip-sample potential curve can be recovered within the $z-$
interval covered by the cantilever motion if the amplitudes and
phases of all the higher harmonics of the cantilever motion are
known. This method is very elegant because, in principle, the
higher harmonics can be measured in real time which obliterates
the need to take time consuming $\Delta f (z)$ or $\Delta f (A)$
spectra. D\"{u}rigs method is particularly promising for
small-amplitude operations, because then the first few harmonics
at $2f,3f ...$ already contain characteristic information about
the tip sample potential.

\subsection{Average tunneling current for oscillating tips}

When the tip of the cantilever and the sample are both conductive,
simultaneous STM and FM-AFM operation is possible, i.e. the tunneling
current $I_{t}$ as well as the frequency shift can be recorded while scanning
the surface. In most cases, the bandwidth of the tunneling
current-preamplifier is much smaller than the oscillation frequency $f_{0}$
of typical cantilevers. The measured tunneling current is given by the
time-average over one oscillation cycle. With the exponential distance
dependence $I_{t}(z)=I_{0}e^{-2\kappa _{t}z}$ (see Eq. \ref{It_z}) we find:
\begin{equation}
\langle I_{t}(z,A)\rangle =I_{0}e^{-2\kappa _{t}z}M_{1}^{1/2}(-4\kappa
_{t}A)
\end{equation}
where $M_{b}^{a}(\zeta )$ is the Kummer Function \citep{Abramowitz:1970}. When $%
\kappa _{t}A\gg 1$,

\begin{equation}
\langle I_{t}(z,A)\rangle \approx I_{t}(z,0)/\sqrt{4\pi \kappa _{t}A}.
\label{average_current}
\end{equation}

\begin{figure}[h]
  \centering
  \includegraphics[width=8cm,clip=true]{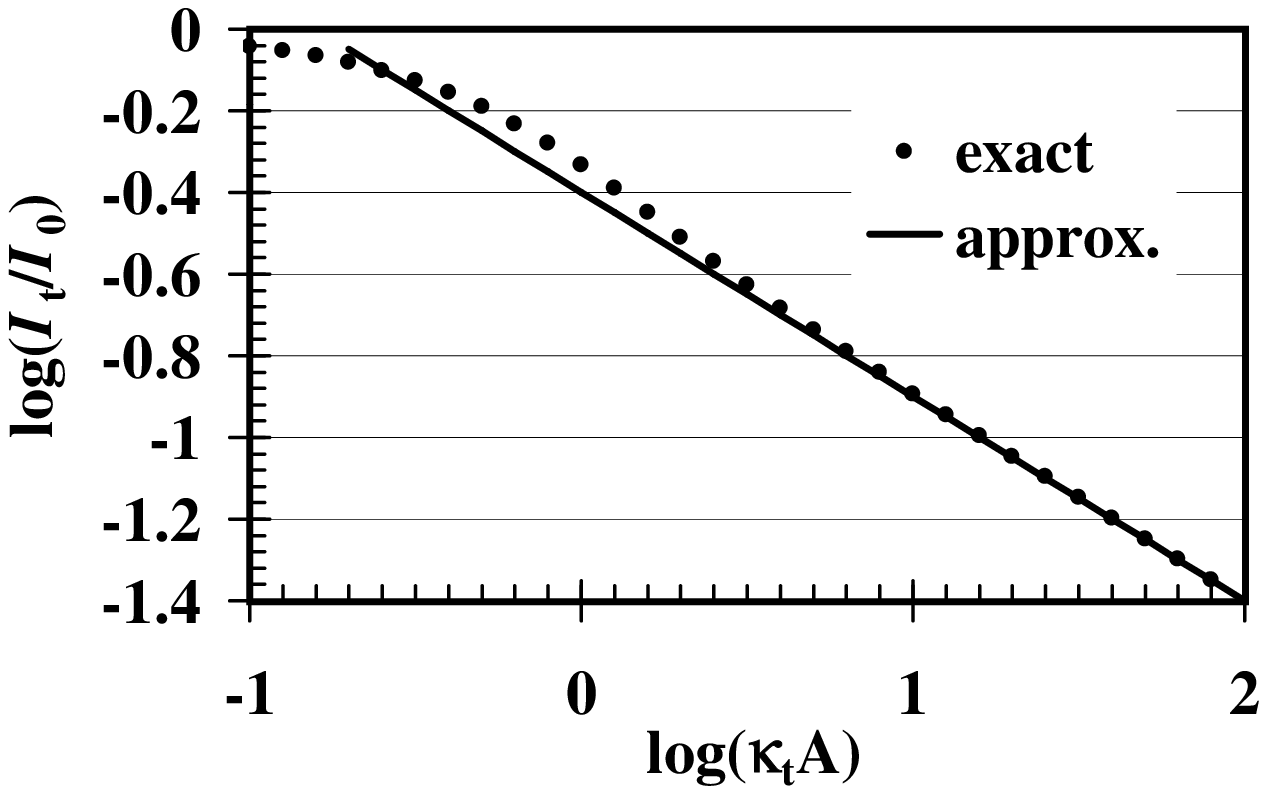}
  \caption{Averaged tunneling current as a function of amplitude for a fixed minimum tip-sample distance.}\label{It_A}
\end{figure}

Figure \ref{It_A} shows the dependence of the tunneling current as a function
 of the product between $\kappa _{t}$ and $A$.
For $A=5$\,nm and $\kappa _{t}=1$\,\AA$^{-1}$, the average tunneling current is $%
\approx 1/25$ of the value when the cantilever does not oscillate. Because the
noise of the current measurement decreases with an increasing average tunneling
current, the use of small amplitudes
improves the quality of simultaneous STM and FM-AFM measurements.

It is noted that Eq. \ref{average_current} is an upper threshold.
When using large amplitudes ($\kappa _t A\gg 1$), the tunneling
current vs. time is a series of Gaussian functions spaced by $1/f$
where $f$ is the oscillation frequency of the cantilever.
Especially when using cantilevers with large eigenfrequencies, the
tunneling current varies very rapidly with time. Because of
slew-rate and bandwidth limitations, typical tunneling
preamplifiers are unable to convert these rapid current variations
in output voltage swings. Thus, the experimental average current can
even become smaller than given by Eq. \ref{average_current}.

\subsection{Damping and dissipative forces}

Conservative tip-sample forces cause a frequency shift. A non-conservative component in the tip-sample force,
that is a hysteresis in the force versus distance graph
\begin{equation}
\Delta E_{ts}(\overrightarrow{x})=\oint\limits_{\Lambda}
\overrightarrow{F}_{ts}(\overrightarrow{x}+\overrightarrow{x}')d\overrightarrow{x}',\label{deltaEts}
\end{equation}%
where $\Lambda$ is the trajectory of the oscillating cantilever,
causes energy loss in the motion of the cantilever. This energy
loss is measurable. The cantilever itself already dissipates
energy (internal dissipation). When the tip of the cantilever is
far from the sample, the damping of the cantilever is due to
internal dissipation and the energy loss per oscillation cycle is
given by:
\begin{equation}
\Delta E_{CL} = 2\pi \frac{E}{Q}
\end{equation}
where $E=kA^{2}/2$ is the energy of the cantilever and $Q$ is its quality factor. When the phase angle
between the excursion of the actuator and the excursion of the cantilever is
exactly $\varphi=\pi/2$, the cantilever oscillates at frequency $f_{0}$ and the driving signal is
$A_{drive}=Ae^{i\pi/2}/Q$. Hence, the driving amplitude and dissipation are connected:
\begin{equation}
|A_{drive}| = |A|\frac{\Delta E_{CL}}{2\pi E}.
\end{equation}
When the tip oscillates close to the sample, additional damping occurs and the driving signal $A_{drive}$
is increased by the oscillator control electronics to $A'_{drive}$ for maintaining a constant amplitude $A$
where
\begin{equation}
|A'_{drive}| = |A|\frac{\Delta E_{CL}+\Delta E_{ts}}{2\pi E}=|A|\left( \frac{1}{Q}
+\frac{\Delta E_{ts}}{2\pi E}\right).\label{Adrive}%
\end{equation}
Equation \ref{Adrive} has an important implication on the optimal
$Q$ factor of the cantilever. While a high $Q$ factor results in
low frequency noise (see Eq. \ref{f_error}), Eq. \ref{Adrive}
shows that the $Q$ value of the cantilever should not be much
higher than the ratio $2\pi E/\Delta E_{ts}$. If $Q$ is much
higher than this value, it is difficult for the oscillator circuit
to maintain a constant amplitude, because small changes in $\Delta
E_{ts}$ require a major correction in the control output $g$.

Measuring the
damping signal yields the dissipation in the approach and retract phases of the oscillating tip where
\begin{equation}
\Delta E_{ts} = 2\pi \frac{E}{Q} \left(\frac{|A'_{drive}|} {|A_{drive}|}-1 \right).
\end{equation}
The ratio $|A'_{drive}|/|A_{drive}|$ is easily accessible in the
dc input ($g$) of the analog multiplier chip in Fig.
\ref{fmprinciple} -- an increase in the tip-sample dissipation
$\Delta E_{ts}$ is reflected in an increased gain signal $g'$ in
the oscillator electronics and $g'/g=|A'_{drive}|/|A_{drive}|$.
Several authors have recorded this signal simultaneously with the
frequency shift and thus measured both elastic and non-elastic
interaction forces simultaneously, see e.g.
\citet{Luethi:1997,Bammerlin:1997,Ueyama:1998,Hug:2002bk}.

Physical origins of dissipation are discussed in
\citet{Duerig:1999a,Abdurixit:2000,Hoffmann:2001a,Gauthier:2002bk,Hug:2002bk,Giessibl:2002a}.

It is noted, that dispersions in the oscillator circuit and in the actuator assembly can lead to artifacts in the
interpretation of damping data, because $|A_{drive}|=|A|/Q$ only holds for $f=f_{0}$.
\citet{Anczykowski:1999} have introduced a method that yields the correct dissipation
energy even for cases where the phase angle between actuator and cantilever is not $\varphi=\pi/2$.

Mechanical resonances
in the actuator assembly are likely to occur at the high resonance frequencies of conventional cantilevers.
These resonances can cause sharp variations of the phase with frequency and thus create artifacts in the
measurement of $\Delta E_{ts}$. A self-oscillation technique for cantilevers (\citet{Giessibl:1997a})
helps to avoid these resonances.

\section{NOISE IN FREQUENCY MODULATION AFM} \label{chapter_noise_in_fmafm}

\subsection{Generic calculation}

\begin{figure}[h]
  \centering
  \includegraphics[width=8cm,clip=true]{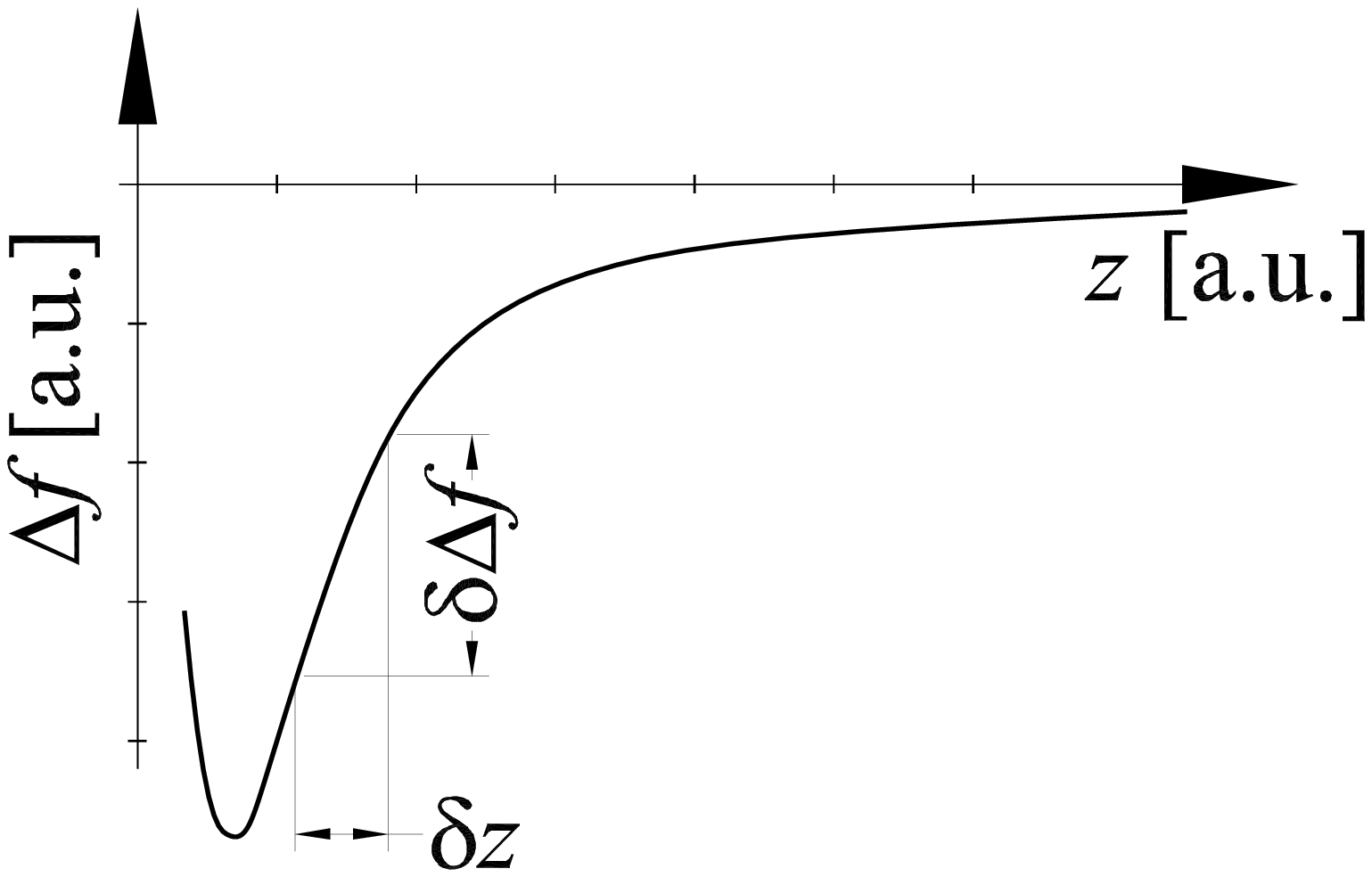}
  \caption{Schematic plot of frequency shift $\Delta f$ as a function of tip-sample distance $z$.
  The noise in the tip-sample distance measurement is given by the noise of the
  frequency measurement $\Delta f$ divided by the slope of the frequency shift curve.}\label{deltaf_z}
\end{figure}

The vertical noise in FM-AFM can be calculated in the same fashion as in the STM
case (see Fig. \ref{current_z}); it is given by the ratio between the noise in
the imaging signal and the slope of the imaging signal with respect to $z$:
\begin{equation}
\delta z=\frac{\delta \Delta f}{|\frac{\partial \Delta f}{\partial z}|}.
\label{z_noise_fmafm}
\end{equation}
Figure \ref{deltaf_z} shows a typical frequency shift versus
distance curve. Because the distance between the tip and sample is
measured indirectly through the frequency shift, it is clearly
evident that the noise in the frequency measurement $\delta \Delta
f$ translates into vertical noise $\delta z$ and is given by the
ratio between $\delta \Delta f$ and the slope of the frequency
shift curve $\Delta f(z)$ (Eq. \ref{z_noise_fmafm}). Low vertical
noise is obviously obtained for a low-noise frequency measurement
and a steep slope of the frequency shift curve. Additional
boundary conditions apply: if the force between front atom and
surface is too large, the front atom or larger sections of tip or
sample can shear off. It is interesting to note, that in FM-AFM
the noise will increase again upon further reducing the tip-sample
distance when approaching the minimum of the $\Delta f(z)$ curve.
Because the frequency shift is not monotonic with respect to $z$,
stable feedback of the microscope is only possible either on the
branch of $\Delta f$ with positive slope or on the one with
negative slope. In FM-AFM with atomic resolution, the branch with
positive slope is usually chosen. However, when using very small
amplitudes, it is also possible to work on the branch with
negative slope (see \citet{Giessibl:2001a}).

It is of practical importance to note that the minimum of the
frequency versus distance curve shown in Fig. \ref{deltaf_z} is a
function of the lateral tip position. Directly over a sample atom,
the minimum can be very deep. However, at other sample sites there
may be small negative frequency shift. Imaging can only be
performed with frequency shift setpoints which are reachable on
every ($x,y$) position on the imaged sample area, otherwise a tip
crash occurs.

\subsection{Noise in the frequency measurement}

Equation \ref{z_noise_fmafm} shows that the accuracy of the
frequency shift measurement determines directly the vertical
resolution in FM-AFM. What is the accuracy of the measurement of
the oscillation frequency of the cantilever? \citet{Martin:1987},
\citet{McClelland:1987} have studied the influence of thermal
noise on the cantilever and \citet{Albrecht:1991} and
\citet{Smith:1995} have calculated the thermal limit of the
frequency noise. Leaving aside prefactors of the order of $\pi$,
all these authors come to a similar conclusion, namely that the
square of the relative frequency noise is given by the ratio
between the thermal energy of the cantilever ($k_BT$) and the
mechanical energy stored in it ($0.5kA^2$), divided by its quality
factor $Q$ and multiplied by the ratio between bandwidth $B$ and
cantilever eigenfrequency $f_0$. Specifically,
\citet{Albrecht:1991} find
\begin{equation}
\frac{\delta f_{0}}{f_{0}}= \sqrt{\frac{k_{B}TB}{\pi k A^{2} f_{0} Q}}.  \label{f_error}
\end{equation}
\citet{Albrecht:1991} support Eq. \ref{f_error} with measurements
on the dependence of $\delta f_{0}$ with $Q$ (Fig. 5 in
\citet{Albrecht:1991}) and $A$ (Fig. 6 in \citet{Albrecht:1991})
and clearly state, that Eq. \ref{f_error} only contains the
thermal cantilever noise and disregards the noise of the
deflection sensor. Correspondingly, the frequency noise becomes
larger than predicted by Eq. \ref{f_error} for large $Q-$ values
in Fig. 5 of \citet{Albrecht:1991}. This deviations are traced to
interferometer noise, e.g. noise in the cantilever's deflection
sensor. \citet{Albrecht:1991} do not provide measurements of
$\delta f_{0}$ as a function of bandwidth $B$. Equation
\ref{f_error} predicts a $B^{0.5}-$dependence. However,
theoretical arguments by \citet{Duerig:1997} and an analysis and
measurements by \citet{Giessibl:2002bk} indicate a
$B^{1.5}-$dependence of frequency noise. The following analysis
shows the reasons for that.

\begin{figure}[h]
  \centering
  \includegraphics[width=8cm,clip=true]{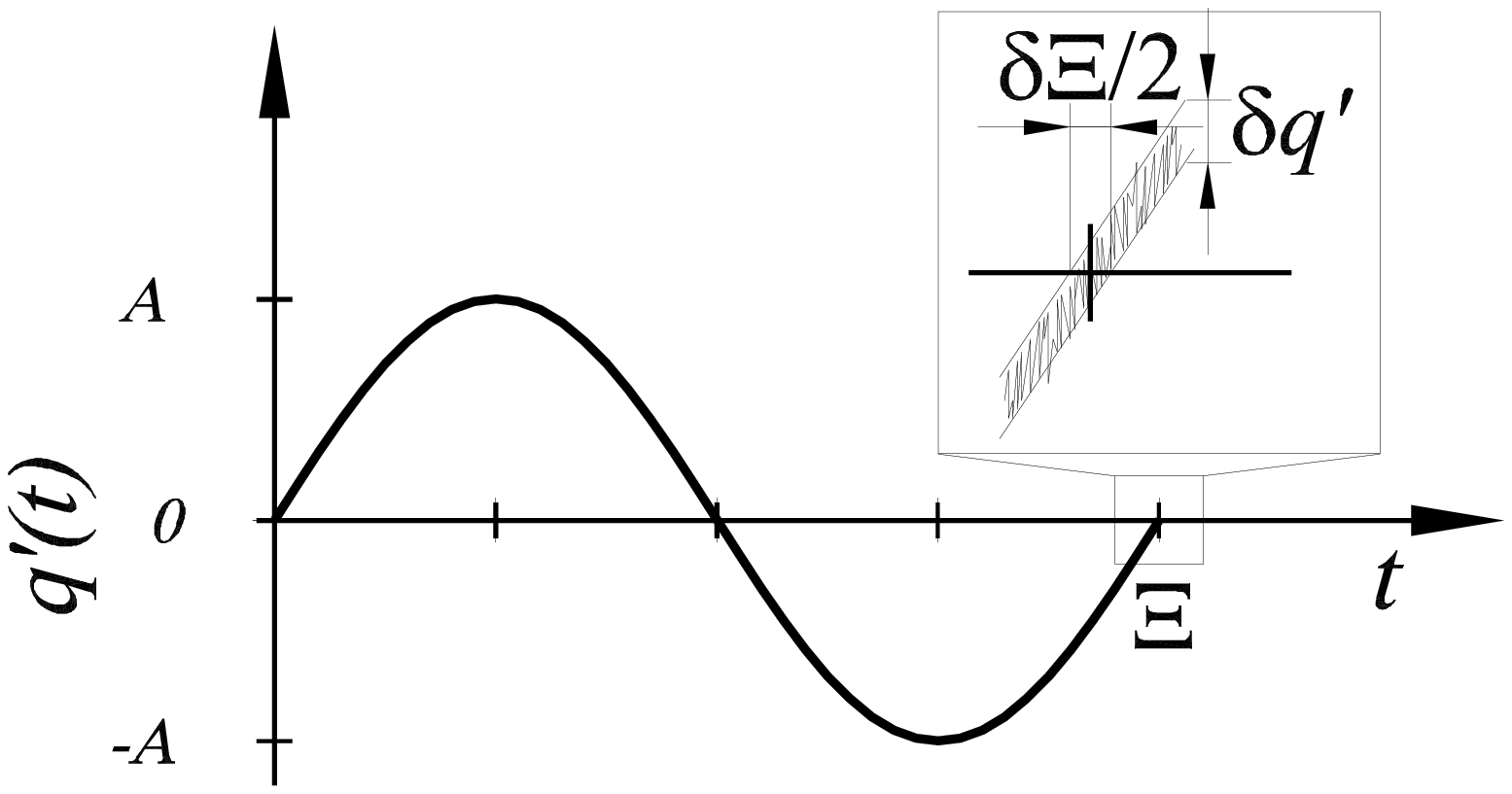}
  \caption{Typical cantilever deflection signal as it appears on an oscilloscope.
  The oscillation frequency is given by the inverse time lag between two
  consecutive zero-crossings with positive velocity.}\label{fmnoise}
\end{figure}

The frequency is given by the inverse of the time lag $\Xi$ between
two consecutive zero--crossings of the cantilever with positive velocity. However,
the deflection of the cantilever $q'$ is subject to a noise level $\delta q'$
as shown in Fig. \ref{fmnoise}.
The deflection noise $\delta q'$ has two major contributions:
thermal excitation of the cantilever outside of its resonance frequency and instrumental noise in the measurement of the
deflection $q'$.
The oscillation period $\Xi$ can only be measured with an rms accuracy $\delta \Xi$. The
uncertainty of the time of the zero-crossing is $\delta \Xi/2$,
where $\delta \Xi/2$ is given by the ratio between the cantilever deflection noise
and the slope of the $q'(t)$ curve:
\begin{equation}
\frac{\delta \Xi}{2}= \frac{\delta q'}{2\pi f_{0}A}.  \label{T_error}
\end{equation}
Because $f_{0}=1/\Xi$, $\delta f_{0}/f_{0}=\delta \Xi/\Xi$ and
\begin{equation}
\frac{\delta f_{0}}{f_{0}}= \frac{\delta q'}{\pi A}.  \label{f_error_gen}
\end{equation}
Equation \ref{f_error_gen} only applies to frequency
changes on a timescale of $1/f_0$. When measuring frequency
variations on a longer timescale, more zero crossings can be used
to determine the frequency change and the precision of the
frequency measurement increases. The output of the frequency
detector (phase-locked-loop, see Fig. \ref{fmprinciple}) typically
has a low-pass filter with bandwidth $B_{FM}\ll f_0$, thus the
effective frequency noise is smaller than the value after Eq.
\ref{f_error_gen} (see Ref. \citet{Giessibl:2002bk}). With $\delta
q' = n_{q'} \sqrt{B_{FM}}$, we find
\begin{equation}
\delta f = \frac{n_{q'}}{\pi A}B_{FM}^{3/2}.  \label{f_error_c}
\end{equation}
The scaling law $\delta f \propto B_{FM}^{3/2}$ has first been found
by \citet{Duerig:1997}. The deflection noise density $n_{q'}$
has two major contributors: a) thermal noise of the cantilever
and b) detector noise. Because the two noise sources are
statistically independent, we find
\begin{equation}
n_{q'} = \sqrt{n_{q'\,thermal}+n_{q'\,detector}}
\end{equation}
with
\begin{equation}
n_{q'\,thermal} = \sqrt{\frac{2k_BT}{\pi k f_0 Q}}  \label{thermal_clnoise}
\end{equation}
after \citet{Becker:1985}. The detector deflection noise density $n_{q'\,detector}$
is determined by the physical setup of the deflection sensor and
describes the quality of the deflection sensor. For practical
purposes, it can be assumed to be constant for frequencies around
$f_0$. Good interferometers reach deflection noise densities of
100\,fm$/\sqrt{\textrm{Hz}}$.

In summary, the frequency noise is proportional to the deflection noise density times $B^{1.5}$ and inversely
proportional to the amplitude.
While Eq. \ref{thermal_clnoise} suggests the use of cantilevers with infinitely high $Q$,
Eq. \ref{Adrive} and the discussion after it imply that
$Q$ should not be significantly larger than the ratio between the energy stored in
the cantilever and the energy loss per oscillation cycle due to the tip-sample interaction.
If $Q$ is much higher than this value, controlling the amplitude of the cantilever can
become difficult and instabilities are likely to occur.
Frequency noise is discussed in greater depth and compared to
experimental noise measurements in \citet{Giessibl:2002bk}.

\subsection{Optimal amplitude for minimal vertical noise}

\begin{figure}[h]
  \centering \includegraphics[width=8cm,clip=true]{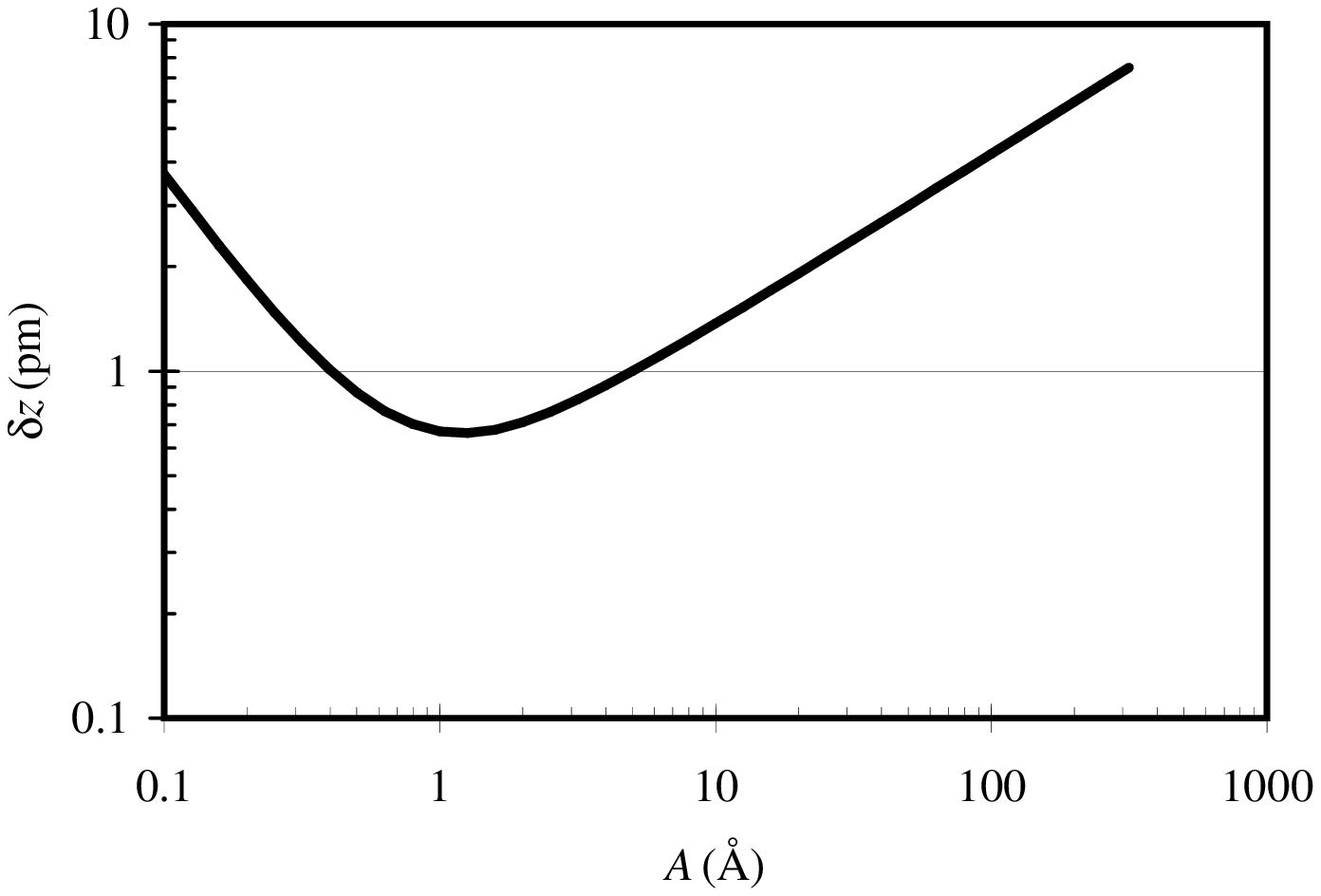}
  \caption{Vertical noise as a function of amplitude for the
  tip-sample potential (Morse type) described in the text. The
  amplitude value where minimal noise results (here
  $A_{optimal}\approx 1$\,\AA{}) is approximately equal to the range
  of the tip-sample force. The absolute noise figure for this optimal
  amplitude is a function of bandwith, noise performance of the
  cantilever deflection sensor and temperature. }\label{znoise}
\end{figure}

Both the nominator and denominator in the generic FM AFM noise
(Eq. \ref{z_noise_fmafm}) are functions of the amplitude -- the
frequency noise is proportional to $1/A$, the slope of the
frequency shift curve is constant at first and drops as $A^{-1.5}$
for large amplitudes. Thus, there is a minimal noise for
amplitudes in the order of the range $\lambda$ of the tip sample
force $F_{ts}$ \citep{Giessibl:1999a}:
\begin{equation}
A_{optimal}\approx \lambda.   \label{optA}
\end{equation}
Here, we calculate the vertical noise for a specific example. We
consider a tip-sample interaction given by a Morse potential with
a depth of -2.15\,eV, a decay length of $\kappa =
1.55\,$\AA$^{-1}$ and an equilibrium distance of $\sigma =
2.35$\,\AA. As a cantilever, we consider a qPlus sensor as shown
in Fig. \ref{qPlus} with $k=1800$\,N/m and $n_{q'}=100$\,fm$/
\sqrt{\textrm{Hz}}$, operated with $B_{FM}=100$\,Hz. Figure
\ref{znoise} shows the vertical noise as a function of amplitude
for a fixed closest tip-sample distance of $z_{min}=4$\,\AA.
Minimum noise occurs for $\log{A/\textrm{m}}=-9.9$, i.e. for $A
\approx 1.26$\,\AA, and for $A=340$\,\AA, the noise is about one
order of magnitude larger. For chemical forces, $\lambda \approx $
1\,\AA. However, operating a conventional cantilever with
amplitudes in the \AA-range close to a sample is impossible
because of the jump-to-contact problem (section
\ref{section_JTC}). The cantilever spring constant $k$ needs to be
at least a few hundred N/m to enable operation with amplitudes in
the \AA-range.

\section{APPLICATIONS OF CLASSIC FREQUENCY MODULATION AFM}
\subsection{Imaging}
\begin{figure}[h]
  \centering \includegraphics[width=8cm,clip=true]{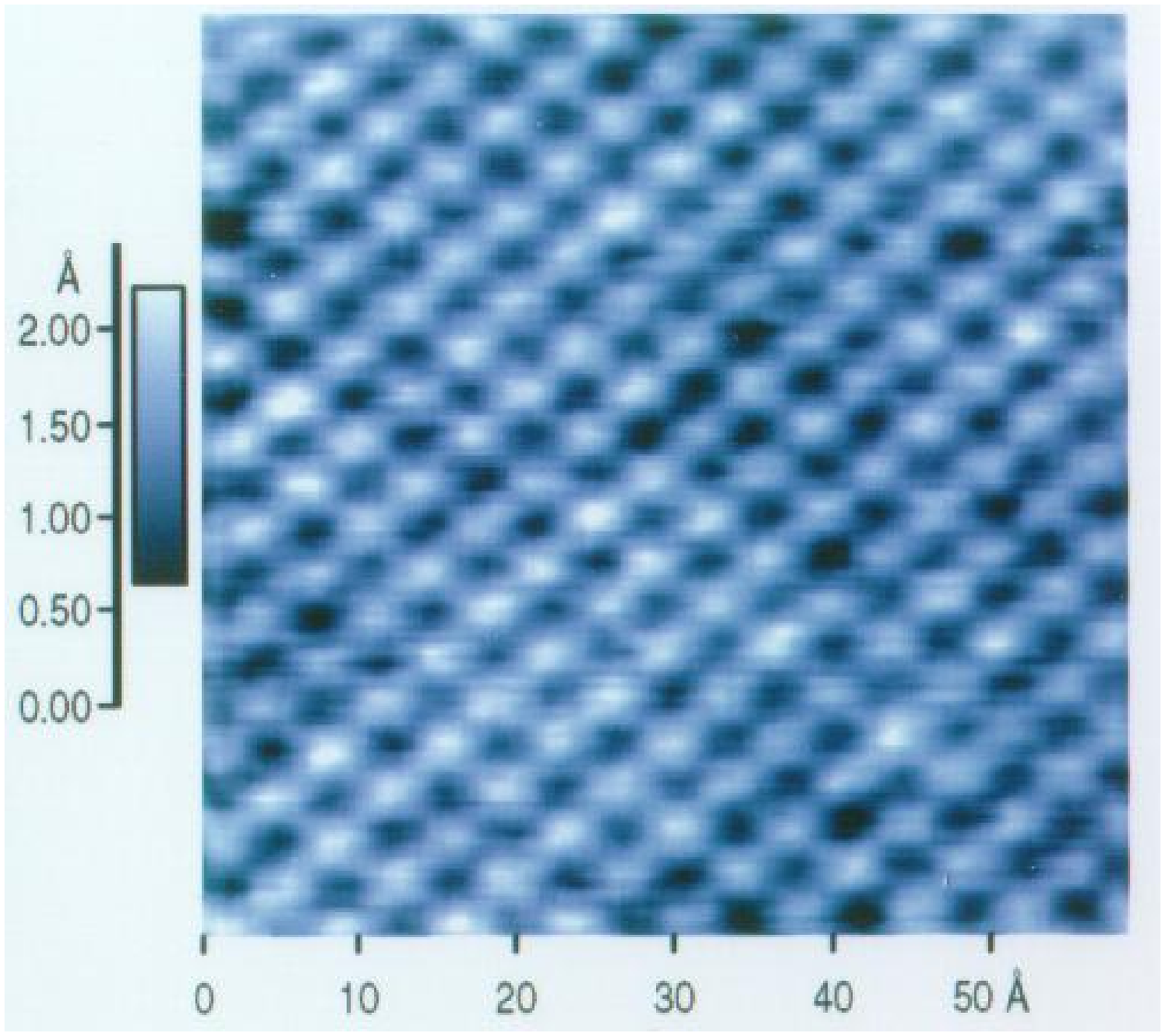}
  \caption{First FM-AFM image of an insulator (KCl) with true atomic
  resolution. Instrument and parameters similar to Fig. \ref{7x7}.
  Source: \citet{Patrin:1995}.}\label{patrin}
\end{figure}

Shortly after the first demonstration of true atomic resolution of
Si by AFM,
\citet{Kitamura:1995,Guethner:1996,Luethi:1996,Nakagiri:1997}
succeeded in imaging Si with atomic resolution using FM-AFM with
similar parameters. In November 1994, \citet{Patrin:1995}
succeeded in imaging KCl, an insulator with FM-AFM (see Fig.
\ref{patrin}). Other semiconductors \citep
{Sugawara:1995,Morita:2002bk3}, more ionic crystals
\citep{Bammerlin:1997,Reichling:1999,Bennewitz:2002bk,Reichling:2002bk},
metal oxides
\citep{Fukui:1997,Raza:1999,Barth:2002bk,Hosoi:2002bk,Pang:2002bk,Fukui:2002bk},
metals \citep{Loppacher:1998,Orisaka:1999}, organic monolayers
\citep{Gotsmann:1998,Yamada:2002bk}, adsorbed molecules
\citep{Sugawara:2002bk,Sasahara:2002bk} and even a film of Xenon
physisorbed on graphite \citep{Allers:1998} have been imaged with
atomic resolution. FM-AFM can also be used for high-resolution
Kelvin probe microscopy by studying the influence of electrostatic
forces on the image \citep{Arai:2002bk,Kitamura:1998}.
Classic frequency modulation AFM provided a new tool to study
problems which were not accessible by STM. \citet{Sugawara:1995}
has imaged defects in InP. While InP can be imaged by STMs, the
bias voltage which is required in STM caused the defects to move.
By FM-AFM, a zero-bias operation is possible which allowed to
study the defects without moving them by the electric field.
\citet{Yokoyama:1999} at the same group imaged Ag on Si by FM-AFM.
\begin{figure}[h]
  \centering
  \includegraphics[width=8cm,clip=true]{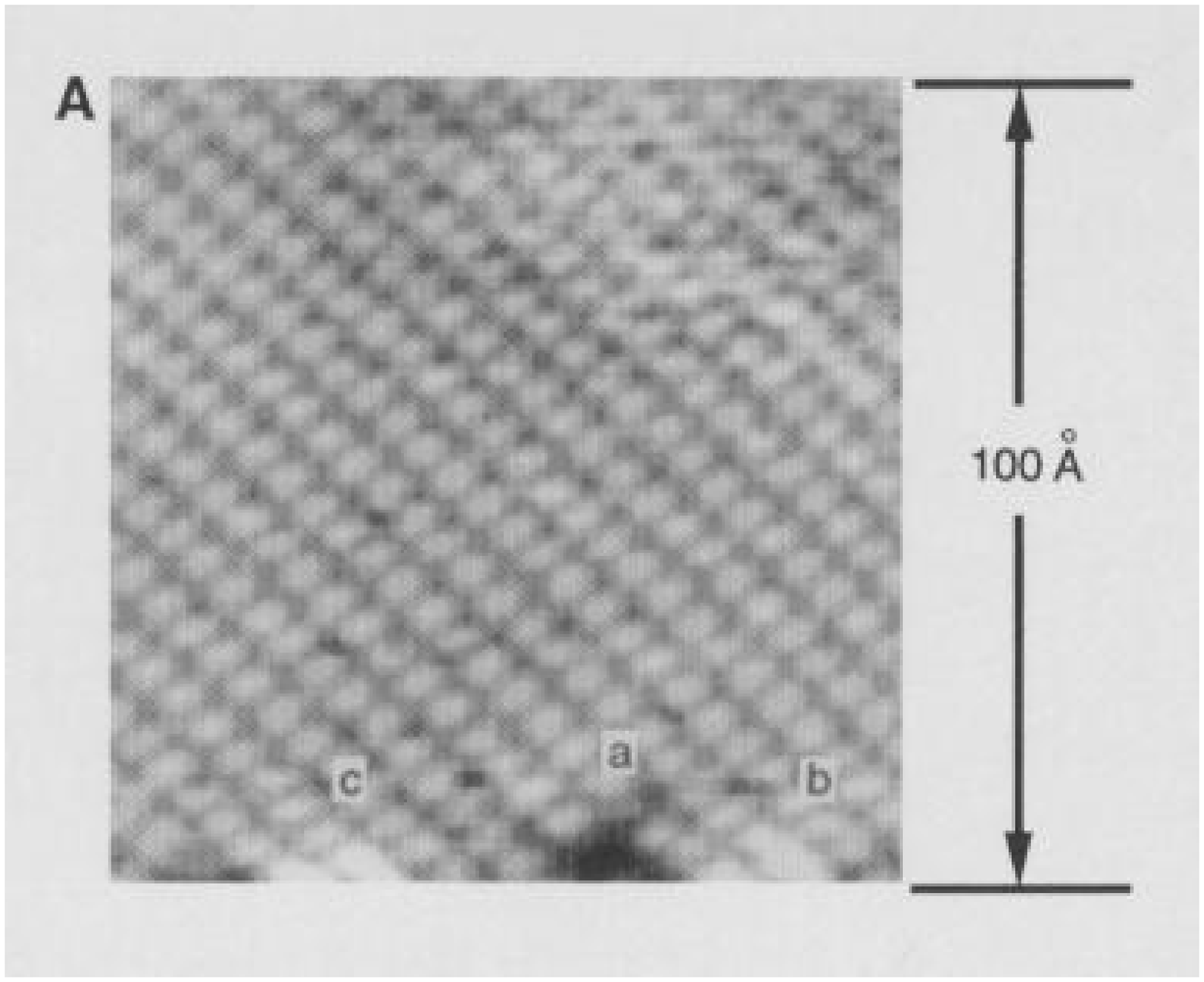}
  \caption{Non-contact UHV AFM image of the cleaved InP(110) surface. The scan area was 100\,\AA{} by 100\,\AA.
  Experimental conditions: spring constant of the cantilever $k=34$\,N/m, mechanical resonant frequency $\nu_0=151$\, kHz,
  vibration amplitude $A=20$\,nm and frequency shift $\Delta\nu=-6$\,Hz. Atomic defects (a) and adsorbates (b and c) are visible.
  After \citet{Sugawara:1995}.}\label{morita}
\end{figure}
\begin{figure}[h]
  \centering
  \includegraphics[width=8cm,clip=true]{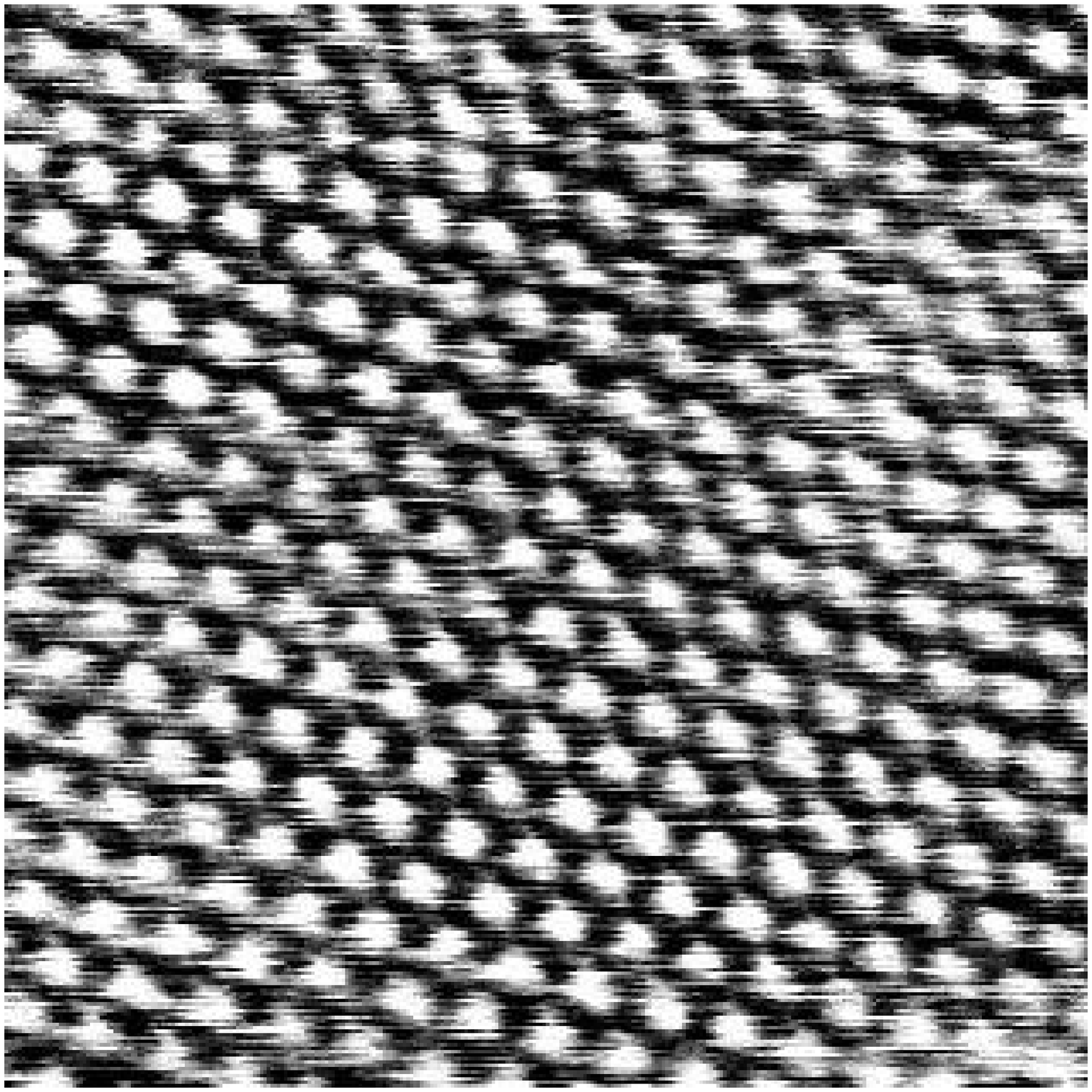}
  \caption{FM-AFM image of a xenon thin film. Image size 70\,\AA $\times$70\,\AA. The maxima correspond to individual Xe atoms.
  Sputtered Si-tip, $f_0 = 160$ kHz, $\Delta f  = -92$ Hz, $A=9.4$\,nm, T = 22\,K, approx. 20\,pm corrugation.
  Source: \citet{Allers:1999b}}\label{allers_xe}
\end{figure}
In Fig. \ref{barth}, the $\alpha$-Al$_2$O$_3$ surface in its
$\sqrt{31}\times\sqrt{31}$\,R\,$+9^\circ$ high temperature reconstruction
is imaged by FM-AFM. This data demonstrates the use of FM-AFM for the
surface science of insulators (see also \citet{Pethica:2001}).
\begin{figure}[h]
  \centering
  \includegraphics[width=8cm,clip=true]{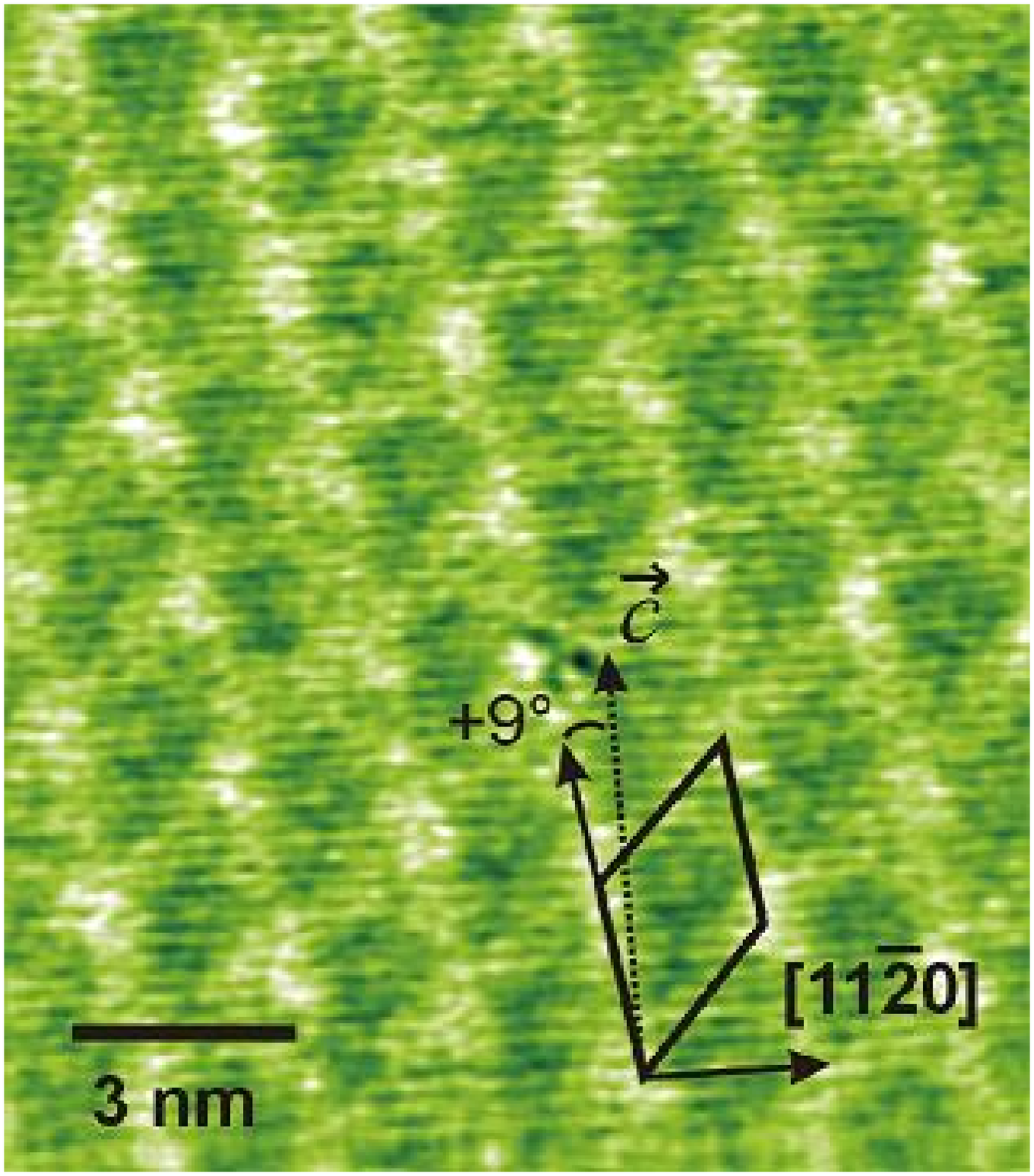}
  \caption{FM-AFM image of an Al$_2$O$_3$ surface.
  Si-tip, $f_0 = 75$ kHz, $\Delta f  = -92$ Hz, $k=3$\,N/m, $A=76$\,nm, ambient temperature.
  Source: \citet{Barth:2001}}\label{barth}
\end{figure}

\subsection{Spectroscopy}

At room temperature, lateral and vertical thermal drift usually
prevents to perform spectroscopy experiments directly over a
specific atom, and frequency versus distance measurements suffer
from thermal drift \citep{Giessibl:1995}. However, at low
temperatures, it is possible to perform spectroscopic measurements
\citep{Allers:2002bk}. \citet{Hoelscher:1999a} have performed
frequency versus distance measurements with a silicon cantilever
and a graphite sample with several different amplitudes and used
their deconvolution algorithm to calculate the tip sample
potential.

\begin{figure}[h]
  \centering
  \includegraphics[width=8cm,clip=true]{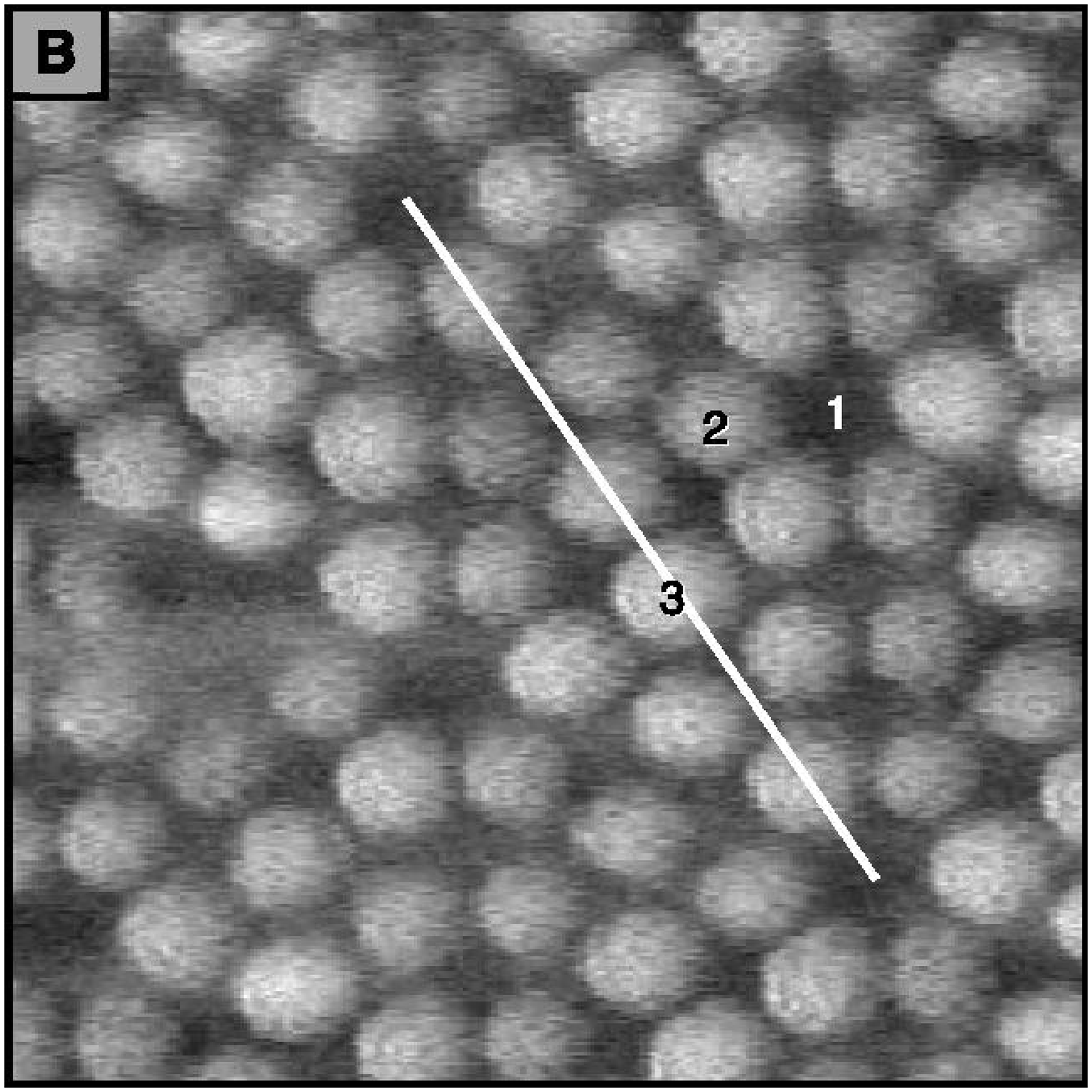}
  \caption{$6$\,nm$\times6$\,nm constant frequency shift image ($\Delta f=-38$\, Hz, rms error 1.15\,Hz, scan speed 2\,nm/s).
  The labels 1, 2, and 3 indicate the position of frequency distance measurements in Fig. \ref{si_spec_graph}. Source: \citet{Lantz:2001}.}\label{si_spec_map}
\end{figure}

\begin{figure}[h]
  \centering
  \includegraphics[width=8cm,clip=true]{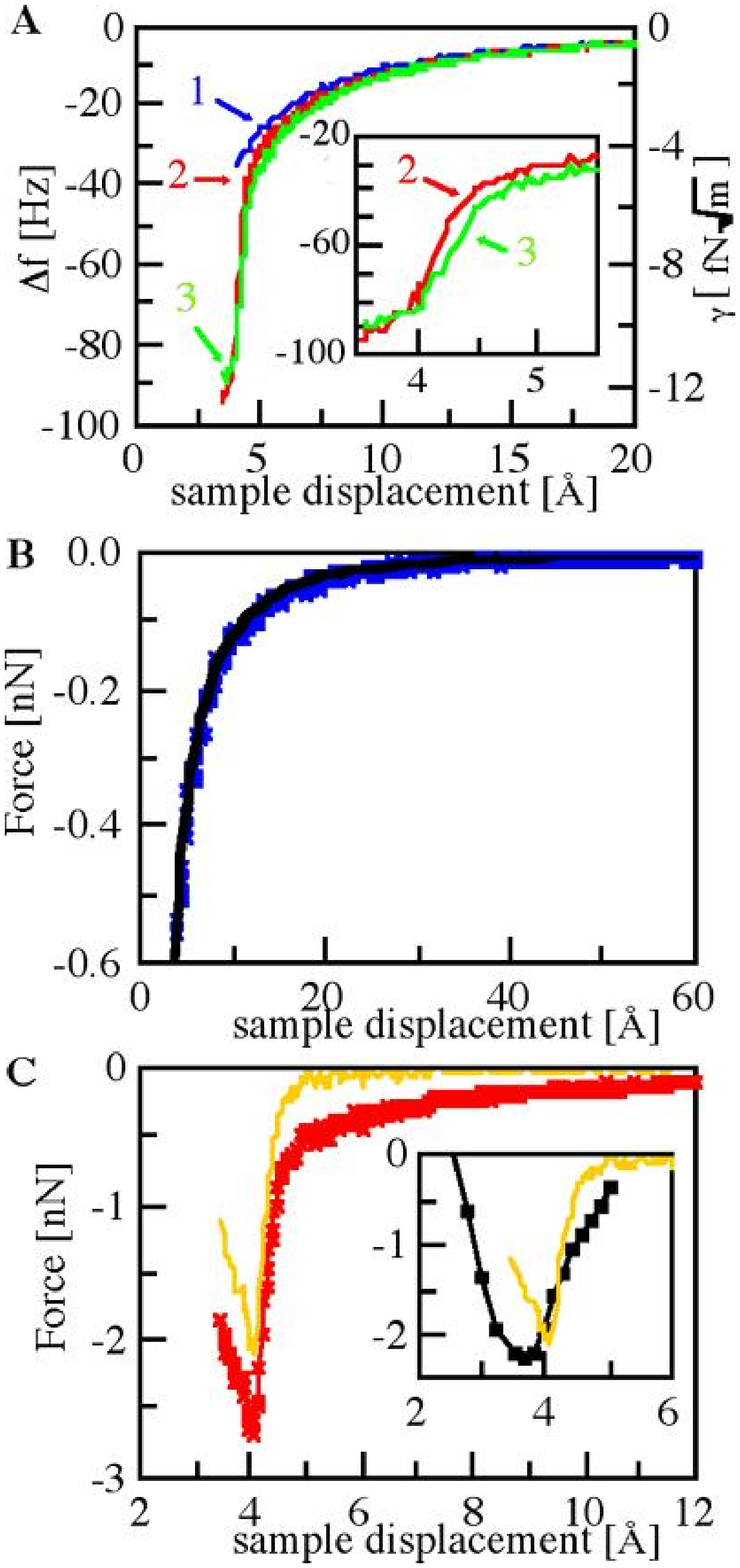}
  \caption{Frequency shift versus distance data measured above the positions labeled 1,2 and 3 in Fig. \ref{si_spec_map}. Source: \citet{Lantz:2001}.}\label{si_spec_graph}
\end{figure}

In 2001, \citet{Lantz:2001} have performed spectroscopy over specific
atomic sites on the silicon surface. Figure \ref{si_spec_map} shows
three distinct sites in the silicon $7\times 7$ unit cell, where
frequency shift data was collected. Figure \ref{si_spec_graph} shows
the corresponding frequency shift data and the corresponding forces,
calculated with the algorithm proposed by \citet{Duerig:1999b}.
This is a significant breakthrough, because the measured forces
are mainly caused by the interaction of two single atoms
(see also the perspective by \citet{deLozanne:2001}).

\section{NEW DEVELOPMENTS}

\subsection{Dissipation measurements and theory}
Already in 1991, \citet*{Denk:1991} used FM-AFM and recorded the
drive signal required to maintain a constant cantilever amplitude.
In the distance regime covered by their early experiment, they
found that the major dissipation mechanism is due to ohmic losses
of currents which are induced by the variable capacitance (due to
oscillation) of the tip-sample assembly in connection with a
constant tip-sample bias voltage. They obtained dissipation images
on semiconductor heterostructures with a feature size of some 10\,
nm and coined the term \lq scanning dissipation microscopy\rq{}.
\citet{Luethi:1997} have recorded the damping signal in atomic
resolution experiments on silicon. Today, a number of theories
have been proposed to explain the energy loss in dynamic AFM
\citep{Gauthier:1999,Duerig:1999a,Sasaki:2000,Kantorovich:2001}. Recently, also
dissipative \emph{lateral} forces have been studied, see
\citet{Pfeiffer:2002} and below.

\subsection{Off-resonance technique with small amplitudes} \label{Pethica_method}
Pethica has early identified the problem of the long-range
background forces and has searched for a way to minimize them (AFM
challenge number 3, see \ref{section_AFM_challenges}). In dynamic
force microscopy, the contribution of various force components
$F_i$ with a corresponding range $\lambda_i$ to the imaging signal
is a function of the cantilever oscillation amplitude $A$. For
$A\gg\lambda$, the imaging signal is proportional to $\sum_i
{F_i\sqrt{\lambda_i}}$, while for $A\ll\lambda$, the imaging
signal is proportional to $\sum_i {F_i/\lambda_i}$, see section
\ref{delta_f_calculation}. Thus, for small amplitudes the imaging
signal is proportional to the force gradient and the weight of
short-range forces is much larger than the weight of long-range
forces. This has been used in an off-resonance technique
by \citet*{Hoffmann:2001}. In this technique, a tungsten
cantilever with $k\approx 300$\,N/m is oscillated at a frequency
far below its resonance frequency with an amplitude $A_0$ of the
order of 0.5\,\AA{}. When the cantilever comes close to the
sample, the oscillation amplitude changes according to
\begin{equation}
A = \frac{A_0} {1+k_{ts} / k}   \label{A_Pethica_method}
\end{equation}
with a tip-sample stiffness $k_{ts}$. Two other AFM challenges,
namely the instability problem and the $1/f$-noise problem are
also solved because of the stiff cantilever and the dynamic mode.
Conceptually, this small amplitude off-resonance technique is very
attractive due to its simplicity in implementation and
interpretation. A lock-in technique can be used to measure $A$,
which improves the signal to noise ratio. The quality of the
images presented is so far not as good as the quality of classic
or small amplitude FM-AFM data, possibly because the scanning
speed is slow and thermal drift is a problem. Ongoing work has to
show whether the image quality issues are just due to technical
imperfections or more fundamental reasons. Atomic dissipation
processes \citep{Hoffmann:2001a} and force versus distance data
\citep{Oral:2001a} have been measured with the technique.

\subsection{Dynamic mode with stiff cantilevers and small amplitudes}
Intuitively, the amplitudes which are used in classic FM-AFM are
much too large: if a silicon atom was magnified to the size of an
orange, the average distance of the cantilever used in classic
non-contact AFM mode would amount to 15\,m. The necessity of such
large amplitudes has been outlined in section
\ref{section_AFM_challenges}. Intuitively, it was clear that
greater sensitivity to short-range forces is achieved with small
amplitudes. It was even planned to use the thermal
amplitude \citep{Giessibl:1994b} to enhance short-range force
contributions. However, empirical findings showed that because of
the stability issues outlined above, large amplitudes had to be
used with the relatively soft cantilevers that were available.
Similar detours were taken in the development of the
STM in several aspects - the first STMs were insulated from
external vibrations by levitation on superconducting magnets and
the first STM tips were fabricated using complicated mechanical
and chemical preparation techniques, while later STMs used much
simpler systems for vibration insulation and tip preparation, see
\citep[p. 59]{Binnig:1997}.

\begin{figure}[h]
  \centering
  \includegraphics[width=8cm,clip=true]{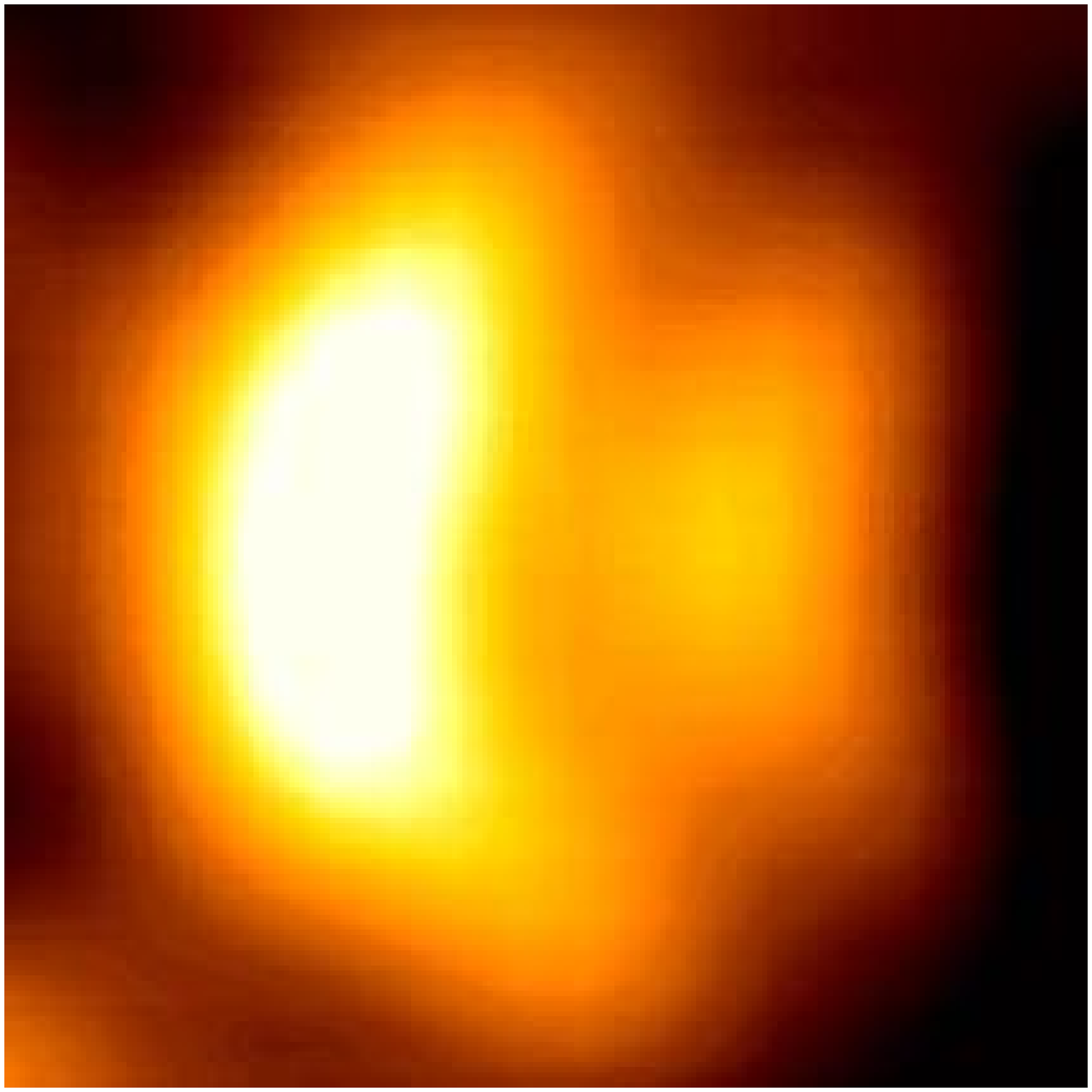}
  \caption{Image of a single adatom on Si (111)-($7\times7$).
  A 3$sp^3$ state points towards the surface normal on the Si (111) surface,
  and the image of this atom should be symmetric with respect to the $z-$axis.
  Because images in AFM are a convolution of tip and sample states, and the sample state is well known
  in this case, the tip state is most likely to be two 3$sp^3$ states
  originating in a single Si tip atom, see \citet{Giessibl:2000c}.
  Image size: 6.6\,\AA{} $\times$ 6.6\,\AA{} lateral, 1.4\,\AA{} vertical.}\label{siatom}
\end{figure}

\begin{figure}[h]
  \centering
  \includegraphics[width=8cm,clip=true]{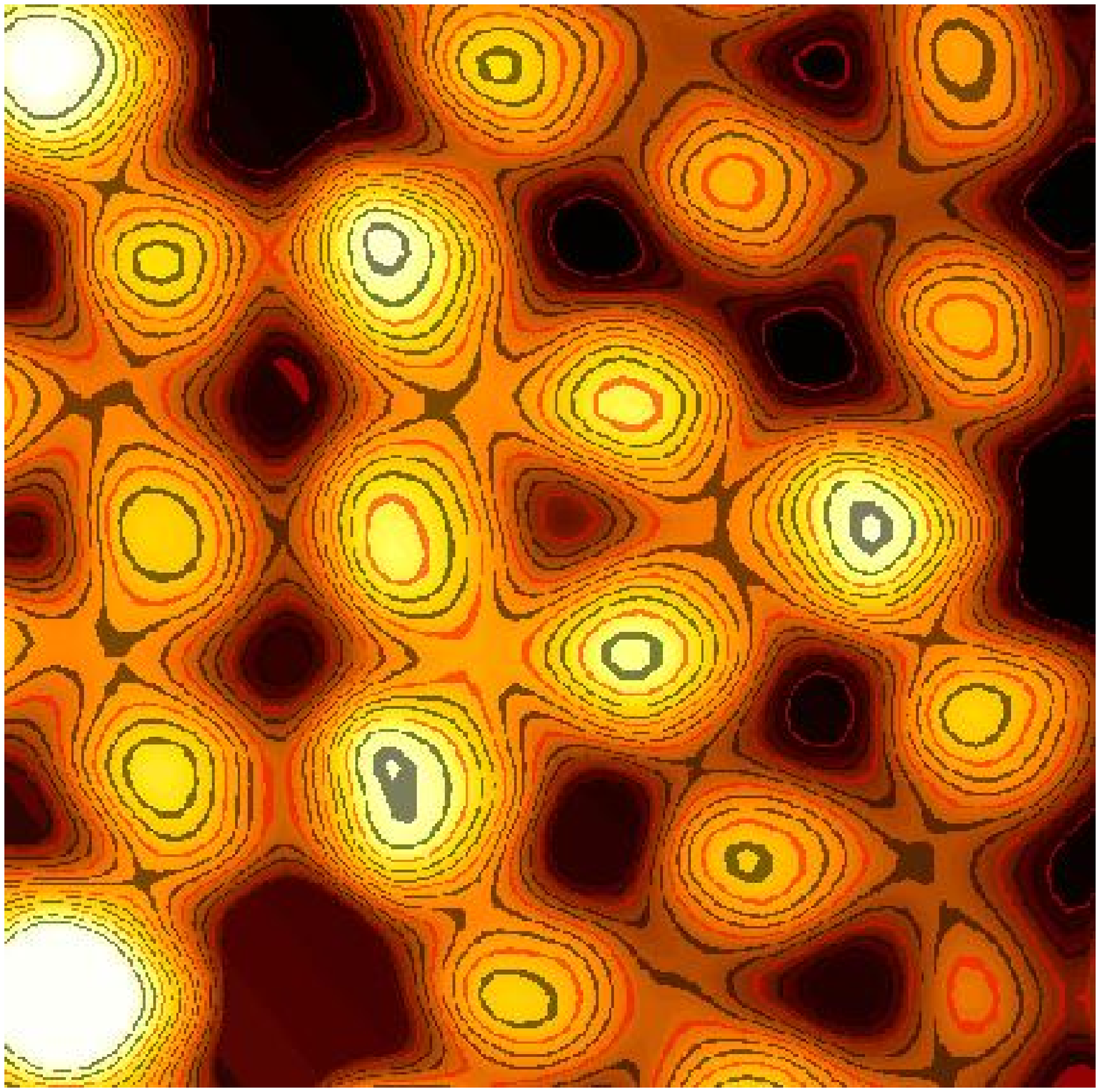}
  \caption{Image of a Si(111)-(7$\times$7) surface imaged with a qPlus sensor. Parameters: $k=1800$\,N/m,
  $A=2.5$\,\AA, $f_0=14772$\,Hz, $\Delta f=+4$\,Hz, $\gamma=28$\,fN$\sqrt{\textrm{m}}$.
  Image size: 40\,\AA{} lateral, 1.4\,\AA{} vertical. Source: \citet{Giessibl:2001d}}\label{111tip7x7.eps}
\end{figure}

In FM-AFM, it was finally shown that small amplitudes do work -
however only if extremely stiff cantilevers are used. After the
theoretical proof of the benefits of using small amplitudes
with stiff cantilevers \citep{Giessibl:1999a}, we
were trying to convince manufacturers of piezoresistive
cantilevers to make devices with $k\approx 500$\,N/m -- without
success. However, the theoretical findings gave us enough
confidence to modify a quartz tuning fork to a quartz cantilever
sensor with a stiffness of roughly 2\,kN/m (qPlus sensor, see Fig.
\ref{qPlus} and \citet{Giessibl:1996,Giessibl:1998}). Already the
first experiments were successful, yielding AFM images of silicon
with excellent resolution. For the first time, clear features
within the image of a single atom were observed (see Fig.
\ref{siatom}). The structure of these images was interpreted to
originate from the orbitals of the tip atom, the first observation
of charge structure within atoms in real space
\citep{Giessibl:2000c}.
Figure \ref{siatom} is an image of a single silicon adatom.
Silicon adatoms display a single $\textrm{sp}^{3}$\,dangling bond
sticking out perpendicular from the surface. Thus, the image of
this atom is expected to be spherically symmetric with respect to
the vertical axis. We interpret the image as being caused by an
overlap of two $\textrm{sp}^{3}$\,dangling bonds from the tip with
the single dangling bond from the surface, for a detailed
description see Refs. \citet{Giessibl:2000c,Giessibl:2001d}. On
the subatomic level, the image is sensitive to the chemical
identity and the structural surroundings of the front atom of the
tip. First attempts to engineer tips with a known symmetry are
under way \citet{Giessibl:2001a}. Figure \ref{111tip7x7.eps} is an
image of a Si(111)-(7$\times$7) surface imaged with a qPlus sensor
with a [111] oriented Si tip (see section \ref{qPlus}) with
extremely small amplitudes (2.5\,\AA) and even positive frequency
shifts, i.e. repulsive forces. The tip was found to be extremely
stable compared to [001] oriented tips.

It is noted, that the claim of subatomic resolution capability is
under debate. \citet{Hug:2001} proposed that the experimental
observations of subatomic resolution could be artifacts due to
feedback errors. However, \citet{Giessibl:2001b} concluded that
analysis of the feedback signals rules out feedback artifacts. So far,
subatomic resolution has not been reported using classic non-contact
AFM. The small amplitude technique with very stiff cantilevers allows
to achieve tip-sample distances close to the bulk distances and
obtains single-atom images with nontrivial internal structures
(subatomic resolution) on silicon \citep{Giessibl:2001d} and
rare-earth metal atoms \citep{Herz:2003}. The enhanced resolution of
short-range forces as a result of using small amplitudes was confirmed
experimentally by \citet{Eguchi:2002}.

\begin{figure}[h]
  \centering,
  \includegraphics[width=8cm,clip=true]{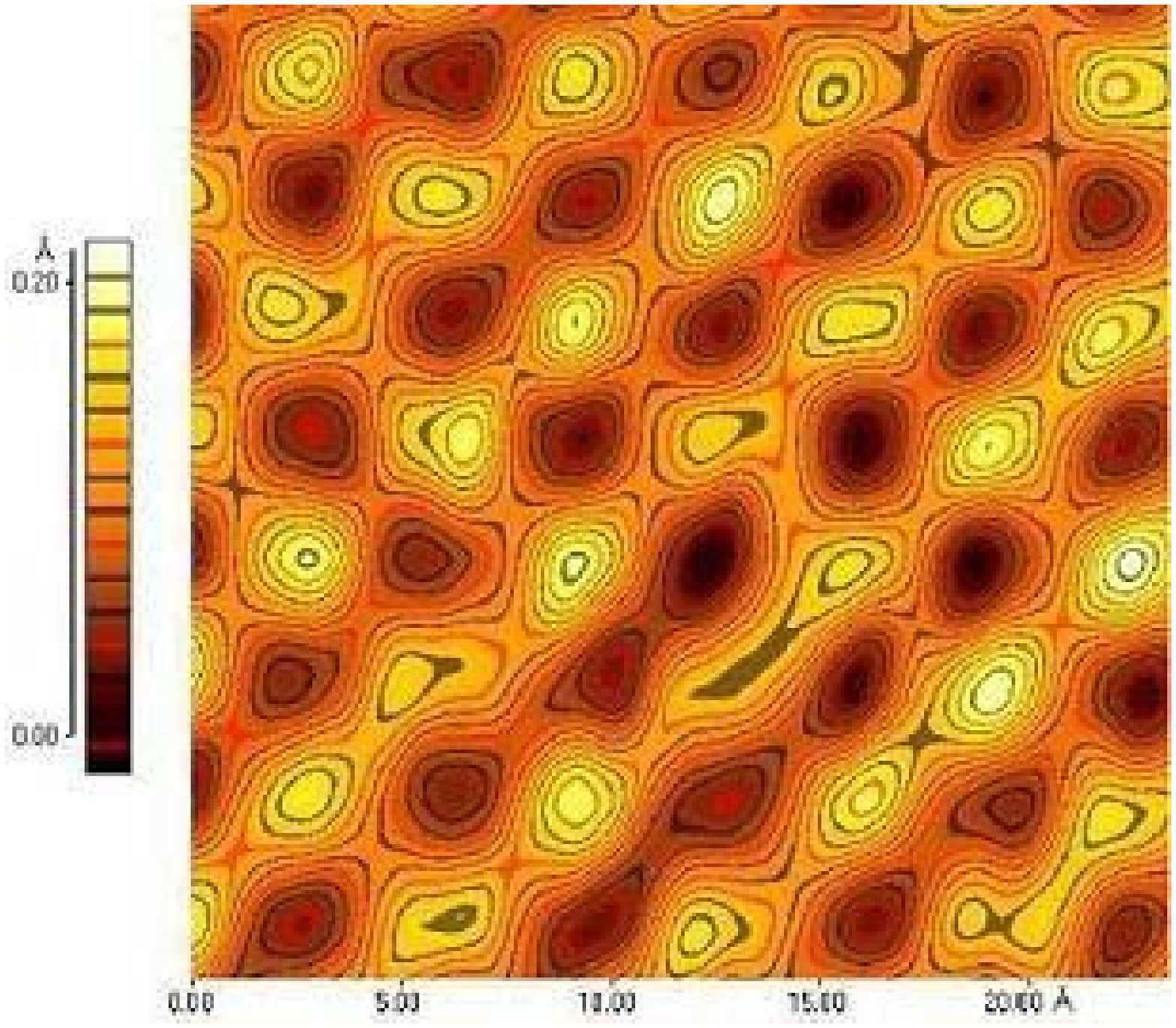}
  \caption{Image of a KCl(001) surface imaged with a qPlus sensor. Parameters: $k=1800$\,N/m,
  $A=9$\,\AA, $f_0=14772$\,Hz, $\Delta f=-4$\,Hz, $\gamma=-13$\,fN$\sqrt{\textrm{m}}$.
  Image size: 23.5\,\AA{} lateral, 0.3\,\AA{} vertical. The contour lines
are spaced vertically by approximately 3\,pm.}\label{kclqplus}
\end{figure}

The capacity of the stiff cantilever -- small amplitude technique
to image standard insulators with moderate short-range forces is
shown in Fig. \ref{kclqplus}, where a KCl (001) surface
is imaged with a qPlus sensor with a silicon tip.

\subsection{Dynamic lateral force microscopy} \label{DLFM}

\begin{figure}[h]
  \centering
  \includegraphics[width=12cm,clip=true]{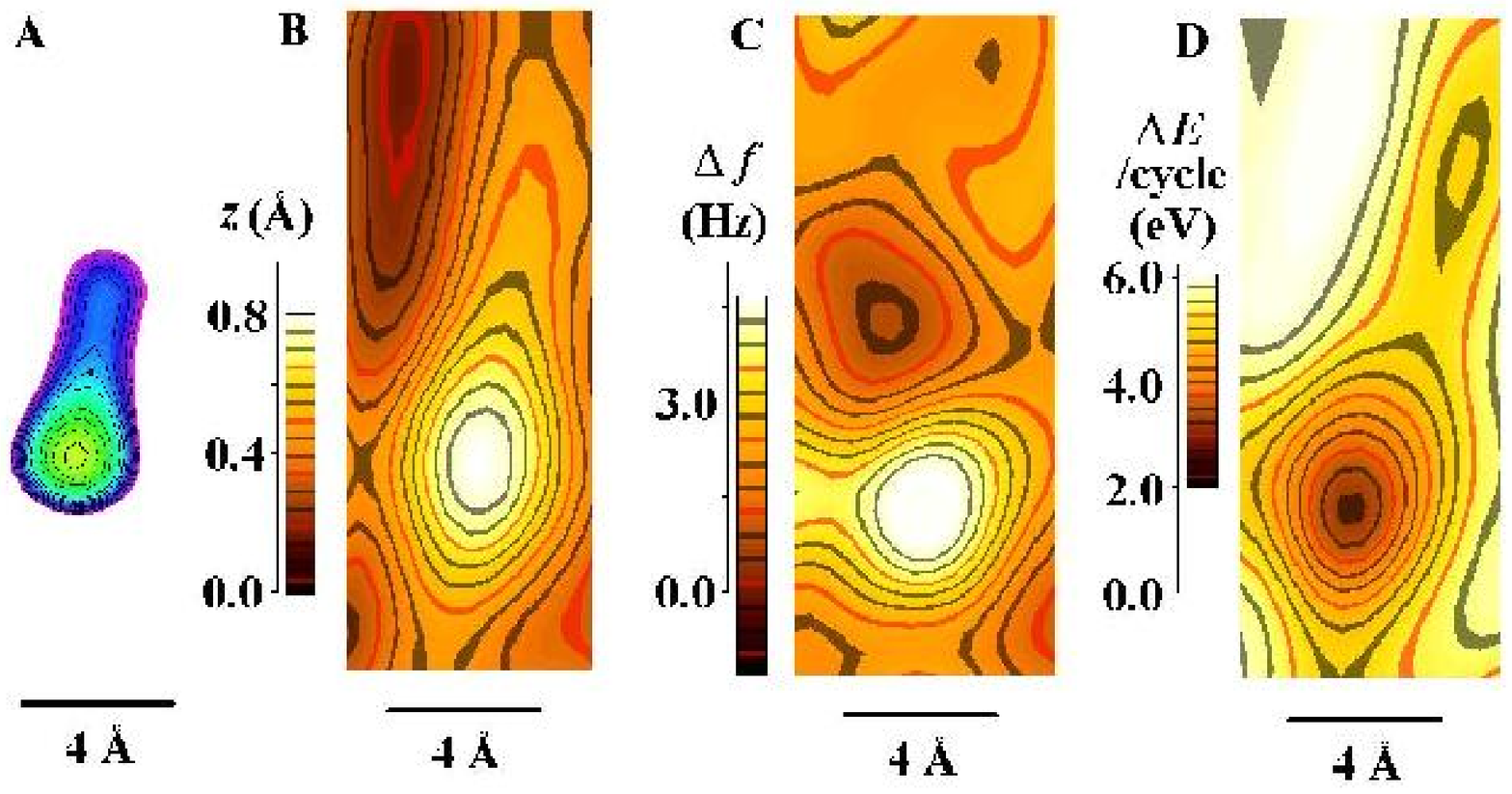}
  \caption{Lateral force microscopy data on a single adatom on Si (111)
  imaged with a qPlus lateral sensor. A) Simulated constant average current
  topographic image, B) Experimental topographic image of a single adatom,
  C) Experimental data of frequency shift, D) Experimental data of
  dissipation energy.}\label{LFMdata}
\end{figure}

Experiments on atomic friction became possible with the invention
of the lateral force microscope, introduced in 1987 by
\citet{Mate:1987}. The resolution power of the lateral force
microscope has been improving steadily, opening many applications
in tribology studies including high-resolution wear studies on KBr
\citep{Gnecco:2002}. However, the observation of single atomic
defects has not been achieved by quasistatic lateral force
microscopy. Because of the similarity of the challenges faced by
normal-force and lateral-force microscopy, the FM method has been
tried and \citet{Pfeiffer:2002} have imaged atomic steps with this
technique, and recently \citet{Giessibl:2002a} have achieved true
atomic resolution with a large-stiffness, small amplitude lateral
FM-AFM (see Fig. \ref{qPluslateral}). In addition to the frequency
shift, the dissipated power between tip and sample has been
measured as the difference between the power required for
maintaining a constant amplitude when the cantilever is close to
the sample and the power required when the cantilever is far away
from the sample, yielding a connection to friction forces. Figure
\ref{LFMdata} shows experimental data on the conservative and
dissipative force components between a single adatom on a Si
surface and a single atom tip. When the cantilever is vibrating
laterally directly over an adatom, almost no extra dissipation
occurs, while when approaching and retracting the tip from the
side, a dissipation of the order of a few eV per oscillation cycle
is measured. The data is interpreted with a theory going back to
\citet{Tomlinson:1929}.

\section{SUMMARY AND CONCLUSIONS}

Imaging flat surfaces with atomic resolution in direct space -
regardless of their electrical conductivity - is standard practice now
thanks to atomic force microscopy. The theoretical understanding of
AFM has been advanced considerably and a direct link between the
experimental observables (frequency shift, damping and average
tunneling current) and the underlying physical concepts (conservative
and dissipative forces) has been established. Forces can be
deconvoluted from frequency shift data easily and low-temperature
spectroscopy experiments show an outstanding agreement of theoretical
and experimental tip-sample forces. AFM yields information about the
strength and geometry of chemical bonds between single atoms.

Four techniques for atomic resolution AFM in vacuum are in use today:
the classic frequency modulation technique with large amplitudes (used
by most experimenters) and soft cantilevers, frequency modulation with
small amplitudes and stiff cantilevers, the off resonance technique
originated by the Pethica group and the amplitude modulation method.
The future will show if one of the techniques will survive as an
optimal method or if all or some techniques will remain in use.
Lateral force microscopy with true atomic resolution has been
demonstrated using the small amplitude/stiff cantilever technique.

\section{OUTLOOK}

While FM-AFM is an established experimental technique, applications in
surface science of insulators are just starting to emerge. AFM is
still more complicated than STM, and these complications appear to
deter many scientists. However, significant progress has been made in
the last years, and exciting results can be expected in the field of
surface science of insulators, where the AFM is a unique tool for
atomic studies in direct space. The possibilities of atomic resolution
AFM are overwhelming: access to the very scaffolding of matter, the
chemical bond. Progress in physical understanding and subsequent
simplifaction of the implementation and interpretation AFM has been
significant over the last years. AFM offers additional observables -
forces and damping - that are even vectors with three spatial
components - the tunneling current in STM is a scalar entity. There
are strong indications that dynamic AFM (and STM) allows the imaging
of features within single-atom images attributed to atomic valence
orbitals. Thus the whole field of classic STM studies could be
revisited with the enhanced resolution technique.

One of the greatest challenges of AFM is the preparation of well
defined tips. Like in the early days of STM, tip preparation is a
black magic with recipes ranging from sputtering to controlled
collisions. Because the tip is closer to the sample in AFM then in
STM, the stability of the front atom is more important in AFM than
in STM. Also, the chemical identity and backbond geometry of the
tip is crucial in AFM. The use of nanotubes as tips appeared to be
promising, however, the observations expressed in Figure 4 of
\citet{Binnig:1987} regarding the necessity of a rigid,
cone-shaped tip are to be heeded especially in AFM, where forces
between tip and sample are larger than in STM.

Atomic manipulation by scanning probe microscopes is an exciting
challenge. Arranging atoms on conducting surfaces has been possible by STM
since a decade
\citep{Eigler:1990,Crommie:1993,Manoharan:2000,Kliewer:2000,MeyerG:1996}.
\citet{Morita:2002bk3} have successfully extracted single atoms
from a Si(111)-($7\times7$) surface in a controlled fashion
with a force microscope.
An exciting extension of this work would be possible if atoms
could also be deposited with atomic precision by AFM, because the
construction of e.g. one dimensional conductors or semiconductors
on insulating substrates would allow to build electronic
circuits consisting only of a few atoms. While it appears
difficult to manipulate atoms with a probe which oscillates to and
from the surface with amplitudes of the order of 10\,nm, the
continuous decrease of amplitudes used in dynamic AFM might allow
to move atoms in a controlled manner by AFM in the future.

\section*{ACKNOWLEDGMENTS}
The author thanks Jochen Mannhart for his longtime collaboration,
enthusiastic support and inspiring discussions. Special thanks to
all current and former colleagues in Augsburg for their
contributions: H. Bielefeldt, Ph. Feldpausch, S. Hembacher, A.
Herrnberger, M. Herz, U. Mair, Th. Ottenthal, Ch. Schiller and K.
Wiedenmann.

Many thanks to G. Binnig, Ch. Gerber and C. F. Quate, who have
contributed in many ways to the authors work on the AFM.

The author is indebted to the following colleagues for supplying
figures and for discussions: T. Akitoshi, T. Albrecht, W. Allers,
A. Baratoff, C. Barth, G. Binnig, U. D\"{u}rig, R. Erlandsson, Ch.
Gerber, H.-J. G\üntherodt, L. Howald, H. Hug, M. Lantz, R.B.
Marcus, E. Meyer, O. Ohlsson, M. Reichling, C. F. Quate, A.
Schwarz, U. D. Schwarz, M. Tortonese, R. Wiesendanger.

L. Howald and D. Br\ändlin from Nanosurf AG have made crucial
contributions to our project by providing prototypes and final
versions of their frequency-to-voltage converters and fruitful
discussions about FM-AFM.

Special thanks to G. Binnig, K. Dransfeld, U. D\ürig, S. Fain, S.
Hembacher and J. Mannhart for editorial suggestions.

This work is supported by the Bundesministerium f\ür Bildung und Forschung (project no. 13N6918).

\bibliography{rmp_cond_mat}

\end{document}